\documentclass[fleqn,usenatbib]{mnras}

\usepackage{newtxtext,newtxmath}
\usepackage{xspace}
\usepackage[T1]{fontenc}

\DeclareRobustCommand{\VAN}[3]{#2}
\let\VANthebibliography\thebibliography
\def\thebibliography{\DeclareRobustCommand{\VAN}[3]{##3}\VANthebibliography}

\newcommand{\teff}{\mbox{$T_{\rm eff}$}\xspace}

\newcommand{\logg}{\mbox{$\log(g)$}\xspace}

\newcommand{\vmic}{\mbox{$v_\mathrm{mic}$}\xspace}
\newcommand{\vsini}{\mbox{$v \sin i$}\xspace}

\usepackage{graphicx}	
\usepackage{amsmath}	
\usepackage{xcolor}
\usepackage{subcaption}
\usepackage{array}
\usepackage{makecell}
\usepackage{csvsimple}
\usepackage{pdflscape}
\usepackage{adjustbox}

\usepackage[normalem]{ulem}

\usepackage{booktabs}

\title[Residual Abundance Signatures]{A Low-Dimensional Chemical Representation of High-Resolution Stellar Disk Spectra: Residual Abundance Signatures}

\author[Aquilina et al.]{
Sarah E. Aquilina$^{1}$,
Melissa K. Ness$^{1}$ and
Sven Buder$^{1}$\\
$^{1}$School of Astronomy and Astrophysics, The Australian National University, ACT 2611, Australia\\
}
\date{Accepted XXX. Received YYY; in original form ZZZ}

\pubyear{2025}

\begin{document}
\label{firstpage}
\pagerange{\pageref{firstpage}--\pageref{lastpage}}
\maketitle

\begin{abstract}
Stellar abundances are generated by complex chemical enrichment pathways yet show strong correlation structure. We test how well high-quality spectra of low-$\alpha$ disk stars can be reconstructed using low-dimensional models. By validating model performance in flux space, we quantify and locate where residual information persists. Using HARPS spectra ($R=115{,}000$, SNR $>100$) and labels derived with \textsc{korg}, we train generative models (\textsc{the cannon}) conditioned on stellar parameters \teff,~\logg,~[M/H],~\vsini,~and~\vmic, and three abundance labels, [Ti/H],~[Y/H],~and~[Nd/H], representing distinct nucleosynthetic families. Evaluating model performance on unseen test stars through flux residuals, the model reproduces the global spectra to a median residual amplitude of 0.3\%. Deviations in 15 absorption features for unlabelled elements are typically $\leq 0.7\%$. Residuals of $\sim1\%$ correspond to abundance differences of 0.01--0.05~dex. Degrading the spectra to medium resolution ($R=28{,}000$) decreases the median residual amplitude from 0.3\% to 0.2\%, redistributing the residual signal over the instrumental line profile. Our publicly available model reproduces features from 29 elements to sub-percent precision across $-1.13 < \text{[M/H]} < 0.29$. This demonstrates that the dominant chemical variation encoded in low-$\alpha$ disk spectra can be captured by a low-dimensional model, while weaker chemical information remains encoded in flux residuals. High-resolution, high-SNR spectra ($R\sim100{,}000$, SNR $>100$) are essential to resolve the hierarchical layers of chemical information beyond this low-dimensional structure. The model provides a benchmark for assessing the chemical information accessible to current surveys like 4MOST and GALAH 2, as well as upcoming facilities like VLT/HRMOS and GMT/G-CLEF. 
\end{abstract}


\section{Introduction}
\label{Section: introduction}

The field of Galactic archaeology has been driven by orbit–age–abundance mapping as well as chemical tagging—using element abundances to identify birth stellar populations that are now kinematically separated \citep{freeman_new_2002} Chemical tagging relies on different progenitor clouds having unique and distinguishable chemical compositions \citep[][]{mitschang_2013}. However, birth sites can have overlapping chemical signatures \citep[eg.][]{blanco_tagging_2015,armillotta_2018,Casamiquela_tagging_2021}. Additionally, the Milky Way disk shows a high "chemical doppelganger rate", which is the rate at which pairs of random field stars are as chemically similar as pairs of conatal stars \citep[eg.][]{ness_doppelgangers_2018,Manea_2024,Manea_2025}. This derives from the rigid coupling of the elements \citep[][]{mead_regression_2025} whereby the elements are highly covariant and a few abundances capture the dominant structure \citep[][]{weinberg_chemical_2022, griffith_2025, ness_homogeneity_2022}. Correlations between elements reduce the effective uniqueness of the fingerprints of individual stellar clusters, and render some elements more useful in discriminating between birth sites than others \citep[][]{blanco_tagging_2015,lambert_2016,armillotta_2018}. However, even small signatures captured in deviations away from low-dimensional abundance models may be key for learning detailed nucleosynthesis \citep{ting_how_2022, griffith_residual_2022, griffith_2025}. Furthermore, inter-element correlations themselves are an important signature of enrichment \citep{zhang_2025} and neutron capture elements in particular offer insights not captured by well-studied iron-peak and $\alpha$-element channels \citep[e.g.][]{casali_gaia_clocks_2020, Manea_2024}. 
\\\\
Due to the highly correlated abundance space and subsequent high precision needed to access each element's independent information, the irreducible dimensionality of Chemical space \citep[or C-space, ][]{freeman_new_2002}, and the precision required to fully capture stellar formation history remain unknown. Although primary abundance trends trace broad population evolution, intrinsic scatter of a population of stars in any element is driven by a combination of generating physical variables. These include stochastic enrichment events and interstellar gas mixing, stellar multiplicity, mass ingestion or transfer and possibly planet host status \citep[although see][]{behmard_2023,rampalli_2024,rampalli_2026,babatsikos_2026}. Defining the operational limit between measuring these signatures and localised scatter that can not be used to trace stellar history remains an ongoing task. Previous works have demonstrated that the overall dimensionality is low, although hierarchical, with progressively smaller amplitudes of information encoded in additional dimensions. Recent studies using linear regression \citep[][]{ness_homogeneity_2022,mead_regression_2025} and k-process models  \citep[][]{weinberg_chemical_2022, griffith_similarity_2021,griffith_kpm_2023} have shown that as few as $2-4$ elemental labels (e.g., Fe, Mg, Y, and Ce to represent iron-peak elements, $\alpha$-elements, light s-process elements and heavy s-process elements respectively) are sufficient to predict the abundances of most other species within the Milky Way's stellar disk to within $0.01-0.05$~dex. However, the precise number of dimensions reported depends on the quantity being measured and the abundance precision achieved \citep[][]{casey_2019}. Although a small number of labels can capture the majority of abundance information, additional dimensions can still encode the lower-amplitude structure and residual correlations between elements \citep{ting_how_2022, griffith_2025}. For example, approaches such as Principal Component Analysis (PCA) may require up to $\sim10$ components to account for smaller-scale abundance variance and reduce residual correlations \citep[][]{ting_2012, Price-Jones_2017}; although not all of these components are likely inherited from abundance information. Additional components often come at the expense of direct physical interpretability or might be necessary to capture possibly deviating patterns beyond the disk.
\\\\
To pursue the idea of using the element composition of stars for chemical tagging, abundance precisions better than $2$--$5\%$ are typically required \citep[][]{mead_regression_2025}. Moreover, this level of precision is necessary to undertake a fundamental assessment of the irreducibility of abundance space. Large medium-resolution surveys such as APOGEE \citep[$R=22\,000$,][]{majewski_apache_2017}, GALAH \citep[$R=28\,000$,][]{DeSilva_2015,buder_galahdr4_2024}, and 4MOST \citep[high-resolution arm at $R=20\,000$,][]{dejong_4most_2019} can achieve this precision for approximately 15 elements \citep[][]{mead_regression_2025}. Smaller samples of high-resolution spectra \citep[e.g., HARPS at $R=115\,000$,][]{pepe_harps_2000} can extend this precision to 20-30 elements. High-resolution spectroscopy is particularly beneficial for resolving subtle variations in neutron-capture elements (e.g., Zr, Y, Ba, La, Ce, Nd, and Eu), which are expected to provide important information for chemical tagging and assessing the intrinsic information captured in each element and its origin \citep{lambert_2016,armillotta_2018,monty_2024,Manea_2025}. Even if the inherent similarity of abundances at a fixed birth radius as well as comparable intra- and inter-cluster abundance scatter limits the absolute effectiveness of chemical tagging, high-precision abundance measurements remain essential. Precise measurements of an ensemble of elements provide stronger constraints for reconstruction of the Galaxy's enrichment history, strengthen connections between observed abundances and their nucleosynthetic sources and can reveal the true information content encoded in stellar abundance measurements.
\\\\
Most studies of chemical dimensionality and information content have been performed using measured elemental abundances. Here, instead, we work directly in spectral space to assess both the information content of element absorption features conditioned on a subset and the generative success of a low-dimensional abundance basis for stellar spectra. Working directly in spectral space provides a more direct approach for examining residual structure beyond low-dimensional generative models of the abundances, while enabling a direct estimate of the spectral quality required, i.e. the signal-to-noise (SNR) needed to detect departures from these conditional models. This approach also reduces assumptions introduced through derived abundances. To do this, we use \textsc{the cannon 2} \citep[][]{casey_cannon_2016}. \textsc{the cannon} \citep{ness_cannon_2015} is a data-driven modeling approach that determines stellar labels from spectra using a reference set of stars with continuum normalised radial velocity spectra and known labels -- these labels include stellar parameters (effective temperature, surface gravity, overall metallicity, microturbulence, \vsini) and individual element abundance labels. Using the derived labels for any spectrum, the best-fit model spectrum can be generated using the model. The underlying assumption is that stars with identical labels will have nearly identical spectra, with small variations arising from observational uncertainties and intrinsic scatter. This generative framework allows us to examine how well models of different label sets reconstruct a stellar spectrum, and via residuals between observed spectra and model, quantify unmodelled information in any given basis. Characterising the information content of stellar spectra in this way can inform future observing strategies and set expectations for the information recoverable from large spectroscopic surveys.
\\\\
To examine the underlying information captured in stellar spectra and evaluate how accurately low-dimensional models reconstruct true spectral features, we utilise high-resolution HARPS observations and chemical abundances from the catalogue of \citet{aquilina_2026}. This dataset comprises 426 red giant stars observed with the HARPS spectrograph at an average SNR of 225. In \citet{aquilina_2026} we derived fundamental stellar parameters ($\teff$, $\logg$, $\text{[M/H]}$, $\vmic$, and $\vsini$) and 22 elemental abundances via $\chi^2$ minimization using the spectral synthesis code \textsc{korg} \citep[][]{wheeler_korg_2023}. Line-by-line measurements were combined to determine a median abundance for each element, achieving an average precision of 0.02~dex across the sample. 
\\\\
The elements in the \citet{aquilina_2026} catalogue span a range of families and include neutron-capture elements identified as crucial for resolving chemical variations: odd-Z elements Na, Al and K; $\alpha$-elements Mg, Si, Ca and Ti; iron-peak elements Sc, V, Cr, Fe, Ni and Zn; s-process elements Sr, Y, Zr, Ba, La, Ce and Nd; r-process elements Mo and Eu. These element families carry varying information due to their different production sites and timescales--Type Ia supernovae \citep{keegans_1asne_2023}, core-collapse supernovae \citep{nomoto_ccsne_2006}, low-intermediate mass asymptotic giant branch (AGB) stars \citep{busso_agb_review_1999,herwig_agb_2005}, neutron star mergers \citep{kobayashi_neutron_2023}, magnetars \citep{patel_magnetar_2025} and Wolf-Rayet stars \citep{peng_wolfrayet_2019}--and metallicity dependencies. Note that most elements have multiple production site and the element families reflect their dominant sources \citep[see][]{kobayashi_origin_2020,johnson_origin_2020}. This provides us with a range of high-precision chemical abundances across element families to use as labels for training \textsc{the cannon}. These precise abundances will allow us to assess the amplitude of additional information remaining after conditioning on a subset of elements directly in spectral space rather than abundance space. \citet{aquilina_2026} also identify the elements in their catalogue that carry the most discriminating information, which we use to inform our choice of training labels.
\\\\
This paper is structured as follows. In Section \ref{Section: method}, we describe our data and implementation of \textsc{the cannon 2} across K = 5-, 8-, and 9-label models. Our K-label model comprises the set of 5 stellar parameters (\teff, \logg, [M/H], \vmic and \vsini) and for the K = 8 and 9 label model, 3 element abundance measurements ([Ti/H], [Y/H] and [Nd/H]) and 4 element abundance measurements ([Ti/H], [Y/H], [Nd/H] and [Eu/H]), respectively. We describe the  details of the reference set from \citet{aquilina_2026} as well as our chosen training and test set inputted into \textsc{the cannon}. The label transfer precision and spectral reconstruction performance are evaluated in Section \ref{sec:results}, where we quantify recovery rates for measured reference lines and identify unmodelled features via residual analysis. We provide additional label and model validation in Appendix \ref{appendix:supplementary label validation} and \ref{appendix:label spectra reconstruction}. In Section \ref{sec:discussion}, we convert residuals at specific absorption features into abundance differences and explore the physical limitations of the 8-label \textsc{cannon} model. We further assess the predictive power of Eu (II) as a label in a 9-label \textsc{cannon} model. Finally, we simulate our model at medium-resolution ($R = 28,000$) of major existing (GALAH) and upcoming (4MOST) surveys to demonstrate how the loss of spectral fidelity masks subtle nucleosynthetic variations resolvable by HARPS.

\section{Data and Methods}
\label{Section: method}

To construct our \textsc{cannon} models, Section \ref{sec:data preparation} first details the processing of stellar spectra from the \citet{aquilina_2026} catalogue to form our reference set. We strictly follow their data preparation steps to ensure our recovered labels and spectra remain directly comparable to this reference. In Section \ref{sec:label selection}, we select the specific parameters and abundances used to train the model. This choice of labels dictates the physical and chemical trends the model can capture, whether through correlations with fundamental stellar parameters or shared nucleosynthetic pathways. To guarantee robust spectral reconstructions, Section \ref{sec: method-training and testing} outlines our validation framework: we select approximately 50\% of the reference stars to train the models—ensuring they completely span the parameter space—and reserve the remaining 50\% for independent testing. Finally, Section \ref{method: the cannon} describes the formal mathematical construction and training of the models.

\subsection{Data Preparation}
\label{sec:data preparation}
The full set of stars that we use to build and test our models consists of 426 red giant stars with derived stellar parameters (\teff, \logg, [M/H], \vmic \& \vsini) and 22 elemental abundances from the catalogue of \citet{aquilina_2026}. We refer to this complete set as the reference set of stars from which we both build and test the model. These targets were originally selected from the \citet{barbieri_esoharps_2023} HARPS catalogue based on Gaia DR3 colour cuts. The HARPS spectra are processed following the methodology of \citet{aquilina_2026}: we use the radial velocity values derived in \citet{aquilina_2026} to shift the individual exposures to the rest frame before interpolating them onto a common wavelength grid spanning 4000~\AA{} to 6900~\AA{} in 0.01~\AA{} steps. On average, each star has 10 observations with SNR = 98 which we stack by summing the flux per pixel and continuum normalise using \textsc{suppnet} \citep{tomasz_suppnet_2022}. The per-pixel flux uncertainty ($\sigma_{\lambda}$) for the combined spectrum of each star is estimated assuming Poisson statistics and scaling to the normalised flux, 

\begin{equation}
    \sigma_{\lambda} = \frac{\mathrm{flux_{norm,\lambda}}}{\sqrt{\mathrm{flux_{\lambda}}}},
\end{equation}

\noindent
where $\mathrm{flux_{\lambda}}$ is the flux per unit wavelength of the combined spectra for each star (where flux units are in Analog to Digital Units) and $\mathrm{flux_{norm,\lambda}}$ is the normalised flux per unit wavelength. The resulting combined spectra per star has an average SNR of 225. 
\\\\
For targets with hundreds of observations (reaching SNR ratios up to 2885 only assuming Poisson noise), the formal flux uncertainties become unrealistically small. We establish a more realistic error floor by empirically measuring the standard deviation of the continuum flux across 0.1~\AA{} continuum regions. This obtains a baseline of 0.002. We adopt a minimal uncertainty of 0.3\% to conservatively encompass this variation in continuum flux. We apply this error floor to all stars by adding a constant term of 0.003 in quadrature to the formal pixel uncertainties ($\sigma_{\lambda}$):

\begin{equation}
\label{eqn:inflated error}
    \sigma_{\mathrm{inflated},\lambda} = \sqrt{\sigma_{\lambda}^2 + 0.003^2}. 
\end{equation}

\noindent
\citet{aquilina_2026} included the same inflation factor for their flux errors when deriving their stellar parameters and element abundances. 
\\\\
We exclude wavelength regions prior to training around prominent telluric features \citep[$5860-6000$~\AA, $6276-6326$~\AA, $6456-6601$~\AA{} and $6866-6900$~\AA,][]{baker_telluric_2020} and the HARPS CCD gap ($5300-5340$~\AA) to ensure the model does not learn from systematic artifacts. We also exclude wavelengths $\lambda < 4300$~\AA{} since we do not have light element labels to reconstruct the prominent molecular bands in this region. While there are numerous valuable neutron-capture element absorption features in this blue wavelength region, the severe line crowding and blending make it difficult to confidently attribute flux residuals between the data and model reconstructed spectra to specific molecular or atomic features. Furthermore, we find that including wavelengths $\lambda < 4300$~\AA{} does not change or improve label or spectral recovery of our models. Instead, the unmodeled features inflate the model scatter in this region ($\sim0.02$), making it approximately double the scatter observed in redder regions (eg., $\sim0.01$ around the Mg I b triplet, see Appendix \ref{appendix:label spectra reconstruction}).

\subsection{Label Selection}
\label{sec:label selection}
We construct \textsc{cannon} models with 5-, 8- and 9- labels. The 5-label model consists purely of stellar parameters ($T_{\mathrm{eff}}$, $\log g$, [M/H], $v_{\mathrm{mic}}$, $v\sin i$) as labels.  The 8-label model includes element abundances [Ti/H] (measured using Ti I lines), [Y/H] (measured using Y II lines) and [Nd/H] (measured using Nd II lines) as labels. The 9-label model includes an additional abundance label [Eu/H] (measured using Eu II lines). These elements are used to trace the elements within their respective families with [Ti/H] to trace $\alpha$-elements, [Y/H] to trace light s-process elements, [Nd/H] to trace heavy s-process elements, and [Eu/H] to trace r-process elements.  We select [Ti/H] as our $\alpha$-element tracer because it exhibits the highest ratio of elemental dispersion to measurement uncertainty (discriminating power) among the $\alpha$-elements analysed by \citet{aquilina_2026}. Although [Y/H] and [Nd/H] do not possess the highest discriminating power within their families, their abundance trends across [Fe/H] in \citet{aquilina_2026} show strong agreement with the main sequence trends used for comparison. We construct an additional 9-label \textsc{cannon} model with [Eu/H] to test its predictive power across our parameter space given previous works have noted the decoupling of r-process elements from other nucleosynthetic families at [Fe/H]~$> -2$~dex \citep[][]{soker_2026,farouqi_2025,mead_regression_2025}.

\subsection{Constructing the Training and Test Sets}
\label{sec: method-training and testing}

\begin{figure}
	\includegraphics[width=\columnwidth]{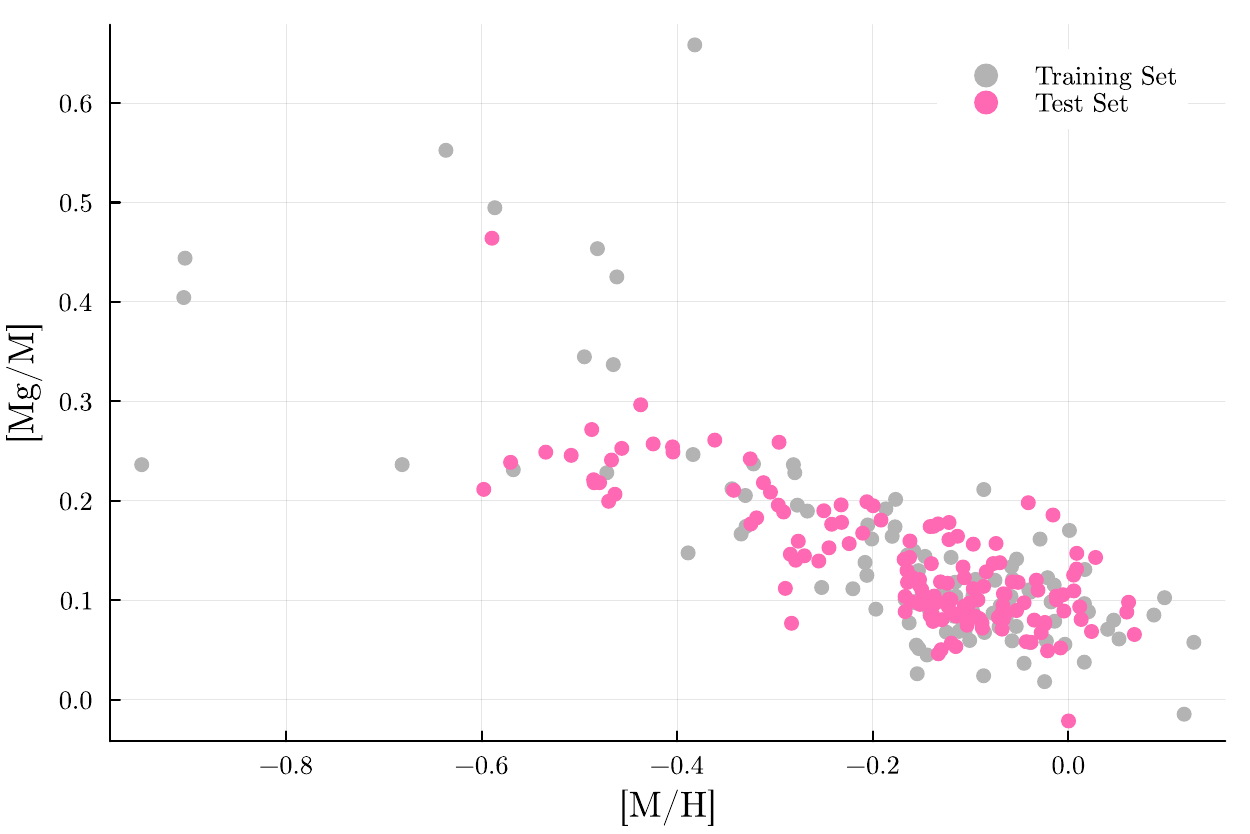}
    \caption{The [M/H]-[Mg/M] plane for the training and test set of stars (as indicated in the legend). This shows that most of the reference set of stars are constrained to the low-$\alpha$ disk.}
    \label{fig:Mg/M trend}
\end{figure}

\begin{figure}
	\includegraphics[width=\columnwidth]{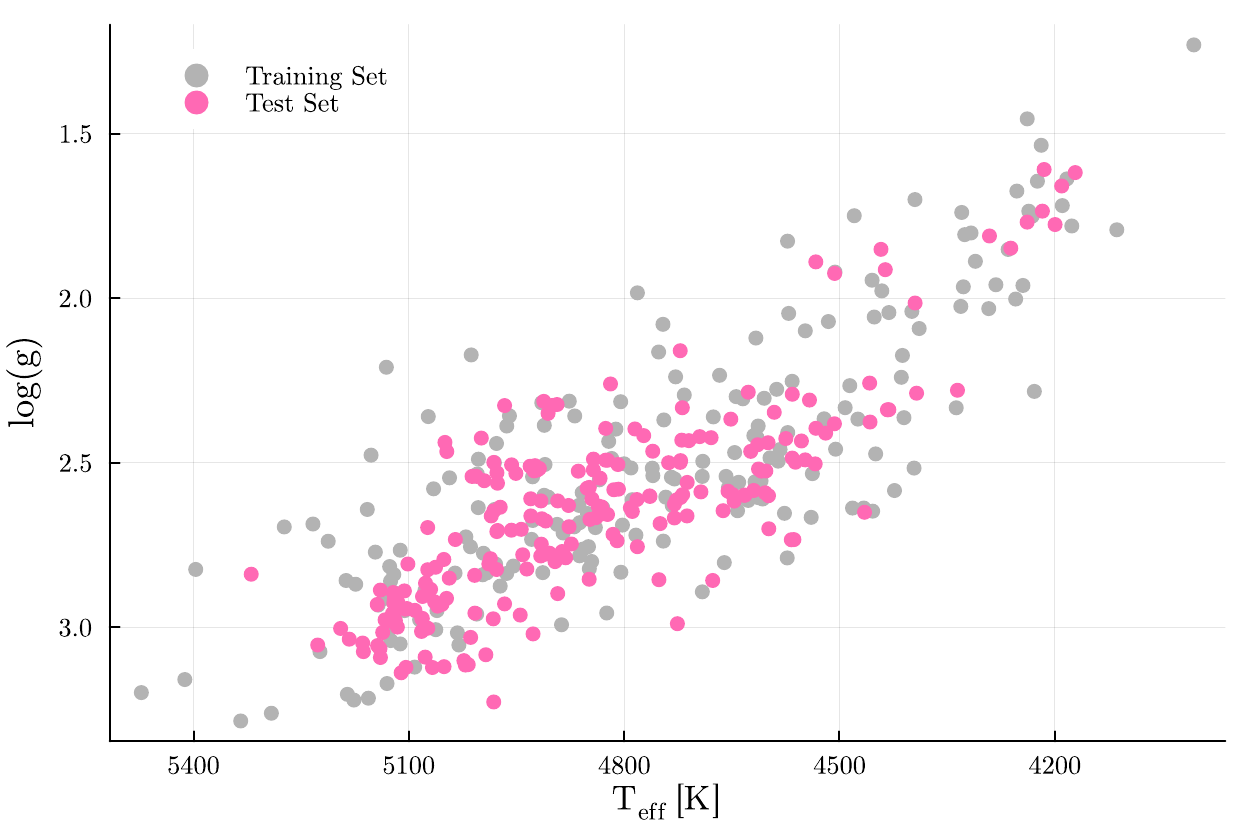}
    \caption{Kiel diagram showing the parameter space spanned by the training and test set across \teff-\logg.}
    \label{fig:set kiel}
\end{figure}

\begin{figure}
	\includegraphics[width=\columnwidth]{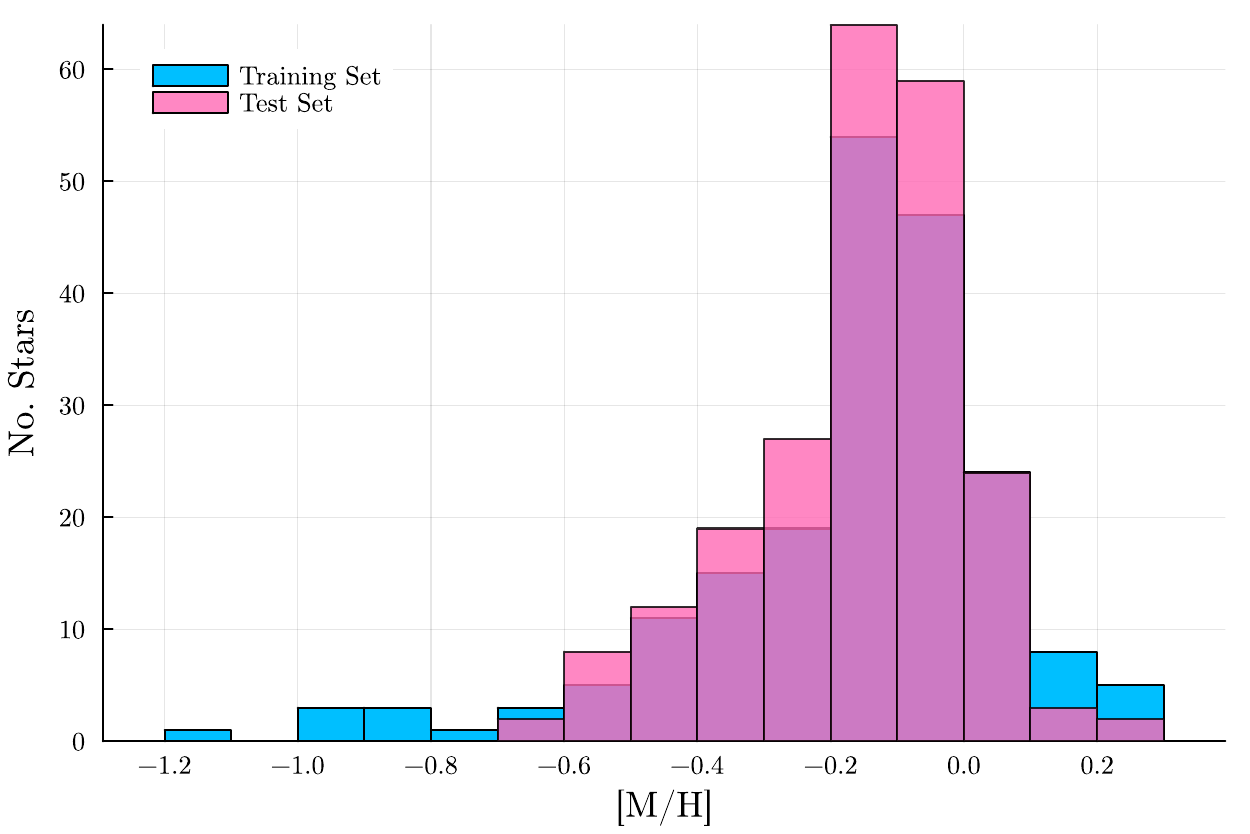}
    \caption{Training and test set distributions for the label [M/H].}
    \label{fig:label histograms mh}
\end{figure}

\begin{figure}
	\includegraphics[width=\columnwidth]{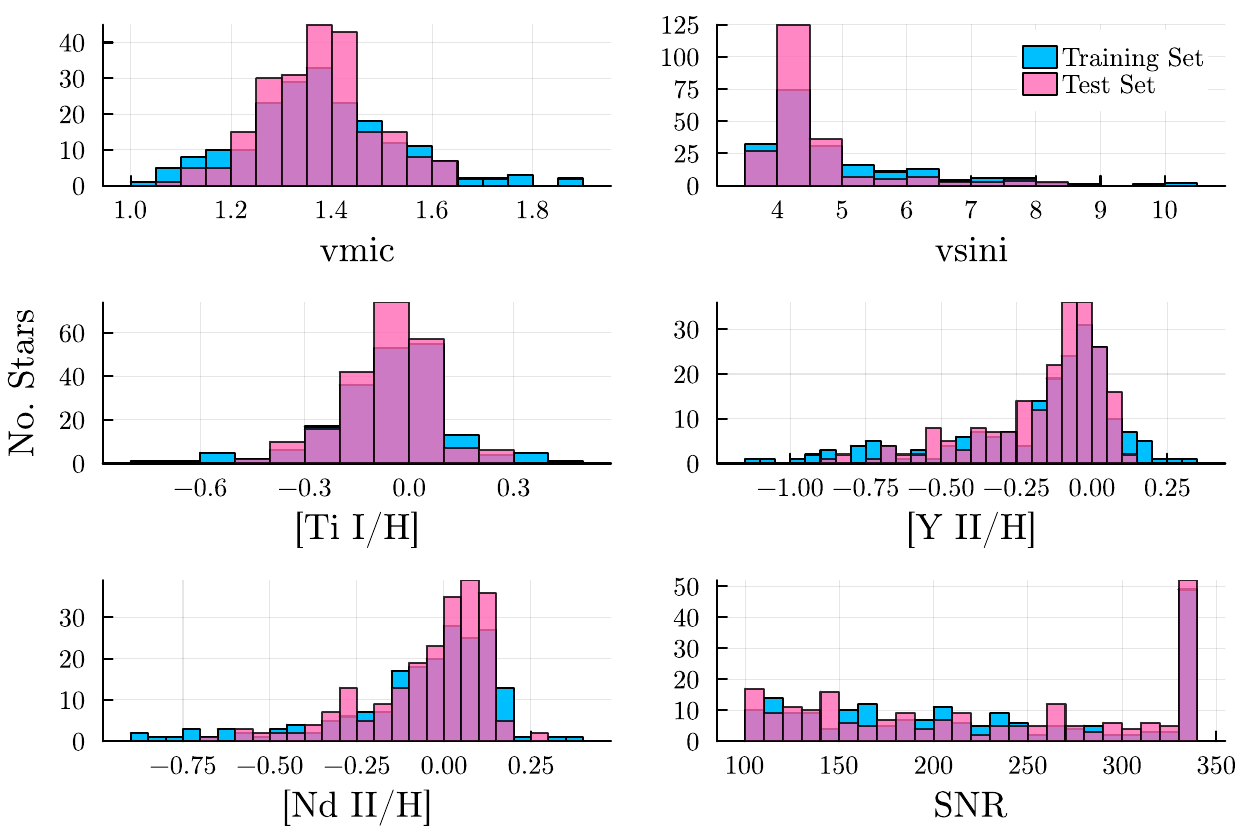}
    \caption{Training and test set distributions for the training labels \vmic, \vsini, [Ti/H], [Y/H] \& [Nd/H]. We also include the SNR coverage of the training and test set, for reference. }
    \label{fig:label histograms abundances}
\end{figure}

To construct \textsc{cannon} models and generate robust reconstructions, we must divide the reference set from \citet{aquilina_2026} into distinct training and test sets. The training set will be used to construct \textsc{the cannon} models, and the test set will be the stars for which we infer labels and reconstruct their spectra. We use the same training and test sets for the 5-, 8- and 9-label models.
\\\\
Prior to defining our training and test set, we remove 7 stars from the overall reference set. These stars exhibit a saw-tooth pattern across their normalised flux that correspond to the individual diffraction orders of HARPS, which could introduce systematics into \textsc{the cannon} models. From the remaining stars in the reference set, we select 199 red giant stars (approximately 50\% of the reference set) to form the training set. This selection was restricted to stars with complete labels, including stellar parameters (\teff, \logg, [M/H], \vmic \& \vsini) and measured abundances for [Ti/H], [Y/H], [Nd/H] and [Eu/H]. To ensure uniform coverage of our parameter space, the training stars are drawn from within evenly spaced bins with widths of $\Delta T_{\mathrm{eff}} = 250$~K, $\Delta \log g = 0.5$~dex, $\Delta$[M/H] $= 0.3$~dex, $\Delta v_{\mathrm{mic}} = 0.3$~$\mathrm{km\,s^{-1}}$, $\Delta v\sin i = 0.3$~$\mathrm{km\,s^{-1}}$, $\Delta$[Ti/H] $= 0.3$~dex, $\Delta$[Y/H] $= 0.3$~dex, and $\Delta$[Nd/H] $= 0.2$~dex. One star is randomly drawn from each bin that is populated. We also show that most of our reference stars lie in the low-$\alpha$ disk in Figure \ref{fig:Mg/M trend}.
\\\\
The test set consists of the remaining 220 stars from the reference set. This group includes stars that either lack a measurement for at least one of the abundance labels or were not selected during the binning process. The parameter space covered by each label in the training set, which fully encapsulates the test set, is summarised in Table \ref{tab:training_space}. The coverage of the training and test set in each label and SNR is shown in Figures \ref{fig:set kiel} to \ref{fig:label histograms abundances}. Note that while the input catalogue provides abundances relative to iron ([X/Fe]), we convert these to Hydrogen-scaled abundances ([X/H]). 

\begin{table}
    \centering
    \begin{tabular}{l c c c l}
        \toprule
        Label & Min & Max & Median $\sigma$ & Unit \\
        \midrule
        SNR & 100 & 333 & --- & --- \\
        \teff & 4006 & 5472 & 21 & K \\
        \logg & 1.23 & 3.28 & 0.04 & dex \\
        $[\mathrm{M/H}]$ & $-1.13$ & $0.29$ & 0.01 & dex \\
        \vmic & 1.01 & 1.88 & 0.07 & $\mathrm{km\,s^{-1}}$ \\
        \vsini & 3.56 & 10.10 & 0.04 & $\mathrm{km\,s^{-1}}$ \\
        $[\mathrm{Ti/H}]$ & $-0.74$ & $0.43$ & 0.01 & dex \\
        $[\mathrm{Y/H}]$ & $-1.15$ & $0.31$ & 0.02 & dex \\
        $[\mathrm{Nd/H}]$ & $-0.90$ & $0.36$ & 0.01 & dex \\
        $[\mathrm{Eu/H}]$ & $-0.63$ & $0.25$ & 0.02 & dex \\

        \bottomrule
    \end{tabular}
    \caption{The labels used for training and the corresponding parameter/label space of the 199 red giant stars in the training set, including the median uncertainty ($\sigma$) for each label.}
    \label{tab:training_space}
\end{table}

\subsection{The Cannon}
\label{method: the cannon}

We utilise \textsc{the cannon} \citep{ness_cannon_2015}, specifically the version by \citet{casey_cannon_2016}. \textsc{the cannon} is a generative model that describes the probability density of the flux at every pixel in the spectrum as a function of the labels. 
\\\\
First, we must train \textsc{the cannon} models using the labels and spectra of the training set described in Section \ref{sec: method-training and testing}. This training set is approximately 50\% of the full reference set. We adopt a quadratic label vector for each reference object $n$, $\boldsymbol{\ell}_n$, which for a set of $K$ labels ($K=5$ for stellar parameters only, or up to $K=9$ when including four abundance labels) takes the form:

\begin{equation}
\boldsymbol{\ell}_n = \left[ 1, l_{n1}, \dots, l_{nK}, l_{n1}^2, \dots, l_{nK}^2, l_{n1} l_{n2}, \dots, l_{n(K-1)} l_{nK} \right].
\end{equation}

\noindent
\\\\
The models predict the flux $f_{n\lambda}$ for stars $n$ at wavelength $\lambda$ as a function of the stellar labels $\ell_n$, plus a noise term:

\begin{equation}
f_{n\lambda} = \boldsymbol{\theta}^T_{\lambda} \cdot \boldsymbol{\ell}_n + noise,
\end{equation}

\noindent
where $\boldsymbol{\theta}_{\lambda}$ is the vector of spectral coefficients at wavelength $\lambda$. The noise is assumed to be Gaussian and combines the measurement uncertainty ($\sigma_{n\lambda}$) with the intrinsic scatter of the model at each wavelength ($s_{\lambda}$):

\begin{equation}
    \mathrm{noise} = (\sigma_{n\lambda}^2 + s_{\lambda}^2)\,\xi_{n\lambda},
\end{equation}

\noindent
where $\xi_{n\lambda}$ denotes a Gaussian random number with zero mean and unit variance. The scatter term captures the variance in the flux that is not explained by the labels or the observational noise, effectively identifying wavelengths where the physical model is insufficient or where unknown systematics exist. 
\\\\
The coefficients $\boldsymbol{\theta}_{\lambda}$ and scatter $s_{\lambda}$ are determined during the training phase by maximising the likelihood function across all reference stars simultaneously for each pixel. The trained model is subsequently applied to the remaining test set stars (the remaining 50\% of stars in the reference set not used during training) to infer their labels and reconstruct their spectra. Finally, although the \citet{casey_cannon_2016} code supports $L_1$ regularisation (sparsity), we choose not to include this term in our objective function.

\section{Results}
\label{sec:results}

As detailed, we construct 3 \textsc{cannon} models: a 5-label with only stellar parameters as labels to provide a baseline, an 8-label with the stellar parameters and three abundance labels (Ti, Y and Nd), and a 9-label model of the stellar parameters with four abundance labels to test the spectral reconstructive power of Eu. We briefly compare these models, and focus our analysis on the 8-label model. For the 8 label model we examine, in detail the label and spectrum recovery. We test the spectra recover both globally and around specific absorption features for 29 elements (with five additional ionisation states) in Sections \ref{sec: results-label accuracy} to \ref{sec: spectrum validation}. Upon examining the 8-label spectral reconstructions we identify five outlier stars which we discuss separately in Section \ref{sec:comp 5 and 8 model}. We include additional label validation in Appendix \ref{appendix:supplementary label validation} where we show the abundance trends of our labels as derived from HARPS in \citet{aquilina_2026} and the trends of the recovered labels from the 8-label \textsc{cannon} model. The model validation where we inspect the linear coefficients of each label and show residuals across the entire spectrum for a typical star in Appendix \ref{appendix:label spectra reconstruction}.  

\subsection{Label Transfer Validation}
\label{sec: results-label accuracy}
\begin{figure*}
	\includegraphics[width=\textwidth]{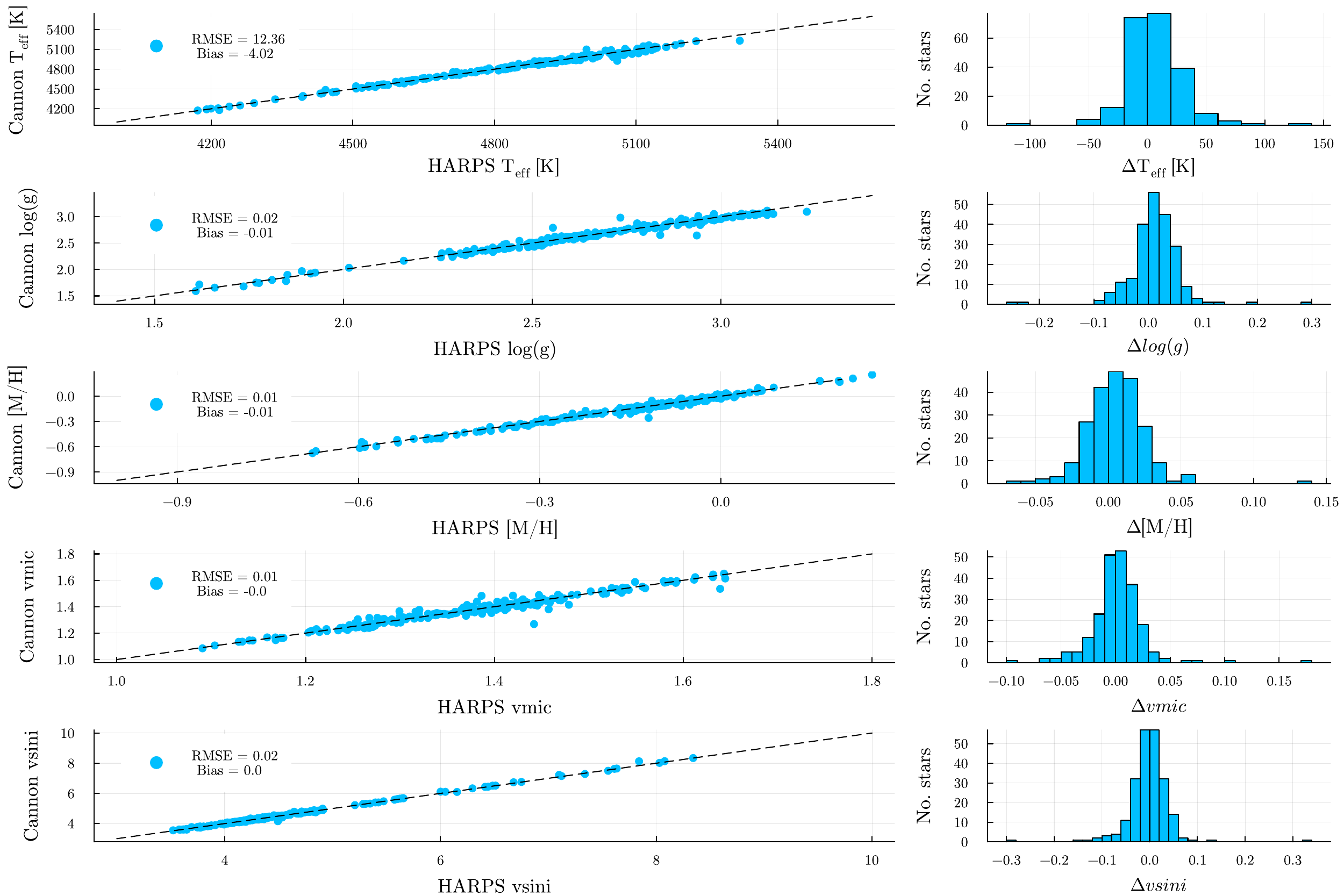}
    \caption{One-to-one plots of the reference set stellar parameter labels and \textsc{the cannon} labels from the 8-label model. We include a distribution of the residuals (HARPS $-$ \textsc{cannon}) to the right. The small global biases for each label indicate that there are no significant systematics and the RMSE values are comparable to measurement uncertainties.}
    \label{fig:param one-to-one}
\end{figure*}

\begin{figure*}
	\includegraphics[width=\textwidth]{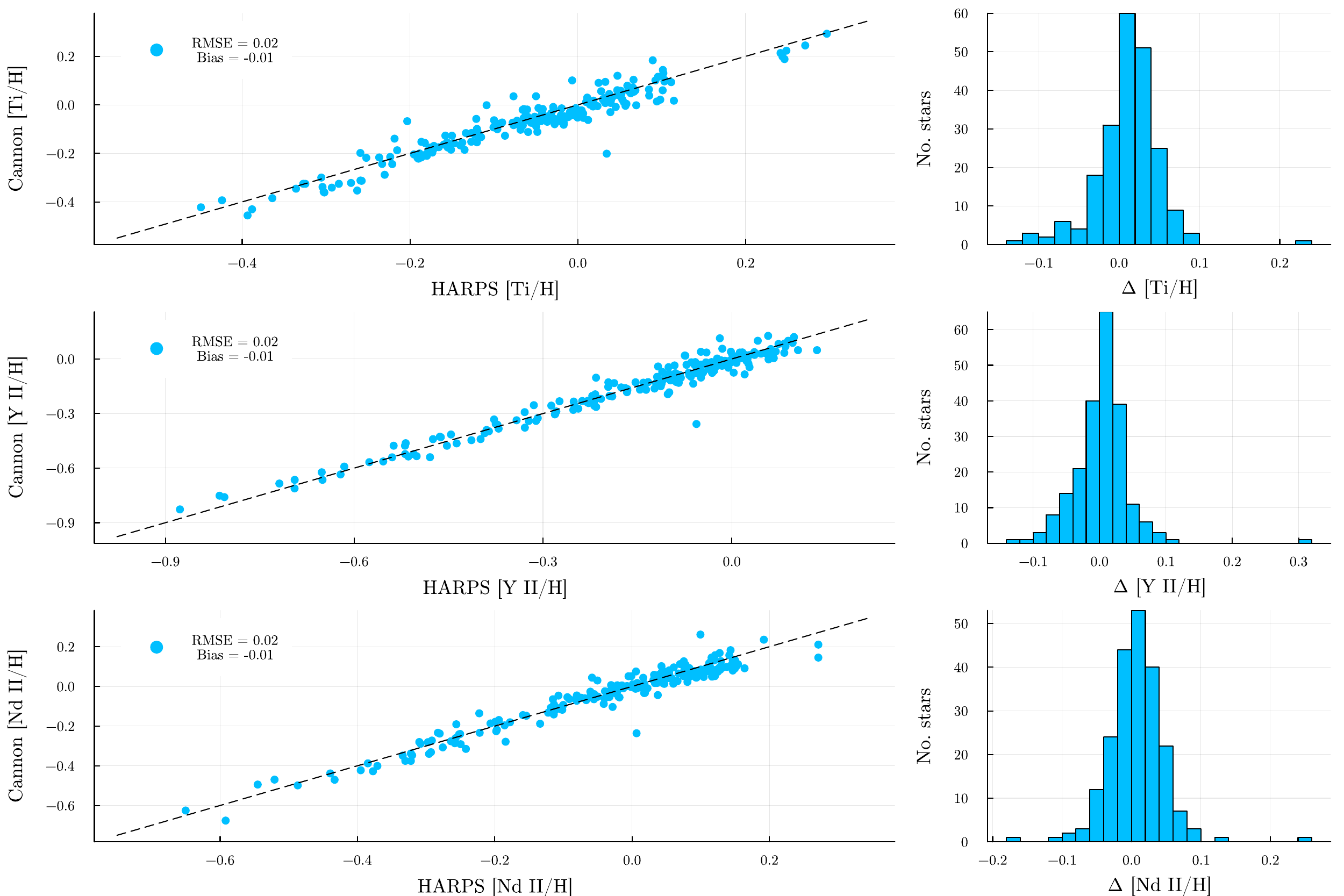}
    \caption{One-to-one plot of the reference set abundance labels and \textsc{the cannon} labels from the 8-label model. We include a distribution of the residuals (HARPS $-$ \textsc{cannon}) to the right. The small global biases for each label indicate that there are no significant systematics and the RMSE values are comparable to measurement uncertainties. Note all statistics are derived using the median.}
    \label{fig:xh one-to-one}
\end{figure*}

We evaluate the performance of the 8-label \textsc{cannon} model by comparing the label values predicted for the 220 test set stars against their reference values from the \citet{aquilina_2026} catalogue. Figures \ref{fig:param one-to-one} and \ref{fig:xh one-to-one} display one-to-one plots between the \textsc{cannon} labels and the reference labels. These Figures include distributions of the difference between the measured and model labels for \teff, \logg, [M/H], \vmic, \vsini, [Ti/H], [Y/H] and [Nd/H]. We quantify the model's performance using the bias and the root median squared error (RMSE) of these residuals ($\mathrm{Label}_{\mathrm{Reference}} - \mathrm{Label}_{\mathrm{Cannon}}$). The global biases for all derived labels are near zero: $-4.02$~K for \teff, $-0.01$~dex for \logg, $-0.01$~dex for [M/H], $0.0$~kms${}^{-1}$ for \vmic, $0.0$~kms${}^{-1}$ for \vsini, $-0.01$~dex for [Ti/H], $-0.01$~dex for [Y/H] and $-0.01$~dex for [Nd/H]. This indicates that the training process introduces no significant systematics and that the model preserves the fundamental chemical abundance scales established by the original high-resolution LTE analysis by \citet{aquilina_2026}. The RMSE for each label stated in Figures \ref{fig:param one-to-one} and \ref{fig:xh one-to-one} are also consistent with the median measurement uncertainties in \citet[][see Table \ref{tab:training_space} for the median uncertainties for each label across stars in the training set]{aquilina_2026}. The RMSE is 12~K for \teff,  0.02~dex for \logg, 0.01~dex for [M/H], $0.01\,\mathrm{km\,s^{-1}}$ for \vmic, $0.02\,\mathrm{km\,s^{-1}}$ for \vsini, 0.02~dex for [Ti/H], 0.02~dex for [Y/H], and 0.02~dex for [Nd/H]. The label recovery means that \textsc{the cannon} model effectively captures the \textsc{korg}-derived label scale (and corresponding physical origins), without constraint; leveraging the full spectral region and subsequent embedded information. We further demonstrate the faithfulness of the 8-label \textsc{cannon} model to recover the reference scale in Appendix \ref{appendix:supplementary label validation}, where we compare the abundance trends of Ti, Y and Nd with respect to [M/H] of the reference set and the recovered labels.

\begin{figure*}
	\includegraphics[width=\textwidth]{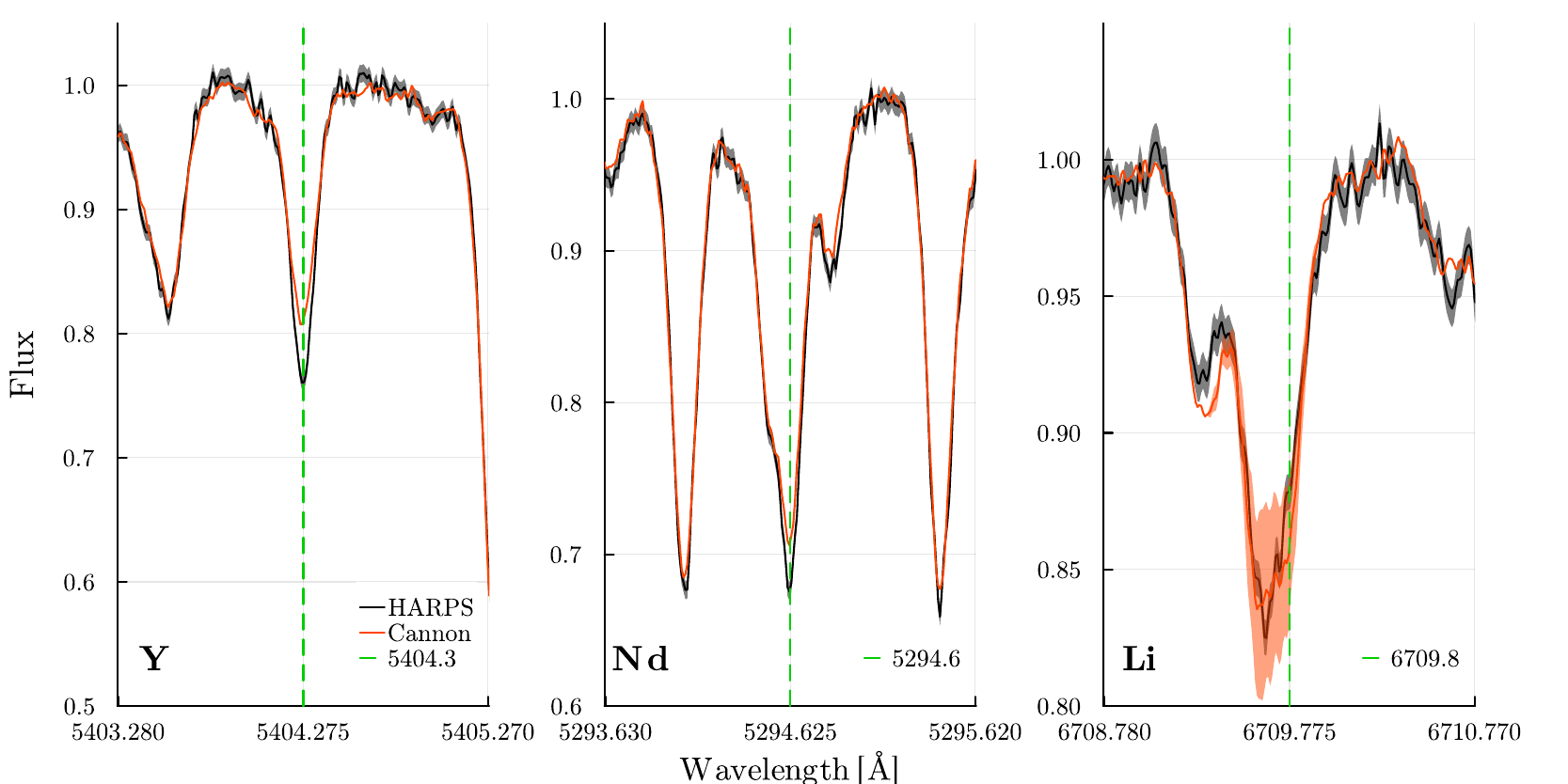}
    \caption{HARPS spectrum (black) with flux uncertainties shown as grey shading and 8-label spectral reconstruction (red) with the model scatter shaded in red around the 5404~\AA{} Y line, 5294~\AA{} Nd line and 6710~\AA{} Li line in HD 78668. The green dashed line indicates the central wavelength of each the line. Although Y and Nd are explicit labels in the 8-label model, the reconstruction fidelity is poor because HD 78668 is a Li-rich giant, a stellar class not represented in the training set.}
    \label{fig:Li giant reconstruction}
\end{figure*}

\noindent
\\\\
One star (HD 78668) deviates from the one-to-one recovery line across abundance labels Ti, Y and Nd (Figure \ref{fig:xh one-to-one}), despite achieving a global fit of $\mathrm{\chi^2_{Reduced}} = 2.9$ (the median reduced $\chi^2$ of the 8-label model is 1). This may be caused by lack of representation in the training stars as HD 78668 is a Li-rich giant with $\mathrm{A(Li)}_{\mathrm{NLTE}} = 1.68$~dex \citep{redbull_2015}. Li-rich giants are known to have a different element correlation structure than typical giant stars \citep[][]{sayeed_2024}. Our training set does not contain Li-rich giants to capture the chemical abundance trends in this population. In addition, we learn element abundances from the full spectrum and their fidelity is determined by the generative model in the relevant abundance windows. In the case of the Li-rich giant, there are discrepancies between the observed and reconstructed spectrum around Y and Nd absorption features. We show examples of these reconstructions around Y,  Nd and Li absorption features in Figure \ref{fig:Li giant reconstruction}. This indicates that HD 78668 is not represented in the reference set. A masked version of \textsc{the cannon} is a possible solution to obtain more accurate element abundance labels in such instances to avoid leveraging correlations. However, here, the purpose of our test is to learn about the successes and failures of the element correlation structure, so building a masked model is not necessary.

\subsection{Comparison of 5- and 8-label Cannon Models}
\label{sec:comp 5 and 8 model}

\begin{figure}
	\includegraphics[width=\columnwidth]{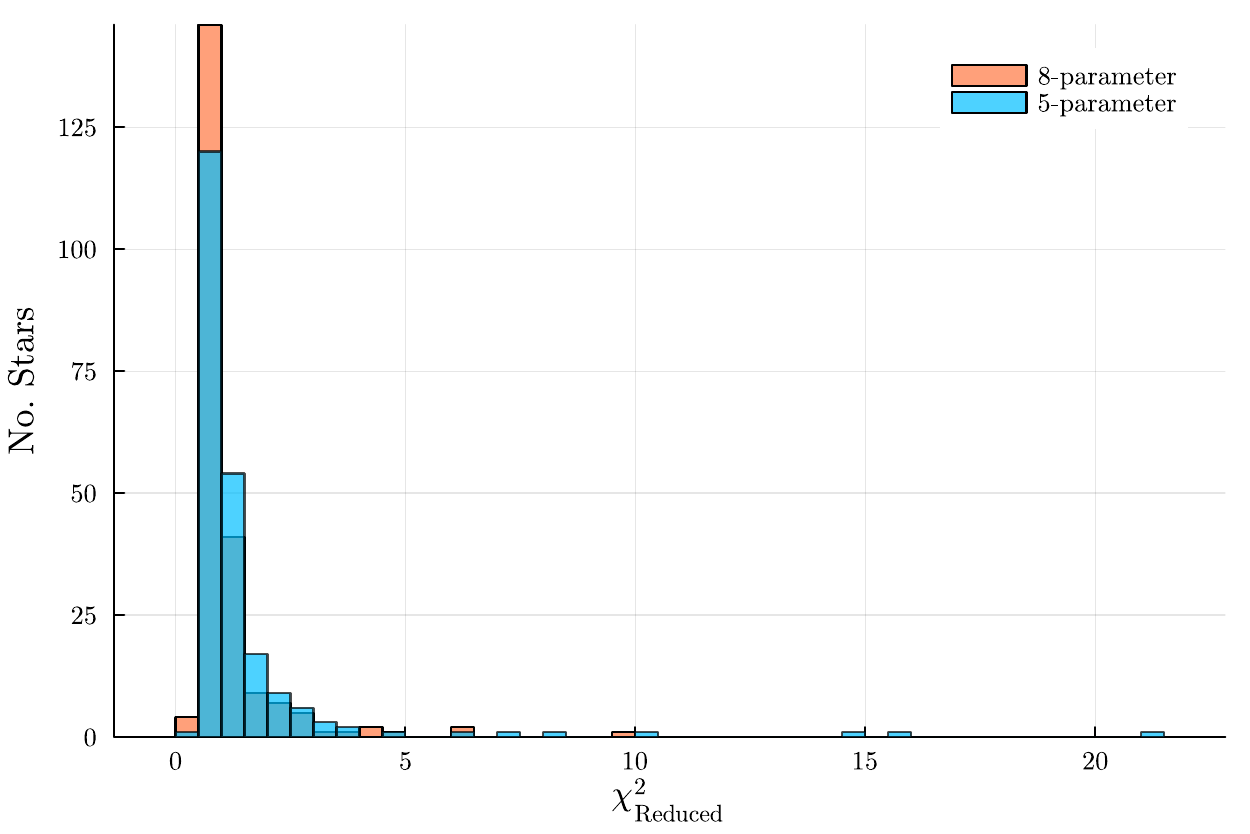}
    \caption{Comparison of the 5- and 8-label model reduced $\chi^2$ distribution (see Equation \ref{eqn:chi2}). The 5-label model has four test stars with $\mathrm{\chi^2_{Reduced}} > 10$. All test stars in the 8-label model are reconstructed with $\mathrm{\chi^2_{Reduced}} < 10$ with 96\% showing $\mathrm{\chi^2_{Reduced}} < 3$.}
    \label{fig:chi2 comparison}
\end{figure}

The $5-9$ label models demonstrate a very good global spectral construction across the HARPS wavelength range (trained across 4300-6900~\AA) of the population, consistent with a median of the reduced $\chi^2$ distribution $\sim$ 1, calculated as:

\begin{equation}
\label{eqn:chi2}
\chi^2_{\mathrm{red}} =
\frac{1}{M-K-1}
\sum_{j=1}^{M}
\frac{
\left(f_{\mathrm{Cannon},j}-f_{\mathrm{HARPS},j}\right)^2
}{
\sigma_{\mathrm{HARPS},j}^2+s_j^2
},
\end{equation}

\noindent
where $M$ is the number of wavelength pixels with data, $K$ is the number of fitted labels,
$f_{\mathrm{Cannon},j}$ is the flux reconstructed by \textsc{the cannon} at
pixel $j$, $f_{\mathrm{HARPS},j}$ is the observed HARPS flux,
$\sigma_{\mathrm{HARPS},j}$ is the corresponding observational flux
uncertainty, and $s_j$ is the model scatter at that wavelength.
\\\\
As shown in Figure \ref{fig:chi2 comparison}, the 5-label model reconstructs four test stars with $\mathrm{\chi^2_{Reduced}} > 10$ (HD 109379, HD 22663, HD 45398 and HD 1690), whereas all test stars are below this threshold in the 8-label model (with 96\% showing $\mathrm{\chi^2_{Reduced}} < 3$). 
The first three stars all have the largest $\mathrm{\chi^2_{Reduced}}$ (6.3, 6.2 and 9.6, respectively) in the 8-label model as well. The poorer recoveries for HD 1690 and HD 45398 in the 5-label model likely occur because they reside at the edges of our parameter grid (\logg = 1.61~dex and \teff = 4171~K, respectively). HD 45398 also has the highest reduced $\chi^2$ in the 8-label model of 9.6. For HD 22663, the 5-label model derives \teff = 4655~K and \logg = 2.8~dex. This is similar to the values from \citet{jofre_giants_2015}, which deviate from our reference set ($T_{\text{eff}} = 4547$ K and $\log g = 2.49$ dex). Furthermore, the 5-label model predicts a metallicity of -0.04 dex, disagreeing with both the reference set (-0.21 dex) and \citet{jofre_giants_2015} (-0.15 dex). However, the 8-label model successfully recovers the reference set's metallicity, \teff, and \logg. The improved performance of the 8-label model in these three cases suggests that explicitly including abundance labels can prevent abundance-driven spectral variation from being absorbed into the inferred stellar parameters. The final poorly-fit star, HD 109379, exhibits anomalous saw-tooth patterns in its residuals and is classified as an outlier.
\\\\
We identify five outliers in the 8-label model that despite showing reasonable reduced $\chi^2$ of $<6$, display a distinct saw-tooth pattern in their flux residuals. The saw-tooth pattern in three of the affected stars correspond directly to the HARPS diffraction orders. The remaining two stars display a localised saw-tooth pattern restricted to wavelengths redder than 5700~\AA{}. The appearance of the HARPS diffraction orders in all five stars confirms that these artifacts are systematic errors introduced during the stitching or normalisation pipeline. However, these stars are still included in our analysis and do not change our results.

\begin{figure}
	\includegraphics[width=\columnwidth]{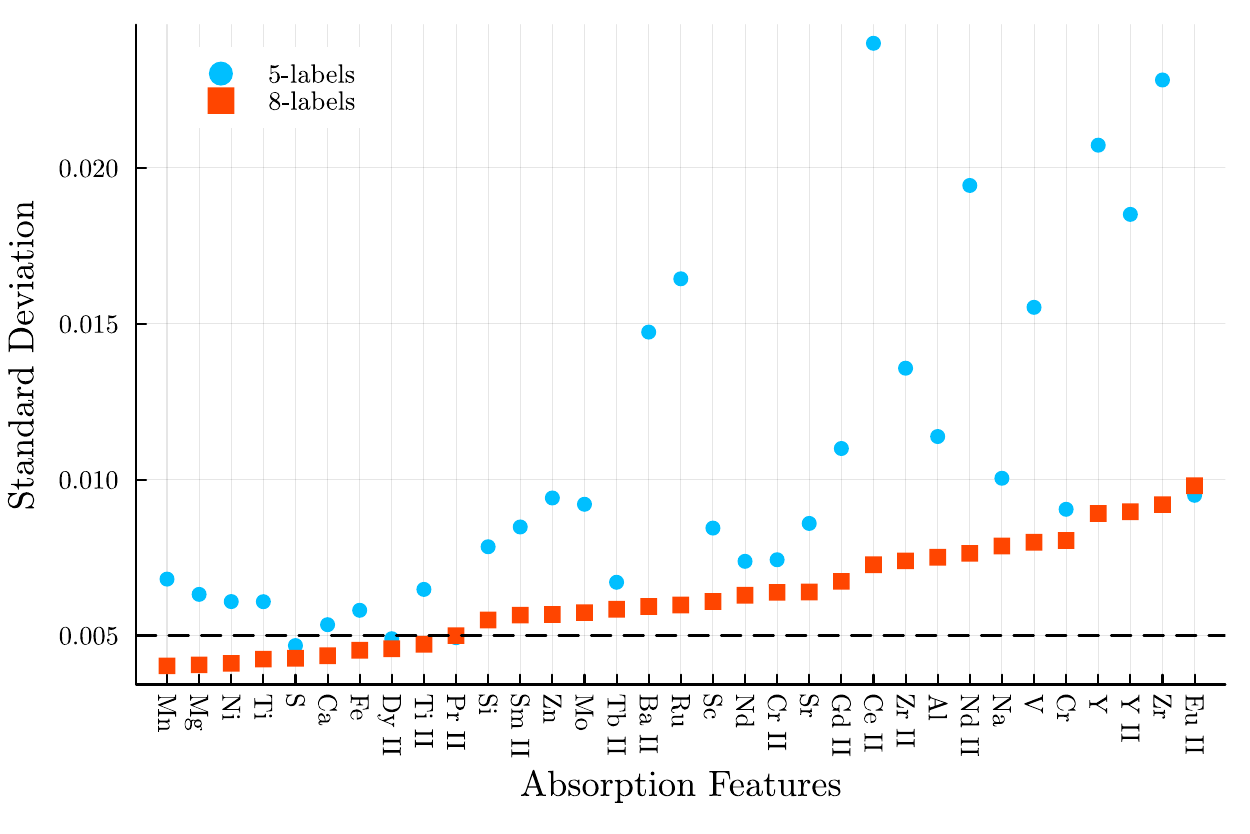}
    \caption{Comparison of the 5- (circles) and 8-label (squares) model standard deviation values of the normalised flux at the center of absorption features of 34 elements (including additional ionisation states) across the 220 test stars. Elements are sorted by their standard deviation in the 8-label model. The horizontal dash line at $\sigma$ = 0.5\% represents the expected noise level. Li is not included in the Figure but has the largest standard deviation of 0.070 in the 8-label model and 0.063 in the 5-label model. This demonstrates the improved performance of the 8-label model around individual features despite the similar reduced $\chi^2$ values of both models.}
    \label{fig:mad comp}
\end{figure}

\noindent
\\\\
To compare the performance of the 5- and 8-label models around specific features, we calculate the median absolute flux residual across all test stars at the centre of 34 absorption features. We refer to this quantity as the MAD, defined as
\begin{equation}
\label{eqn:mad}
    \mathrm{MAD}(\lambda_{\rm line})
    =
    \underset{i}{\mathrm{median}}
    \left|
    f_{{\rm Cannon},i}(\lambda_{\rm line})
    -
    f_{{\rm HARPS},i}(\lambda_{\rm line})
    \right|,
\end{equation}
where $i$ indexes the stars and $f_{\rm Cannon}$ is the generated normalised flux given the best-fit labels and $\lambda_{\rm line}$ is the center wavelength of the element absorption line. We also calculate the standard deviation of the flux residuals across the test stars at each absorption feature. The MAD and corresponding standard deviation values for absorption features of 34 different elements (including additional ionisation states) for the 5- and 8-label \textsc{cannon} models are shown in Figure \ref{fig:mad comp}. The dashed line in the figure indicates the expected noise level of 0.005.
\\\\
We find that overall the MAD values across the 34 absorption features are lower for the 8-label model by up to 0.01 compared to the 5-label model. The only exception is Li, which exhibits a MAD value increase of 0.006. The change in MAD values demonstrates that while both models exhibit a seemingly similar global performance in terms of reduced $\chi^2$ (which includes the scatter term in the calculation), including specific abundance labels significantly improves the local recovery of individual absorption features for the vast majority of elements, as evaluated using the flux residual alone. In particular, the increased predictability of iron-peak elements with the abundance labels demonstrates that [M/H] alone is insufficient to capture the variance of the entire iron-peak family. This is consistent with their diverse nucleosynthetic origins. For example, elements like Sc, V, Co, and Zn are produced in hypernovae \citep[][]{grimmett_neutrino_winds_2020}, distinct from the bulk Fe production \citep[see][and references therein]{kobayashi_origin_2020}. We therefore only discuss the 8-label model in Sections \ref{sec: spectrum validation}. The 9-label model is discussed in Section \ref{sec: discussion-Eu II label}.

\subsection{Analysis of the Generated Spectra for the 8-Label Model}
\label{sec: spectrum validation}
We successfully reconstruct the labels for all test stars using the 8-label \textsc{cannon} model. Here, we validate our spectral reconstructions by exploring the global reproduction of all test stars in Section \ref{sec: global recon} as well as that of individual absorption features in Sections \ref{sec: results-line characterisation} and \ref{sec: results-unmodelled features}. 

\subsubsection{Global Spectral Reconstruction}
\label{sec: global recon}
\begin{figure}
	\includegraphics[width=\columnwidth]{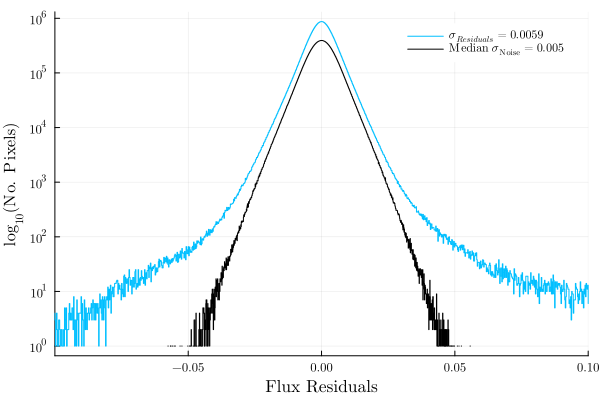}
    \caption{Flux residuals between the HARPS spectra and 8-label \textsc{cannon} reconstructed spectra across all 220 test stars. The standard deviation of 0.006 is only slightly in excess of the expectation from the noise of $\sigma = 0.005$, corresponding to a standard deviation of $\delta = 1.14$ when the flux residuals are normalised by the flux uncertainties (see Equation \ref{eqn:normalised_residuals}).}
    \label{fig:flux residuals all stars}
\end{figure}

To quantify how an accurate label recovery translates to the fidelity of the reconstruction of individual stellar spectra, we examine the model generated spectra for the test objects and their flux residuals relative to the HARPS observations. Across the entire wavelength range, the typical per-star MAD between the HARPS spectra and the 8-label \textsc{the cannon} model is 0.3\%. In Figure \ref{fig:flux residuals all stars} we show the distribution of per-pixel flux residuals across all 220 test stars. The standard deviation of  0.006 for this flux residual distribution corresponds to a standard deviation of 1.14 when the flux residuals are normalised by the flux uncertainties. The normalised residuals, $\delta_{i\lambda}$, are given by

\begin{equation}
\delta_{i\lambda}
=
\frac{
f_{{\rm Cannon},i\lambda}
-
f_{{\rm HARPS},i\lambda}
}{
\sigma_{i\lambda}
},
\label{eqn:normalised_residuals}
\end{equation}

\noindent
where $f_{{\rm Cannon},i\lambda}$ is the flux predicted by the
8-label \textsc{the cannon} model for star $i$ at wavelength $\lambda$,
$f_{{\rm HARPS},i\lambda}$ is the corresponding observed HARPS flux, and
$\sigma_{i\lambda}$ is the observational uncertainty. We deliberately
exclude the intrinsic model scatter inferred by \textsc{the cannon} from
the denominator. Including this term would down-weight regions in which
the model performs poorly, thereby partially masking systematic
discrepancies between the model and the observed spectrum. Our definition
is intended to retain sensitivity to such model inadequacies.
\\\\
Figure \ref{fig:flux residuals all stars} shows that the typical residual amplitude is small (MAD $=0.3\%$), but the overall residual distribution is broader than expected from the observational noise, with $\sigma_{\rm residuals}=0.59\%$ compared to a median noise level of $0.5\%$. The extended wings arise from larger residuals in a subset of wavelength pixels, primarily associated with specific absorption features (see Section \ref{sec: results-line characterisation}). We calculated the expected noise threshold (i.e., the average per-star MAD expected from uncertainty alone) by generating 100 noise-perturbed flux realisations for each model spectrum, drawing the flux at each pixel from the corresponding observational uncertainty. We then subtract the original model spectrum from each noise realisation to obtain residuals expected from noise alone, and calculate the MAD in the same way as for the data. The median per-star MAD across these samplings from the observational uncertainties is 0.3\%, which we adopt as our expected noise level. This expected noise threshold remains the same when including the model scatter with the observational uncertainties.

\begin{figure*}
	\includegraphics[width=\textwidth]{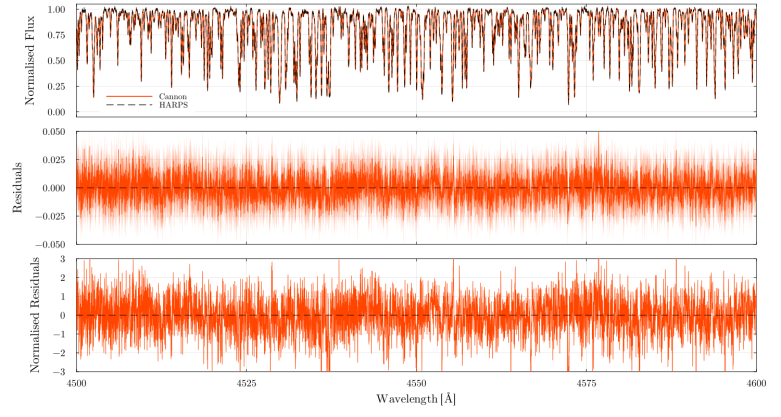}
    \caption{\textit{Top:} Sample spectrum from 4500~\AA{} to 4600~\AA{} of HIP20226 ($\teff = 4722$~K, $\logg = 2.15$~dex, [M/H] $= -0.67$~dex) showing the HARPS data and the successful spectral reconstruction of the 8-label \textsc{cannon} model. \textit{Middle:} Residuals between HARPS and the 8-label \textsc{cannon} model. \textit{Bottom:} Residuals between HARPS and the 8-label \textsc{cannon} model normalised by the flux uncertainties (see Equation \ref{eqn:normalised_residuals}) to demonstrate that most residuals are within 2$\sigma_{\lambda}$. Note that all wavelengths are in the vacuum rest frame.}
    \label{fig:Example spectrum and cannon 1}
\end{figure*}

\begin{figure*}
	\includegraphics[width=\textwidth]{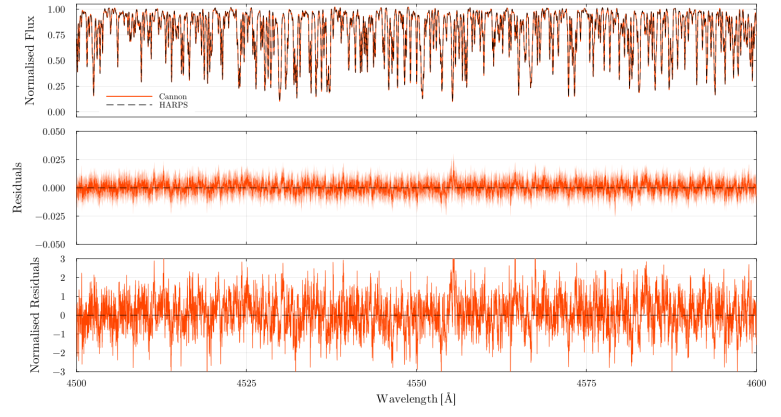}
    \caption{Format identical to Figure \ref{fig:Example spectrum and cannon 1} but instead showing the spectra, residuals and normalised residuals of HD 17793 ($\teff = 5127$~K, $\log g = 2.98$~dex, [M/H] $= -0.06$~dex)}
    \label{fig:Example spectrum and cannon 2}
\end{figure*}

\begin{figure*}
	\includegraphics[width=\textwidth]{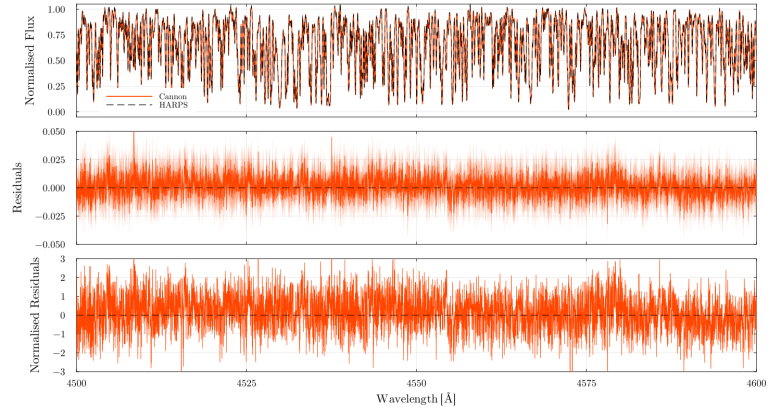}
    \caption{Format identical to Figure \ref{fig:Example spectrum and cannon 1} but instead showing the spectra, residuals and normalised residuals of HD 176704 ($\teff = 4457$~K, $\logg = 2.38$~dex, [M/H] $= 0.22$~dex).}
    \label{fig:Example spectrum and cannon 3}
\end{figure*}

\noindent
\\\\
We show the observed spectra, respective 8-label spectral reconstructions, and the corresponding residuals for three representative test stars in Figures \ref{fig:Example spectrum and cannon 1} to \ref{fig:Example spectrum and cannon 3}: HIP 20226 ($\teff = 4722$~K, $\logg = 2.15$~dex, [M/H] $= -0.67$~dex), HD 17793 ($\teff = 5127$~K, $\log g = 2.98$~dex, [M/H] $= -0.06$~dex), and HD 176704 ($\teff = 4457$~K, $\logg = 2.38$~dex, [M/H] $= 0.22$~dex). Note that all wavelengths are in the vacuum rest frame. These Figures show two things. First, that the low-dimensional representation provides a basis for an accurate reconstruction of the spectra, as residuals are typically sub-percent and at most extend to 2-3\%. Second, that the residual structure indicates there is real and recoverable information away from the low-dimensional model, if sufficient signal to noise is reached. For example, the spike in residuals around the wings of the Ba resonance line at 4555.4~\AA{} in Figure \ref{fig:Example spectrum and cannon 2} corresponds to subtle variations in Ba isotopic abundances \citep[][]{gallagher_2010}. See Appendix \ref{appendix:label spectra reconstruction} for further spectral reconstructions around features of chosen abundance labels for training: Ti, Y, and Nd in HD 17793. HD 17793 has a MAD value below the 0.3\% sample average, while both HIP 20226 and HD 176704 display greater MAD values of 0.49\% and 0.43\%, respectively.

\begin{figure*}
	\includegraphics[width=\textwidth]{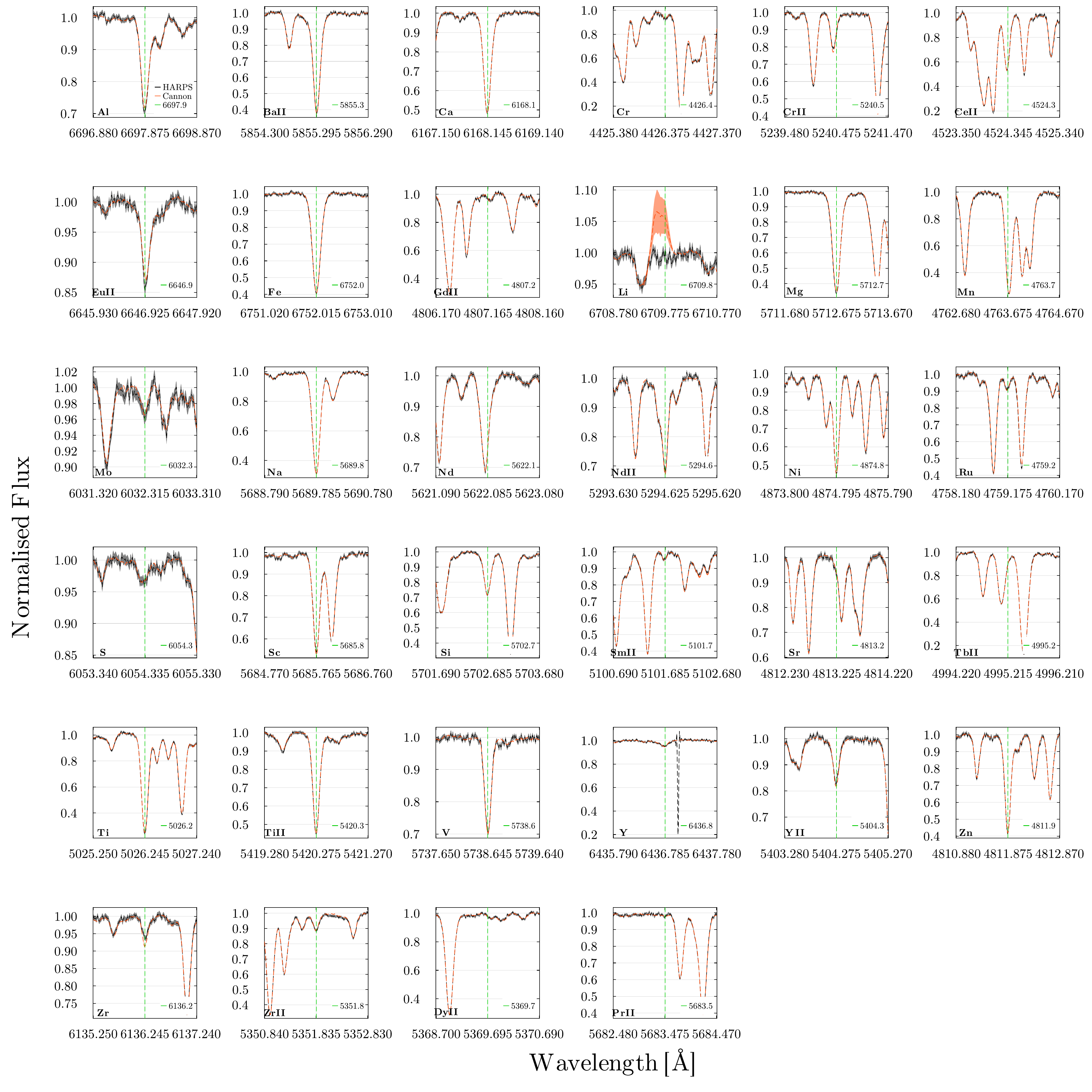}
    \caption{The HARPS spectra and 8-label \textsc{cannon} reconstructed spectra around absorption features of 34 elements (including additional ionisation states) in HIP20226. The black dashed line shows the normalised HARPS spectrum with the flux uncertainties shown as grey shading. The reconstructed \textsc{cannon} spectrum is shown as the red dashed line with the scatter of the model shaded in red. The green dashed line indicates the central wavelength of each element line. Note that all wavelengths are in the vacuum rest frame. Our spectral reconstruction successfully reproduces a majority of absorption features with the most notable deviations occurring for Li.}
    \label{fig:Example line reconstruction}
\end{figure*}

\begin{figure*}
	\includegraphics[width=\textwidth]{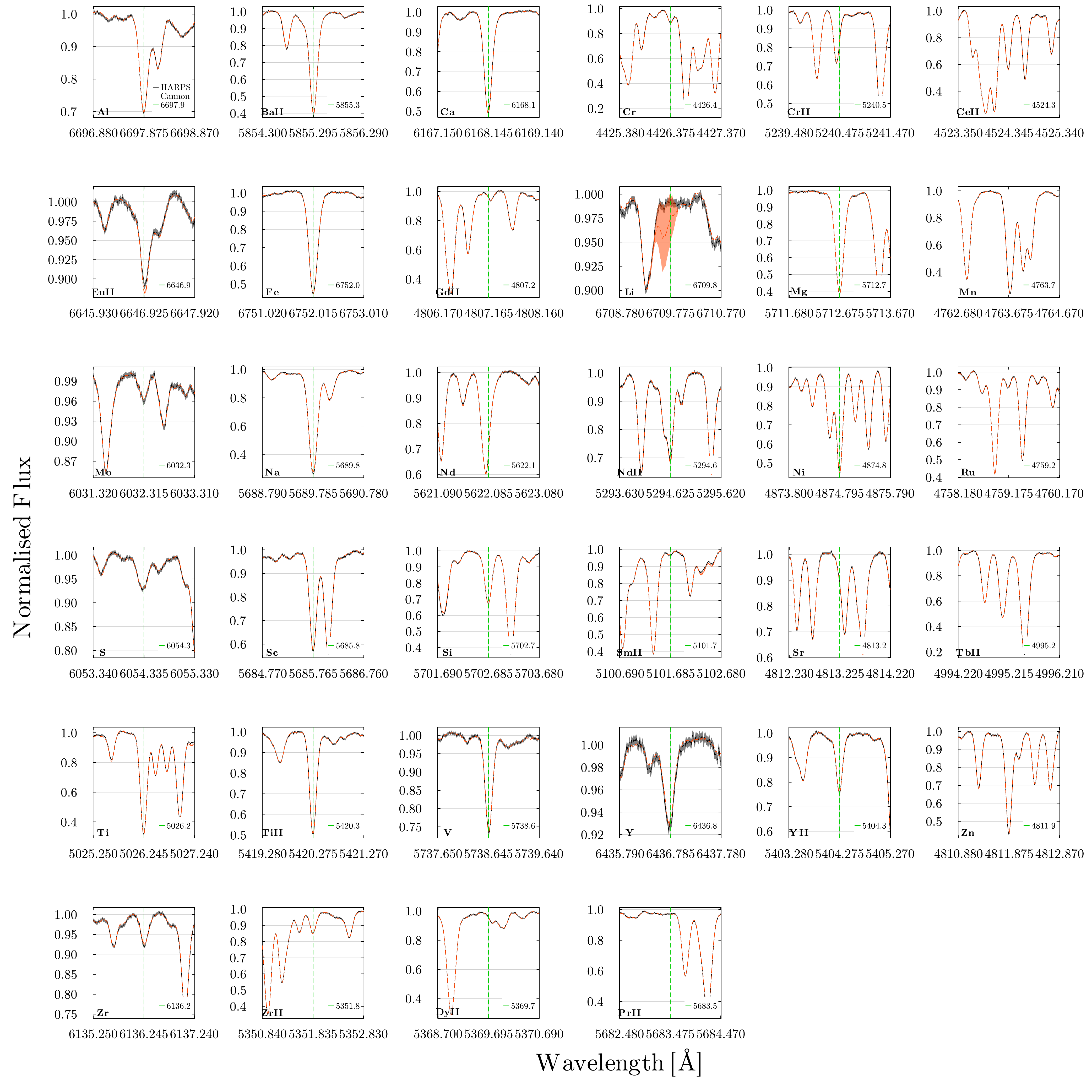}
    \caption{Format identical to Figure \ref{fig:Example line reconstruction} but instead showing spectra of HD 17793.}
    \label{fig:Example line reconstruction 2}
\end{figure*}

\begin{figure*}
	\includegraphics[width=\textwidth]{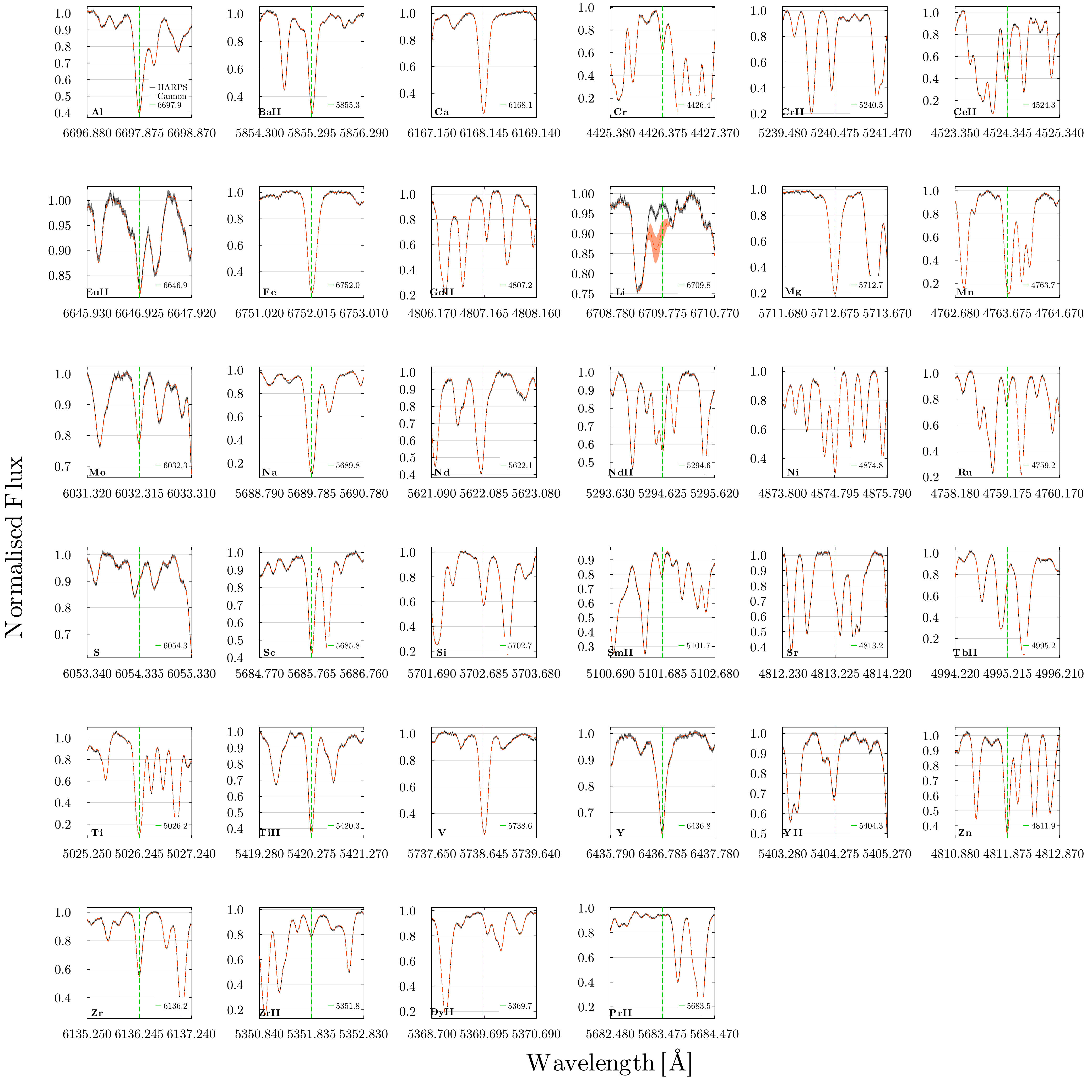}
    \caption{Format identical to Figure \ref{fig:Example line reconstruction} but instead showing spectra of HD 176704.}
    \label{fig:Example line reconstruction 3}
\end{figure*}

\subsubsection{Reconstruction of Individual Absorption Features}
\label{sec: results-line characterisation}

\begin{figure}
	\includegraphics[width=\columnwidth]{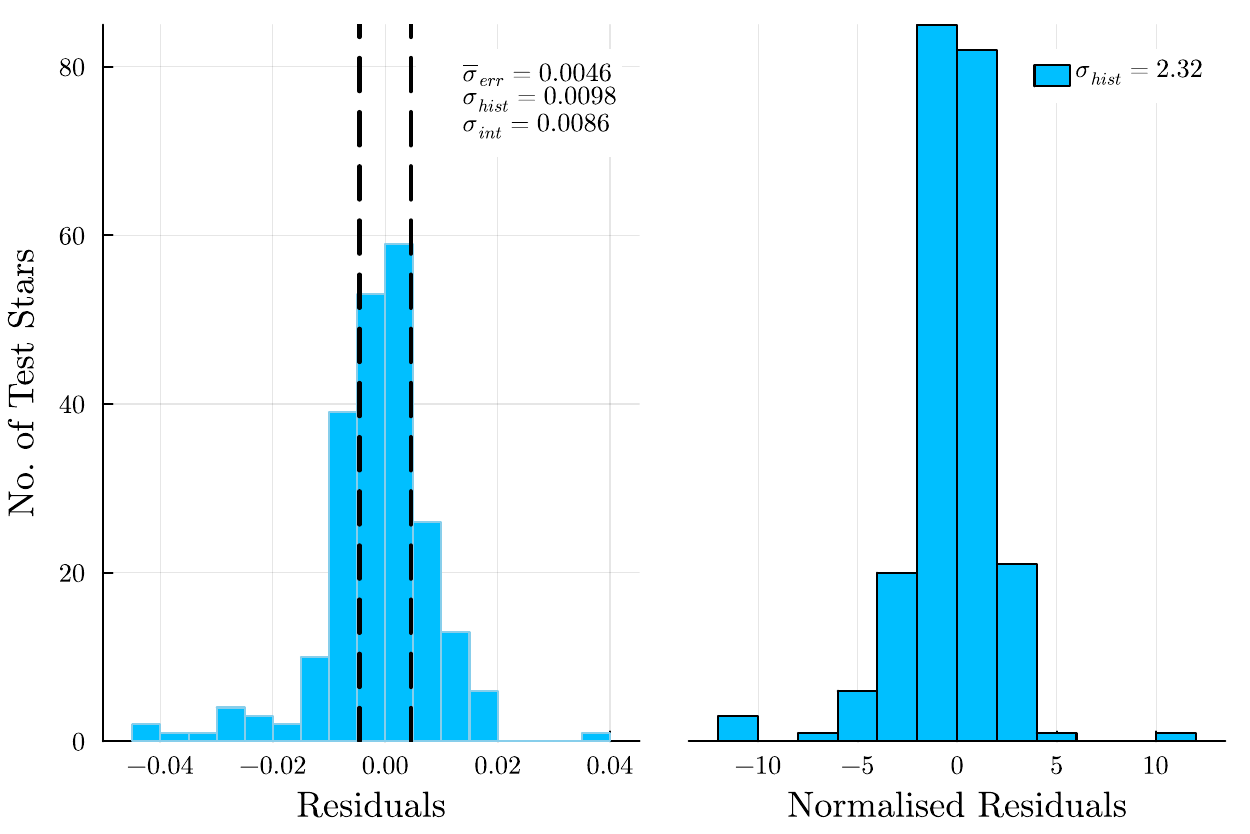}
    \caption{\textit{Left}: Distribution of the test star residuals at the 6646~\AA{} Eu line in the 8-label \textsc{cannon} model. We include the median error, standard deviation of the residuals and intrinsic scatter to quantify the unexplained variance in our model. \textit{Right}: Distribution of test star residuals normalised by the median error (see Equation \ref{eqn:normalised_residuals}) for the Eu II line at 6646~\AA{} in the 8-label \textsc{cannon} model. The intrinsic scatter of 0.0086 exceeds the expected noise level of 0.005, indicating that this Eu line contains additional information not contained in the model labels \teff, \logg, [M/H], \vmic, \vsini, [Ti/H], [Y/H] and [Nd/H].}
    \label{fig:Eu II line hist 8 param}
\end{figure}

\begin{figure*}
	\includegraphics[width=\textwidth]{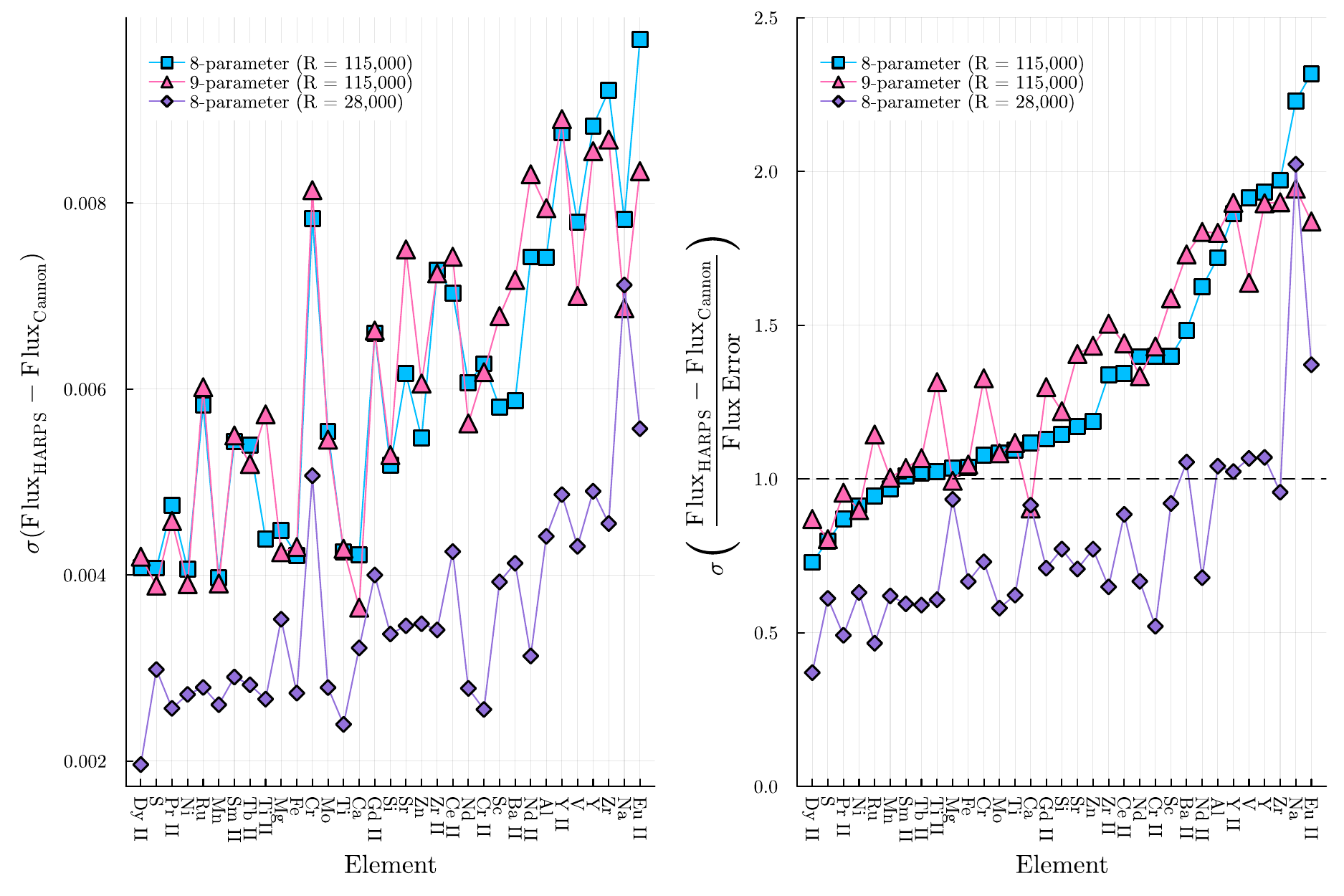}
    \caption{\textit{Left:} Standard deviation of test star residuals around a single line for 28 elements (including additional ionisation states). This is shown as blue squares for the 8-label model, pink triangles for the 9-label model (see Section \ref{sec: discussion-Eu II label}), and purple diamonds for the convolved 8-label model to $R = 28000$ (see Section \ref{sec: discussion-galah resolution}). \textit{Right:} Standard deviation of test star residuals normalised by the flux error (see Equation \ref{eqn:normalised_residuals}) around a single line for 28 elements (including additional ionisation states). Li is not included but has standard deviations in the normalised residuals of 18, 14 and 16 in the 8- (R=115000), 9- and 8-label (R=28000) models respectively. Elements are ordered from smallest to largest standard deviations in the 8-label model. Standard deviations less than 1.0 indicate that the line was recovered better than expected by noise, around 1.0 indicates that the line is recovered at the expected noise limit, and greater than 1.0 indicates that there is residual information not captured by the model. Note that the standard deviations of normalised residuals are calculated from a single line and overall element recovery may be larger across the entire spectrum.}
    \label{fig:line sigma int}
\end{figure*}

The 8-label \textsc{cannon} model is able to globally reconstruct spectra that is on average on the order of the noise, at a median MAD = 0.3\%. However, to investigate which element features are the least accurately reconstructed, we analyse the residuals at specific absorption features. We focus on lines measured in the reference catalogue for 20 elements (Na,~Mg,~Al,~Si,~Ca,~Sc,~Ti,~V,~Cr,~Fe,~Ni,~Zn,~Sr, Y,~Zr,~Mo,~Ba,~Ce,~Nd ~and~Eu) and 9 additional elements (Li, S, Mn, Ru, Pr, Sm, Gd, Tb and Dy). We choose absorption features for these 9 elements using the same line criteria as \citet{aquilina_2026} for lines $\lambda \geq 4750$~\AA{} in the \textit{Gaia} ESO line list \citep[][]{heiter_linelist_2021} and lines $\lambda < 4750$~\AA{} in the \textsc{linemake} line list \citep[][]{placco_linemake_2021}. In total, there are 95 lines across 29 elements after excluding telluric regions. Note that K and La are measured in the reference catalogue but have lines inside these telluric regions and are therefore excluded from the analysis. All wavelengths discussed are in the vacuum rest frame. We provide examples of the model's performance in Figures \ref{fig:Example line reconstruction} to \ref{fig:Example line reconstruction 3}, which display the reconstructed spectra from the 8-label model and the HARPS spectra around one line per element for stars HIP 20226, HD 17793 and HD 176704 (The same stars shown in Figures \ref{fig:Example spectrum and cannon 1} to \ref{fig:Example spectrum and cannon 3}). Appendix \ref{appendix:label spectra reconstruction} includes supplementary reconstructions for the abundance labels Ti, Y, and Nd in HD 17793 to validate the model in a typical test set star.
\\\\
To quantify how well each spectral feature is recovered, we calculate the flux residual at the line core. To contextualise the amplitude of this residual given the flux uncertainty, we also normalise this residual by dividing by the flux uncertainty i.e. see Equation \ref{eqn:normalised_residuals}. The central wavelength of each line is determined using \textit{Gaia}-ESO line list \citep[][]{heiter_linelist_2021}. Figure \ref{fig:Eu II line hist 8 param} shows an example distribution of the residuals (left) and normalised residuals (right) for the Eu line at 6646~\AA{} across all test stars. The standard deviations of these distributions ($\sigma_{\mathrm{hist}}$) serve as our metric for ranking the recovery of each element line. This ranking is illustrated in Figure \ref{fig:line sigma int} (left), where lines are ordered by increasing $\sigma_{\mathrm{hist}}$ based on the 8-label model's normalised residuals. For comparison, we also plot $\sigma_{\mathrm{hist}}$ for both the 9-label model and a convolved 8-label model ($R = 28,000$) to demonstrate the change in residual amplitude when explicitly including Eu as an abundance label or varying the spectral resolution, respectively. For the normalised residuals in Figure \ref{fig:line sigma int} (right), $\sigma_{\mathrm{hist}} < 1.0$ indicates the line is recovered better than expected from noise (indicating our noise may  be slightly overestimated), $\sigma_{\mathrm{hist}} \approx 1.0$ indicates recovery exactly at the expected noise limit, and $\sigma_{\mathrm{hist}} > 1.0$ implies the presence of residual variance not captured by the model.

\begin{figure}
	\includegraphics[width=\columnwidth]{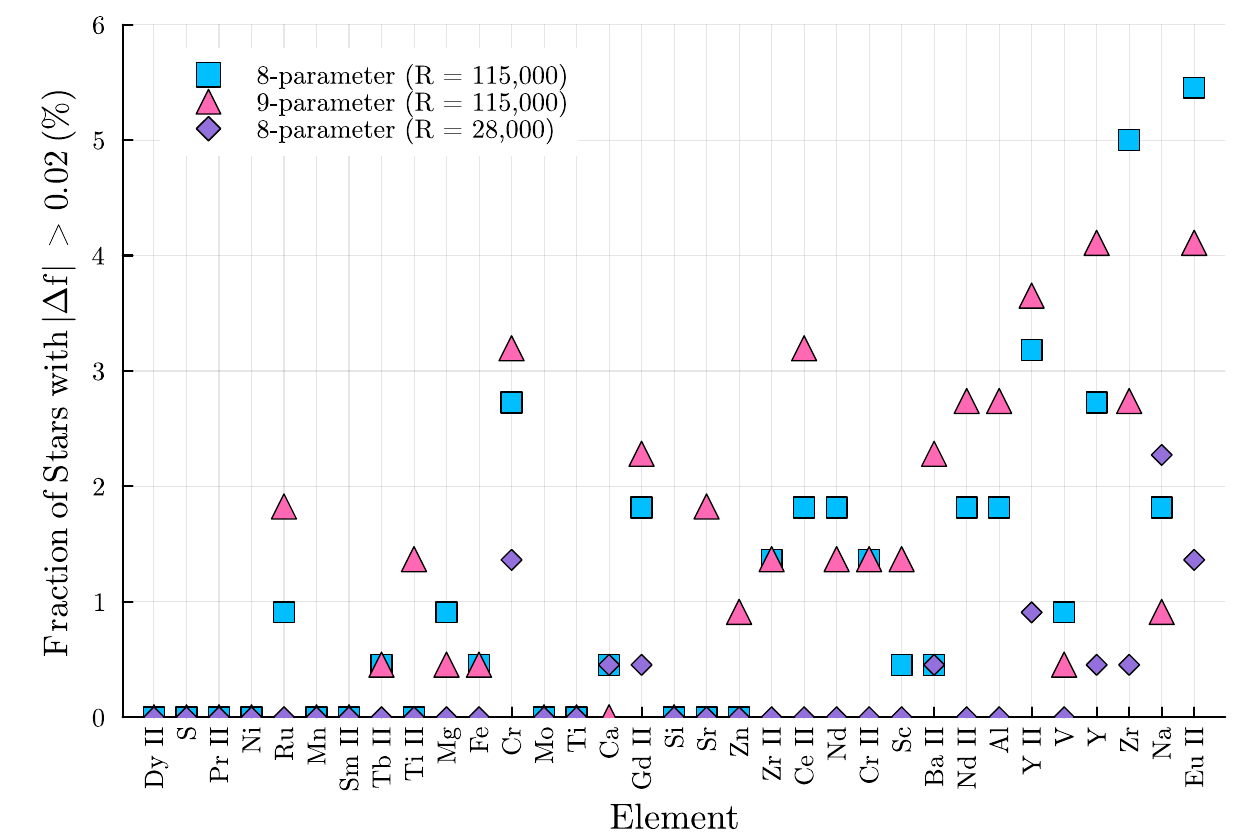}
    \caption{Percentage of test stars with absolute flux residuals greater than 0.02 around a single line for each of the 29 elements (including additional ionisation states). This is shown as blue squares for the 8-label model, pink triangles for the 9-label model (see Section \ref{sec: discussion-Eu II label}), and purple diamonds for the convolved 8-label model to $R = 28000$ (see Section \ref{sec: discussion-galah resolution}). Elements follow the same ordering as Figure \ref{fig:line sigma int}. Li is not included but has 54.1\%, 56.2\% and 47.3\% of test stars with absolute flux residuals greater than 0.02 in the 8- ($\mathrm{R}=115000$), 9- and 8-label ($\mathrm{R}=28000$) models respectively.}
    \label{fig:line 0.02 residuals}
\end{figure}

\noindent
\\\\
Overall, we find that the 8-label model successfully recovers line fluxes across the 95 lines with a median normalised residual standard deviation ($\sigma_{\mathrm{hist}}$) of 1.2. The model performs exceptionally well for 18 species (Mg, Si, S, Ca, Ti, Cr, Mn, Fe, Ni, Zn, Sr, Mo, Ru, Pr, Sm, Gd, Tb and Dy), achieving a normalised residual standard deviation ($\sigma_{\mathrm{hist}}$) of less than 1.2 for the representative lines in Figure \ref{fig:line sigma int}. Conversely, Na, Li, Zr and Eu have the poorest recovery with normalised standard deviations $>2.0$. Li is not included in Figure \ref{fig:line sigma int} but has standard deviations in the normalised residuals of 18, 14 and 16 in the 8- (R=115000), 9- and 8-label (R=28000) models respectively. We discuss these elements with poorer reconstructions in Section \ref{sec: results-nlte and uncorrelated}. Note that the standard deviations of normalised residuals are calculated from a single line for reference and different lines of the same element may show different amplitudes of residuals, sensitive to line depth. 
\\\\
We note that the flux residuals for elements that are included as training labels are of similar amplitude to the element absorption features not included as training labels. This is not unexpected because we fit the full spectrum, rather than masking the spectrum to regions associated with individual labels. The model therefore learns how the full set of training labels map onto flux variations across all wavelengths, and precise spectrum reconstruction is not guaranteed to be confined to absorption features associated with the labelled elements. Conversely, spectral features from elements not explicitly included as labels can also be reproduced when their abundances co-vary with the training labels in the reference sample. We observe no change in residual amplitude (as parameterised by Equation \ref{eqn:normalised_residuals}) across metallicity for all 95 absorption features. Examining the linear model coefficients (see Appendix \ref{appendix:supplementary label validation}) verifies that the model is learning information from the expected lines (eg., Y from Y lines).
\\\\
We analyse the fraction of test stars with absolute absorption feature residuals greater than 0.02 for our representative element lines in Figure \ref{fig:line 0.02 residuals}. This threshold identifies unmodelled information that is easily accessible at $\text{SNR} > 100$ given a $2\sigma$ detection limit. We find that across our 34 representative element lines (29 different elements with 5 additional ionisation states), $<6\%$ of test stars exhibit absolute residuals $>0.02$ for these same lines. This is with the exception of Li for which 54\% of test stars have absolute residuals greater than 0.02 in the 8-label model. For absorption features with a normalised residual standard deviation $<$ 1.2 only $\sim 2.7\%$ of stars exceed 0.02 flux residual amplitude. For 11 elements (Si, S, Ti, Mn, Ni, Zn, Sr, Mo, Pr, Sm, and Dy) there are no test stars above our 0.02 threshold. The absorption features for Sr, Mo, Pr, Sm, and Dy, are weak and blended and may contain r-process contributions not traced by our chosen labels, however the amplitude of this information is captured at the level of $<$ 0.02 in flux residual. Conversely, the variance in elements such as Si, S, Ti, Mn, Ni, and Zn is much more likely to be inherently captured by the model.

\subsubsection{Poorly Modelled Absorption Features}
\label{sec: results-unmodelled features}
\begin{figure*}
	\includegraphics[width=\textwidth]{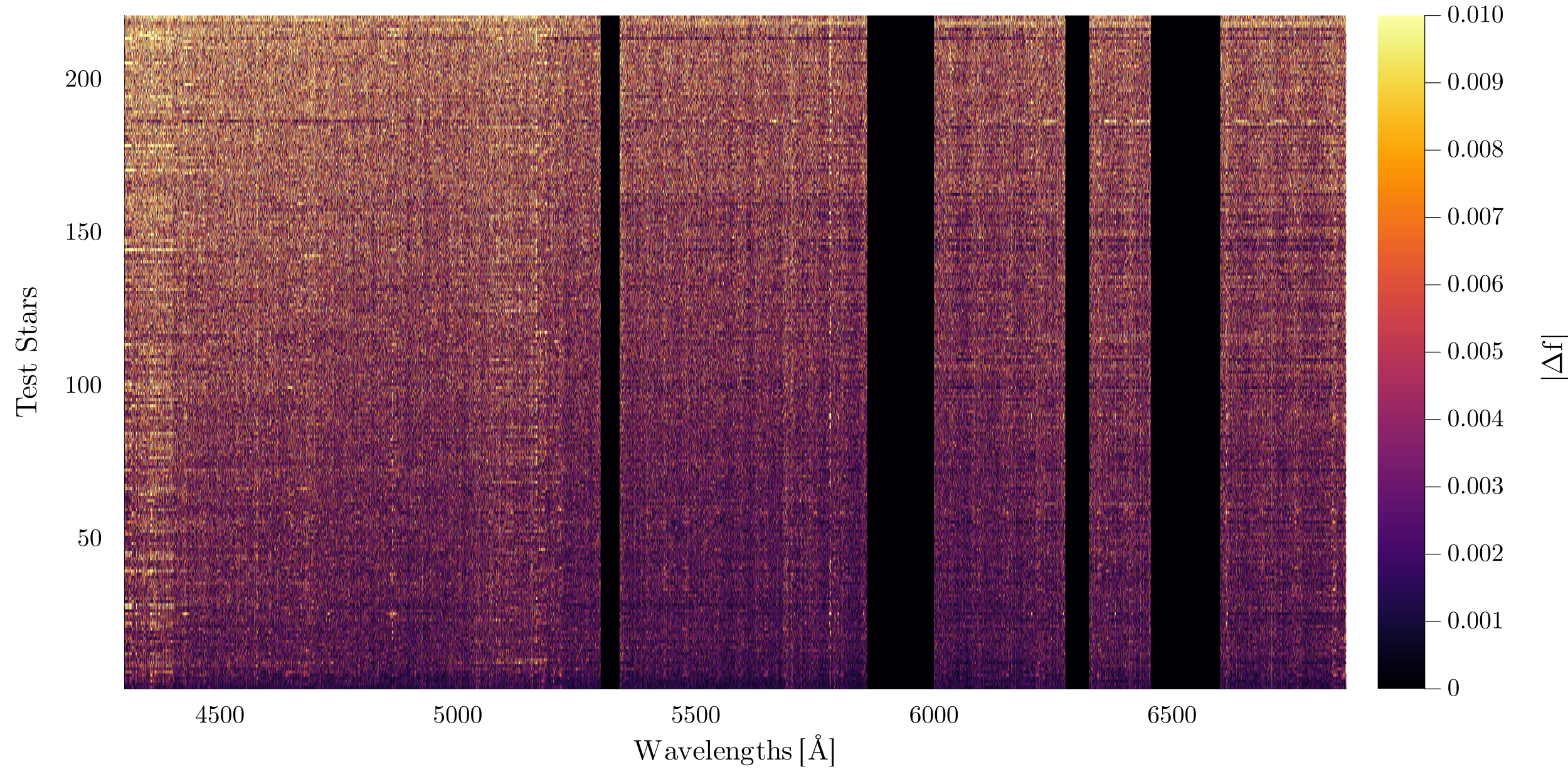}
    \caption{Flux residuals for all test set stars between the HARPS spectra and 8-label \textsc{cannon} model as a function of wavelength. Stars are ordered by $\sum_{\lambda} |\Delta f_\lambda|$. Black regions correspond to tellurics and the HARPS CCD gap ($5300-5340$~\AA{}) removed prior to training. Note that all wavelengths are in the vacuum rest frame.}
    \label{fig:residual heatmap}
\end{figure*}

In Section \ref{sec: results-line characterisation}, we analyse how well the 8-label \textsc{the cannon} model reproduces 95 elemental absorption features whose corresponding abundances were not included as training labels. These features are generally reproduced at the sub-percent level, with only a small subset showing residuals approaching or exceeding 2\%. We now explore residuals across the entire wavelength region to discover and quantify the most poorly modelled features.
\\\\
To identify regions of systematic discrepancy between the HARPS data and 8-label \textsc{cannon} model, we analyse the flux residuals at each wavelength across the test set. Figure \ref{fig:residual heatmap} highlights the absorption features that exhibit the highest flux residuals across our valid wavelength region (in the vacuum rest frame), with all test stars ordered by their total absolute residual flux ($\sum_{\lambda} \vert{}\Delta f_\lambda\vert{}$). Note that black regions correspond to tellurics and the HARPS CCD gap at $5300-5340$~\AA{} \citep[][]{barbieri_esoharps_2023}, which we exclude from training. All wavelengths mentioned are also in the vacuum rest frame. All spectral features are identified using \citet{hinkle_atlas_2000}. The most significant inconsistencies ($MAD > 0.3\%$, see Section \ref{sec: global recon}) are found in features of Li, Cu, CN, MgH and diffuse interstellar bands listed in Table \ref{tab:idetification of unmeasured residuals}. The median MAD and median standard deviations in Table \ref{tab:idetification of unmeasured residuals} across all test stars are calculated from the median of the absolute flux residuals of each spectrum at the central line wavelength.

\begin{figure}
	\includegraphics[width=\columnwidth]{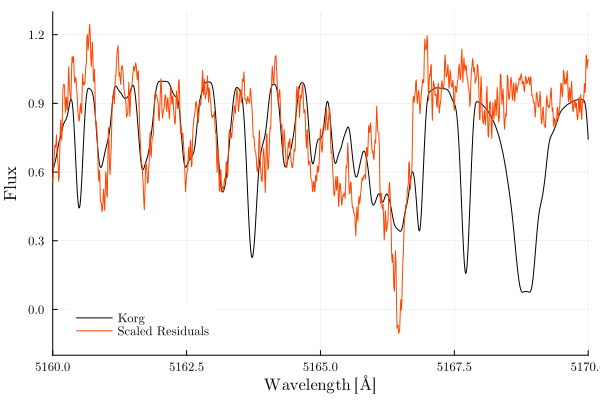}
    \caption{Comparison of scaled flux residuals of HD 17793 to a \textsc{korg} spectrum generated using only molecules in the \textit{Gaia} ESO line list and the corresponding stellar parameters of the star at 5160--5170~\AA. The flux residuals were multiplied by -25 and translated by +0.9 to demonstrate that peaks and troughs in the residuals correspond to molecular lines.}
    \label{fig:cross-correlation residuals}
\end{figure}

\noindent
\\\\
The most prominent ($MAD > 0.3\%$) unmodelled features displayed in Figure \ref{fig:residual heatmap} and listed in Table \ref{tab:idetification of unmeasured residuals} correspond to chemical species not included in our training labels. We observe notable residuals at 4355.7~\AA{}, 5166.5~\AA{} and 6614.6~\AA{}, which correspond to CH, MgH, and CN molecular bands, respectively. Additionally, we observe elevated flux residuals across our parameter space around CH bands at $\lambda < 4400$~\AA{}, $\mathrm{C_2}$ bands around 5100~\AA{} and a CH band at 5480.62~\AA. To demonstrate this relationship, Figure \ref{fig:cross-correlation residuals} compares the scaled flux residuals of HD 17793 around the Mg b triplet against a synthetic \textsc{korg} spectrum generated with the corresponding stellar parameters and only molecular transitions from the Gaia ESO line list. For direct comparison, the flux residuals are scaled by a factor of -25 and offset by +0.9. The peaks and troughs in the residuals align with the synthetic MgH bands, whereas regions corresponding to Fe and Cr atomic lines exhibit flat residuals. This contrast confirms that the localised increase in our flux residuals are in some instances driven by unmodelled molecular features, which could be exploited to test for variations in isotopic abundances (see Section \ref{sec: results-nlte and uncorrelated}).

\noindent
\\\\
Deviations around molecular bands is likely as our model lacks the light-element abundances (C, N, H) required to constrain these features. We also detect significant residuals at the Li doublet near 6709.8~\AA{}. While we do lack light-elements to effectively model Li, the Li abundance in red giant stars is also sensitive to mixing process \citep[][]{aguilera_2023,mori_mixing_2021,schwab_flash_mixing_2020}. The largest scatter in our Li flux residuals also occurs at \logg = 2.4, corresponding to the centre of the red clump where stars are known to show inherent Li variations \citep[][]{sayeed_Li_2024}]. This means that the Li abundance varies independently of any abundance labels, resulting in a high model uncertainty and scatter in Figure \ref{fig:linear coefficients X/H} around this Li feature.
\\\\
When examining the residuals between the HARPS spectra and 8-label model reconstructions, we observe broad discrepancies coincident with Diffuse Interstellar Bands (DIBs) at 5782~\AA{}, 5787, 5799.5~\AA{}, 6205~\AA{}, and 6615~\AA{} \citep{vogrincic_dibs_2023,chen_dibs_2013}. The former two wavelengths have been previously identified as DIBs by \citet{Ho_2017}, appearing as peaks in the linear coefficients of their \textsc{cannon} model. These features appear as broadened peaks in Figure \ref{fig:residual heatmap} because the interstellar features remain fixed in the observer frame. This results in smearing in the stacked residuals that reflects the radial velocity distribution of the training set \citep[][]{Ho_2017}. The DIB at 5782~\AA{} is the largest interstellar deviation in Figure \ref{fig:residual heatmap} with a residual dispersion of 1.5\%. The residuals between the reconstructed spectra and data also shows a median MAD value of $0.50\%$ at this line across all test stars. \textsc{the cannon} cannot predict DIBs since DIB strengths do not correlate with stellar parameters and abundance labels. \textsc{the cannon}'s inability to model these features can be exploited to identify DIBs \citep[see][]{mckinnon_data-driven_2023,Ness_review_2018,tchernyshyov_2018,buder_galahdr2_2018}.
\begin{table}
    \centering
    \begin{tabular}{l c c c l}
        \toprule
        Elements & Wavelength (\AA{}) & MAD (\%) & Std (\%)\\
        \midrule
         CH & 4355.7 & 0.53 & 1.00 \\
         MgH & 5090.6 & 0.66 & 1.08 \\
         MgH & 5115.5 & 0.32 & 0.60 \\
         MgH & 5139.8 & 0.36 & 0.62 \\
         MgH & 5142.4 & 0.33 & 0.59 \\
         MgH & 5166.5 & 1.16 & 1.90 \\
         Cu & 5702 & 0.42 & 0.65 \\
         DIB & 5782 & 0.50 & 1.50 \\
         DIB & 5798 & 0.33 & 0.60 \\
         DIB & 5799.5 & 0.35 & 0.57 \\
         DIB & 6205 & 0.36 & 0.82 \\
         CN & 6614.6 & 0.45 & 0.70 \\
         DIB & 6615 & 0.39 & 1.03 \\
         CN & 6685 & 0.39 & 0.67 \\
         Li & 6709.8 & 2.19 & 7.01 \\
        \bottomrule
    \end{tabular}
    \caption{Lines identified with \citet{hinkle_atlas_2000} that have significant residuals that were not measured by \citet{aquilina_2026}. We consider residuals to be significant if their median MAD values are greater than the noise level of $0.3\%$. Note that all wavelengths are in the vacuum rest frame.}
    \label{tab:idetification of unmeasured residuals}
\end{table}

\section{Discussion}
\label{sec:discussion}

We construct three \textsc{cannon} models: a 5-label model comprising basic stellar parameters (\teff, \logg, [M/H], \vmic, and \vsini), an 8-label model incorporating specific abundance labels ([Ti/H], [Y/H], and [Nd/H]), and a 9-label model that additionally includes [Eu/H]. We evaluate the performance of this model by inferring labels and generating model spectra for 220 red giant stars in the low-alpha disk with -1.13$<$ [M/H] $<$0.29~dex. The 8-label model generally recovers the reference labels within their measurement uncertainties, with a small tail extending slightly beyond the expected uncertainties. The inclusion of abundance labels also increases the confidence of the model. The 8-label model has a median model scatter of 0.003 for the 9\% of pixels with scatter greater than $10^{-6}$, while the 5-label models has a median model scatter of 0.004 for the 24\% of pixels with scatter greater than $10^{-6}$. The 8-label model also has smaller median MAD values around most absorption features. Across the entire wavelength range, on average the flux residuals between the 8-label model and the HARPS spectra for all test stars yield a MAD of 0.3\%, consistent with the typical median noise level (Section \ref{sec: global recon}). However, in detail, we demonstrate that individual spectral features across 29 elements retain additional unmodelled information. The amplitude of these residuals dictates the necessary measurement precision and SNR to access this variance. This variance beyond the low-dimensional model may hold key information regarding the underlying physical processes of element origins and subsequent mixing. We find that accessing this unmodelled information at the line core requires spectra with $\mathrm{SNR} \gtrsim 100$; only up to 6\% of test stars show residual amplitudes around specific absorption features exceeding 0.02, with the Eu line at 6646~\AA\ showing the highest fraction of residuals above this threshold. We highlight that this experiment does not measure the intrinsic dimensionality of stellar abundance space. Instead, it tests the extent to which the spectral information in this sample can be predicted from a low-dimensional chemical representation, and quantifies the information that remains: first in terms of flux residual amplitude, relevant to signal-to-noise requirements. This residual information can subsequently be translated into a corresponding abundance variation, providing a physical scale for the remaining chemical information.
\\\\
In Section \ref{sec: abundance diff}, we convert the flux residuals we measure away from the low dimensional model at the center of absorption features, into abundance differences. By measuring abundances from the reconstructed spectra following the procedure of \citet{aquilina_2026}, we explicitly quantify how model residuals translate into true abundance variance. In Section \ref{sec:correlations}, we analyse residual correlations and examine absorption features not explicitly tied to our training labels, demonstrating that these amplitudes capture independent chemical variance. These residual correlations could extend our understanding of the origin of each element beyond their classifications into nucleosynthetic families. Finally, to explicitly test the correlation between Eu and other elements, we evaluate the 9-label model in Section \ref{sec: discussion-Eu II label}. In Section \ref{sec: discussion-galah resolution}, we convolve the 8-label reconstructions and HARPS spectra to a medium resolution ($R = 28,000$) to assess how resolving power impacts the retrievable elemental information content.

\subsection{Converting Residuals Between Model and Observations to Abundance Differences}
\label{sec: abundance diff}
\begin{figure*}
	\includegraphics[width=\textwidth]{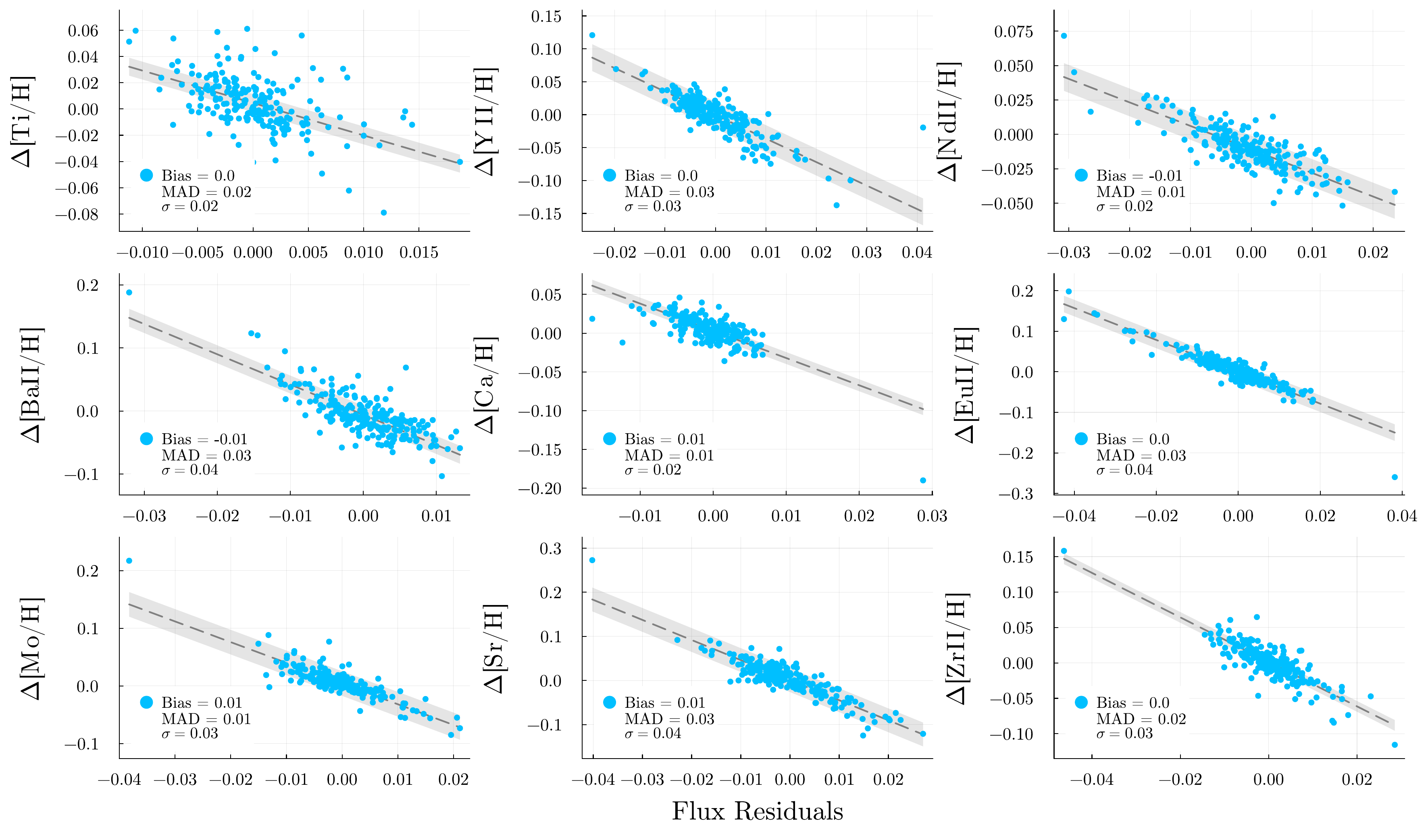}
    \caption{Mapping flux residuals to abundance differences for the 220 test stars. The abundance differences between those derived from the HARPS spectra and the 8-label \textsc{Cannon} model are shown against the corresponding flux residuals for three absorption features of each fitted abundance label (Ti 5026~\AA, Y 6797~\AA, Nd 5294~\AA) and six additional elements (Ba 5855~\AA, Ca 6168~\AA, Eu 6646~\AA, Mo 5508~\AA, Sr 4963~\AA, Zr 5351~\AA). The dashed line shows the fitted relation between flux residual and abundance difference, illustrating how a mismatch in flux translates into an abundance offset for each feature. The shaded region around the black dashed line is the median measurement uncertainty using values from \citet{aquilina_2026} propagated from flux uncertainty.}
    \label{fig:abundance diff}
\end{figure*}

We convert the flux residuals into abundance differences to quantify the physical impact of the discrepancies between the HARPS data and the 8-label \textsc{cannon} model. We restrict this analysis to absorption lines that have measured abundances in the catalogue of \citet{aquilina_2026} and are shown in Figures \ref{fig:Example line reconstruction} to \ref{fig:Example line reconstruction 3}. To ensure that any resulting abundance differences are driven strictly by the reconstructed flux, we adopt their exact line-by-line abundance derivation procedure using their stellar parameters for each star. This involves fitting synthetic spectra to the reconstructed spectra of each test star with \textsc{korg}'s \textsc{fit\_spectrum} $\chi^2$ minimisation method \citep[][]{wheeler_korg_2023}. We input the \textit{Gaia} ESO line list \citep[][]{heiter_linelist_2021} for lines $\lambda \geq 4750$~\AA{} and the \textsc{linemake} line list \citep[][]{placco_linemake_2021} for lines $\lambda < 4750$~\AA{}.  
\\\\
Figure \ref{fig:abundance diff} maps the derived abundance differences against the flux residuals for three absorption features for each abundance label (Ti 5026~\AA{}, Y 6797~\AA{}, Nd 5294~\AA) and six element lines (Ba 5855~\AA, Ca 6168~\AA, Eu 6646~\AA, Mo 5508~\AA, Sr 4963~\AA, Zr 5351~\AA). Using the abundance differences presented in Figure \ref{fig:abundance diff}, we calculate the scatter in abundance around the model prediction for each element as
\begin{equation}
    \sigma_{\Delta\mathrm{[X/H]}}
    =
    \mathrm{std}\left(\Delta\mathrm{[X/H]}\right).
\end{equation} 

\begin{figure}
	\includegraphics[width=\columnwidth]{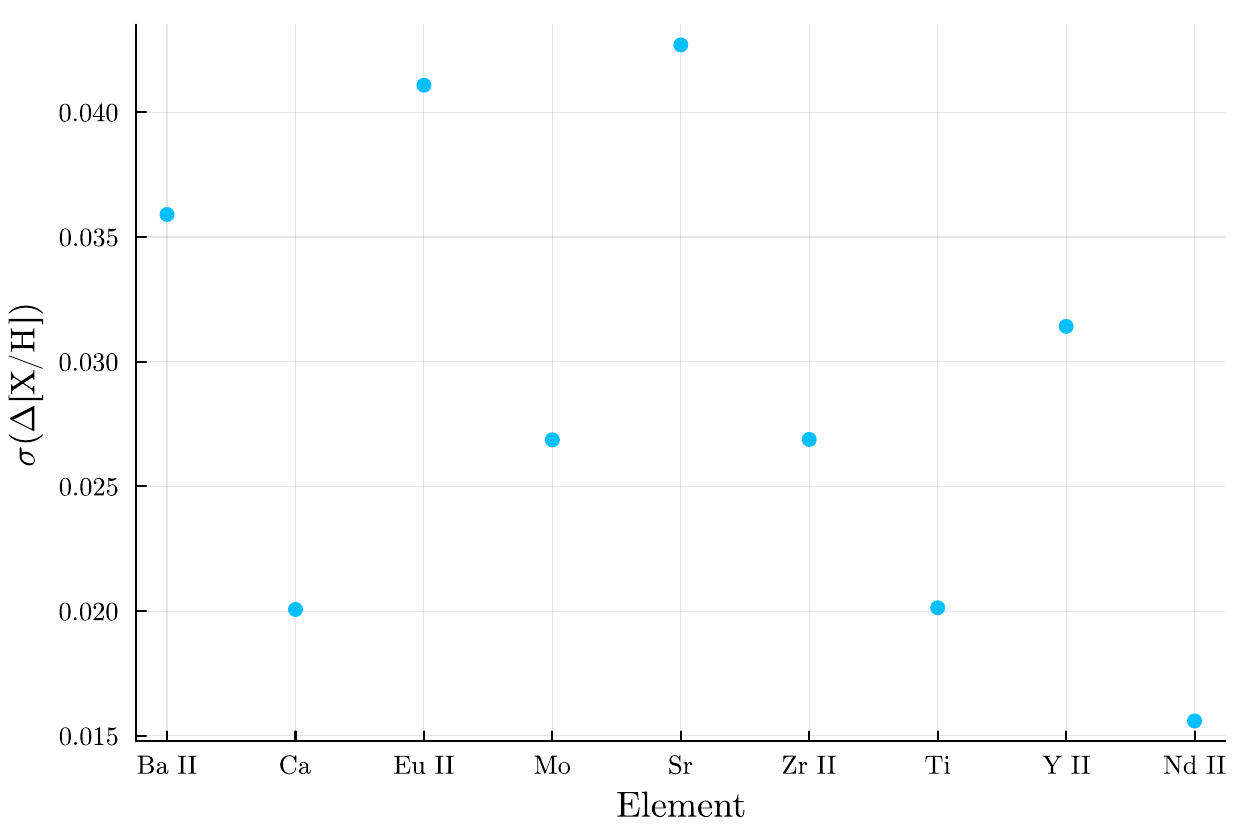}
    \caption{Standard deviation of the abundance differences derived in Figure \ref{fig:abundance diff} for three absorption features of each abundance label (Ti 5026~\AA{}, Y 6797~\AA{}, Nd 5294~\AA) and six additional elements (Ba 5855~\AA, Ca 6168~\AA, Eu 6646~\AA, Mo 5508~\AA, Sr 4963~\AA, Zr 5351~\AA).}
    \label{fig:std abundance diff}
\end{figure}

\noindent
The standard deviation of these abundance differences are shown in Figure \ref{fig:std abundance diff} for each absorption feature. The monotonic relationship between the small test set flux residuals and their corresponding abundance differences measures how spectral amplitude connects to abundance information not captured by the 8-label \textsc{cannon} model. Across all nine species, the model shows minimal systematic bias (approximately 0.0~dex) and overall scatter ($\text{MAD} \leq 0.03$~dex or $\sigma \lessapprox 0.04$~dex). We find that small flux residuals for our explicitly modelled labels can induce abundance variations up to $\sim$0.1 dex, although more typically reside around 0.01~dex to 0.05~dex.
\\\\
Each element will have a different abundance sensitivity to flux variation which we effectively quantify for these residual amplitudes at these stellar parameters. First, examining the elements used as training labels: in Figure \ref{fig:abundance diff}, Nd exhibits a maximum abundance deviation of 0.07~dex for a normalised flux residual of 0.03 although the abundance scatter is modest at $\sigma$ = 0.02 dex (and consistent with The Cannon's validation test). Despite the Ti and Y features also displaying small normalised flux residuals, a normalised flux residual of 0.01 corresponds to an abundance differences of up to 0.06~dex. The scatter for the Ti, Y and Nd abundance differences in Figure \ref{fig:std abundance diff} of 0.020, 0.031 and 0.016, respectively, are consistent with the median measurement uncertainties from the reference set. The median measurement uncertainties for each element line in \citet{aquilina_2026} are indicated by the shaded region in Figure \ref{fig:abundance diff} around the dashed fitted relation.
\\\\
We examine the abundance differences for the remaining 6 absorption features of non-label elements in Figure \ref{fig:abundance diff}. The residual flux differences correspond to a scatter in the element abundances ranging from 0.015 -- 0.043~dex for the elements Nd, Ca, Ti, Mo, Zr, Y, Ba, Eu and Sr (ordered from smallest to largest). Figure \ref{fig:std abundance diff} shows all training and test elements with the scatter of the abundance corresponding to the residual flux from the low-dimensional model: Ba, Ca, Eu, Mo, Sr, Zr, Ti, Y, Nd. However, we also find some stars show large deviations away from the low-dimensional model in some elements. The most extreme deviations in both flux and abundance consistently originate from two specific stars: HD 109379 (previously identified as an observational outlier due to saw-tooth residual patterns in Section \ref{sec:comp 5 and 8 model}) and HD 78668 (a Li-rich giant exhibiting an anomalous chemical correlation structure, Section \ref{sec: results-label accuracy}). These outliers highlight a critical physical scaling: for Eu and Sr lines, a 0.04 variation in flux translates to an abundance difference of 0.27~dex. This discrepancy is an order of magnitude larger than their respective median measurement uncertainties measured using \textsc{korg}, of 0.02~dex and 0.03~dex \citep{aquilina_2026}. The potential for subtle flux variations (0.04) to induce vast abundance differences (up to 0.27 dex) underscores the critical need for high-fidelity spectra.

\noindent
\\\\
We can compare the scatter in abundance around our model predictions with that from the 4-abundance regression model of \citet{mead_regression_2025}. Their model, analogous to ours but constructed directly in abundance space, predicts elemental abundances from stellar parameters, metallicity, and three elemental abundances. It was constructed for solar-metallicity main-sequence stars using the precision HARPS abundance measurements of \citet{Bedell_2018}. The two measures of elemental abundance scatter are not strictly equivalent: \citet{mead_regression_2025} infer an intrinsic abundance scatter after accounting for measurement uncertainties, whereas we measure the directly observed scatter in abundance residuals around the \textsc{cannon} model predictions. Our measurements may therefore include contributions from spectral noise and residual imperfections in the model, and should not be interpreted as a direct estimate of the same intrinsic scatter. Nevertheless, the characteristic scale is similar. Our Ba scatter of 0.032~dex is slightly higher than their $\sim0.02$~dex, while the scatter across all six elements common to the two studies remains at the few-hundredths-of-a-dex level. This suggests that the amount of abundance information remaining beyond a low-dimensional chemical description is comparably small across the two analyses, despite the different methodologies, evolutionary states, and the substantially wider metallicity range considered here ($-1.13 < \text{[M/H]} < 0.29$).

\subsection{Interpreting Unmodelled Spectral Variance}
\label{sec:correlations}

To test if the residual amplitudes in absorption features are likely to capture meaningful nucleosynthetic information or are more likely stochasticity, we examine the correlations between the flux residuals of 95 absorption features across 34 elements (including additional ionisation states). A lack of any residual correlation would indicate that all residual chemical information is independent, and suggest that residuals are  potentially a measure of stochasticity in the ISM. This test shows that correlations do persist in the residual amplitudes of the absorption features. Such correlations may indicate shared production pathways or formation timescales for elements. It may also be inherited from systematics shared between specific absorption features. 

\subsubsection{Residual Correlations between Absorption Features}
\label{sec:residual correlations}

We show the residual correlations (parameterised with the Pearson correlation coefficient) between 95 absorption features in Figure \ref{fig:residual correlations} across all 220 test stars, based on the 8-label \textsc{cannon} model. We see residual correlations of up to 0.80 between absorption features of elements within the same nucleosynthetic families. We find correlations of up to 0.44 between the light s-process elements (Sr, Y, Zr) and the (partial) r-process element Gd. We also find anti-correlations of up to -0.39 between these light s-process and the r-process element Eu. Similar anti-correlations between Eu and light s-process elements have been reported in previous studies \citep{roederer_2022, mead_regression_2025}. Additionally, the positive correlation we see between Gd and (Sr, Y, Zr) is in agreement with prior work \citet{mead_regression_2025}. This correlation implies that Gd could be partly produced in intermediate mass AGB stars \citep[possibly contributing to its s-process contributions, see ][]{prantzos_nc_contributions_2020} or via processes of similar timescales, whereas Eu may be formed by other production sites at a slower rate. Furthermore, Tb is anti-correlated with Gd (-0.35) and V at 6041~\AA{} (-0.38) but correlated with Cr at 5240~\AA{} (0.41) and Nd at 5622~\AA{} (0.51). Given that Tb lacks a strong correlation with Eu (up to 0.1 in magnitude) and has opposing correlations to Eu (with the exception of Gd), the correlations suggest either the presence of shared systematics among these features or a distinct production site for Tb that follows timescales similar to Nd. Sm shares this anti-correlation with V (-0.31) and may share these systematics or production sites with Tb. In contrast, Ru shows the strongest correlations with Ce (up to 0.46), Nd (0.37) and Mo (0.36), indicating that Mo and Ru are produced on a similar timescale or at a similar astrophysical site as the heavy s-process elements. However, residual correlations are less than 0.25 in Figure \ref{fig:residual correlations} for r-process elements Dy and Er, likely emphasising the existence of multiple r-process production sites rather than a single dominant source.

\subsubsection{Reconstruction Fidelity of Unmodelled Nucleosynthetic Pathways}
\label{sec: results-nlte and uncorrelated}

\begin{figure*}
	\includegraphics[width=\textwidth]{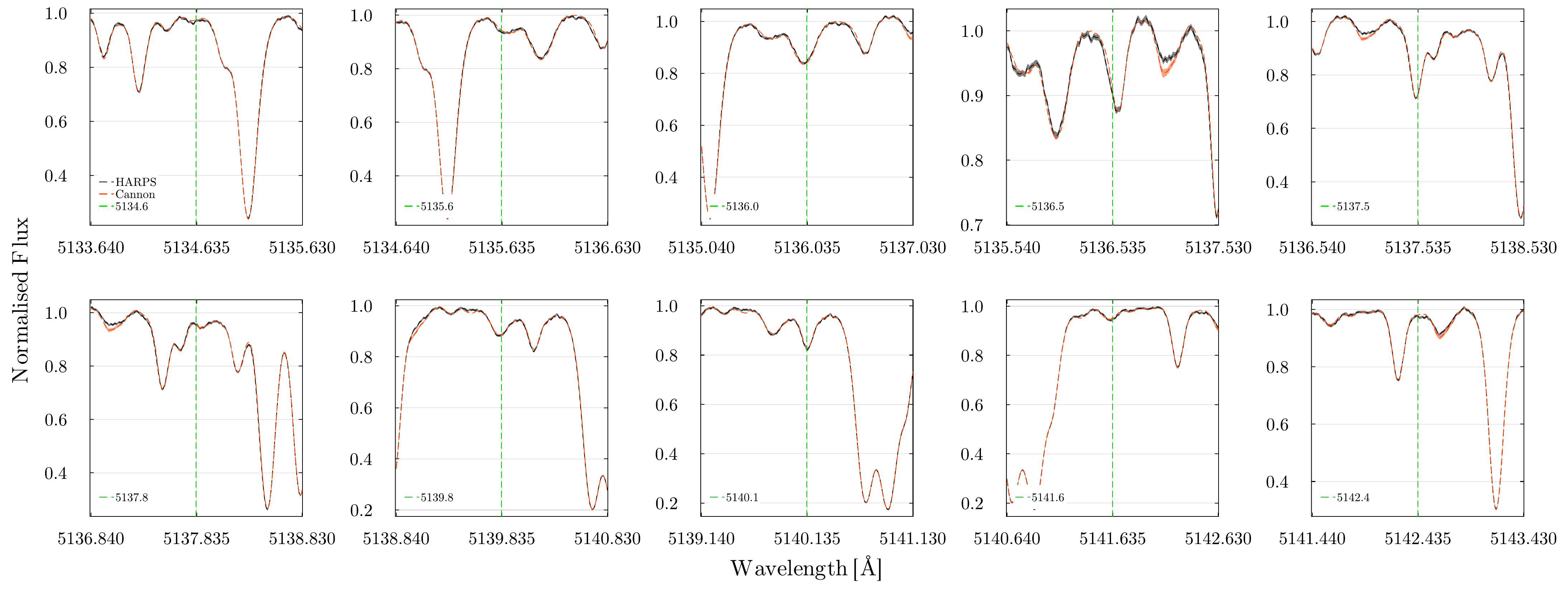}
    \caption{Observed HARPS spectrum (black) and reconstructed spectrum (red) from our 8-label \textsc{cannon} model for 10 MgH lines in HD 17793 ($\teff = 5127$~K, $\logg = 2.98$~dex, [M/H] $= -0.06$~dex) that are commonly used to measure isotopic ratios between $\mathrm{^{24}Mg}$, $\mathrm{^{25}Mg}$ and $\mathrm{^{26}Mg}$. The green dashed line indicates the central wavelength of each line. Note that all wavelengths are in the vacuum rest frame. We successfully reconstruct seven of these MgH lines within the expected noise despite the diverse origins of these Mg isotopes.}
    \label{fig:Example Mg isotopes}
\end{figure*}

The specific spectral features successfully recovered by any \textsc{cannon} model depends on the chosen training labels. Elements that do not correlate strongly with the provided abundance labels (ie., they are predominantly formed via different nucleosynthetic processes) will not be reproduced to the same fidelity as those that do. We discuss this limitation for Na and Eu absorption features, which exhibit the largest residuals among our 95 analysed features (excluding Li, see Sections \ref{sec:comp 5 and 8 model} and \ref{sec: results-unmodelled features}), and whose production is not included in our chosen abundance labels. We further test whether or not this limitation hinders the recovery of individual absorption features that blend isotopic contributions from multiple production sources.
\\\\
The standard deviation of the normalised flux residuals in Figure \ref{fig:line sigma int} is 2.23 for the Na line, indicating that the residual distribution is approximately 2.2 times broader than expected from measurement uncertainties alone. This lower recovery for Na is perhaps unsurprising, since we do not include any light element or odd-Z element labels to capture the distinct nucleosynthetic pathways of these element families \citep[][]{owusu_2026,kobayashi_origin_2020,Pagel_evolution_1997}. The model's ability to partially recover Na for some test stars is likely driven by the [M/H] label, which captures its metallicity-dependent production. Similarly, we do not explicitly trace r-process element production in the 8-label \textsc{cannon} model, but are able to partially reconstruct Eu across our parameter space as shown in Figures \ref{fig:Example line reconstruction} to \ref{fig:Example line reconstruction 3}. We portray the recovery rate for 1 of the 2 measured Eu lines (6646~\AA{}) as the distribution of residuals and normalised residuals (see Equation \ref{eqn:normalised_residuals}) in Figure \ref{fig:Eu II line hist 8 param}. The standard deviation of the normalised residual distribution is 2.32, substantially larger than the value of unity expected from observational uncertainties alone, indicating significant residual structure not captured by the 8-label model. The low fidelity demonstrated by the normalised residuals is again perhaps unsurprising,  as Eu is an r-process element whose specific nucleosynthetic variance is not explicitly captured by our abundance labels. The partial reconstruction of Eu may be because it is weakly correlated with $\alpha$-elements and anti-correlated with Fe in the \citet{aquilina_2026} catalogue. Previous works have also demonstrated these relationships between Eu and $\alpha$-element with metallicity \citep[][]{battistini_neutron_2016,delgado_mena_chemical_2017,mead_regression_2025,carrasco_2025}. The model therefore seems to be able to leverage our stellar parameter labels and $\alpha$-like correlations to predict a baseline for Eu.
\\\\
The model reconstructs the Mg line at 5712~\AA{} within uncertainties ($\sigma = 0.9$ in Figure \ref{fig:line sigma int}). This is likely due to the correlation between Mg and our [Ti/H] label \citep{aquilina_2026}. However, the production of Mg isotopes is more complex. While $\mathrm{^{25}Mg}$ and $\mathrm{^{26}Mg}$ are co-produced with $\mathrm{^{24}Mg}$ in massive stars, they also form in the outer carbon layers of intermediate-mass AGB stars \citep{carlos_2018,vangioni_2019}. Figure \ref{fig:Example Mg isotopes} compares the HARPS spectra to our 8-label reconstruction of HD 17793 around ten MgH lines used by \citet{mckenzie_2024} to measure $\mathrm{^{24}Mg}$, $\mathrm{^{25}Mg}$ and $\mathrm{^{26}Mg}$. These ten features are generally recovered well with only two features (5139.8~\AA{} and 5142.4~\AA{}) demonstrating median MAD values across all test stars at slightly above the noise level of 0.3\% across all test stars in Table \ref{tab:idetification of unmeasured residuals}. The recovery of these Mg lines is likely due to a minimal diversity in the Mg isotopic ratios of the reference set. As a result, the broader correlation between Mg and Ti is generally sufficient to capture the combined spectral contributions of the various Mg isotopes.

\subsubsection{Residual Trends across the Milky Way Disk}

Here we examine residuals of our 95 absorption  across radius and height from the plane to test if there is any physical correlation with unmodeled spectral variance. We bin the data for each of the 95 absorption features by Galactic radius (3 bins at 7.5~kpc, 8.2~kpc and 8.9~kpc) and distance from the Galactic plane, $\vert{}z\vert{}$ (2 bins at 0.05~kpc and 0.3~kpc). These parameters are obtained using the parallax and coordinates from \textit{Gaia} DR3 \citep[][]{gaia_2023}. We consider a feature to exhibit a notable spatial trend if the absolute flux residual difference between the first and last bin exceeds 0.006 for Galactic radius or 0.004 for $\vert{}z\vert{}$. Based on this criteria, 8 features show trends with Galactic radius (Ce 4524 \AA, Li 6710 \AA, Fe 6616 \AA, Fe 5800 \AA, Ti 5490 \AA, Y 5548 \AA, Cr 4426 \AA, and Sc 5688 \AA) and 8 show trends with $\vert{}z\vert{}$ (Li 6710 \AA, Fe 6422 \AA, Fe 6616 \AA, Fe 6636 \AA, Fe 6786 \AA, Zr 4963 \AA, Al 6698 \AA, and Cr 4426 \AA). 
\\\\
For both (R,z) coordinates, the majority of these trends are positive with amplitude differences ranging from $\Delta = 0.004$ to $0.008$. The abundance of Li is known to vary across the Galactic disk \citep{bandyopadhyay_2022, sun_2025}. We see this in our data as opposing residual amplitudes between the first (+0.005 at R = 7.5~kpc) and third (-0.007 at R = 8.9~kpc) radial bins. In addition to Li 6710 \AA, Fe 6616 \AA, and Cr 4426 \AA,  display the largest magnitude differences in residual amplitudes of $\Delta = 0.011--0.026$ between R = 7.5~kpc and 8.9~kpc bins. \textsc{the cannon} over-predicts the line depths of the Fe 6616 \AA, and Cr 4426 \AA\ absorption features across our test set; this may be because they reside in DIB regions. The presence of DIBS could drive the radial dependence of the amplitude of these features. Finally, only Ce (with Galactic radius) and Zr residuals (with $\vert{}z\vert{}$) exhibit negative trends in their flux residuals across radius. Previous studies have noted weak, debated spatial trends for Ce and Zr abundances \citep[see][and references therein]{stonkute_2021, hijarrubia_2024}. Our residual trends indicate that where our model fails to fully capture the independent information in Ce not correlated with the training labels is tied to physical changes in Ce enrichment across radius. 

\subsection{The Predictive Power of Eu}
\label{sec: discussion-Eu II label}
\begin{figure}
	\includegraphics[width=\columnwidth]{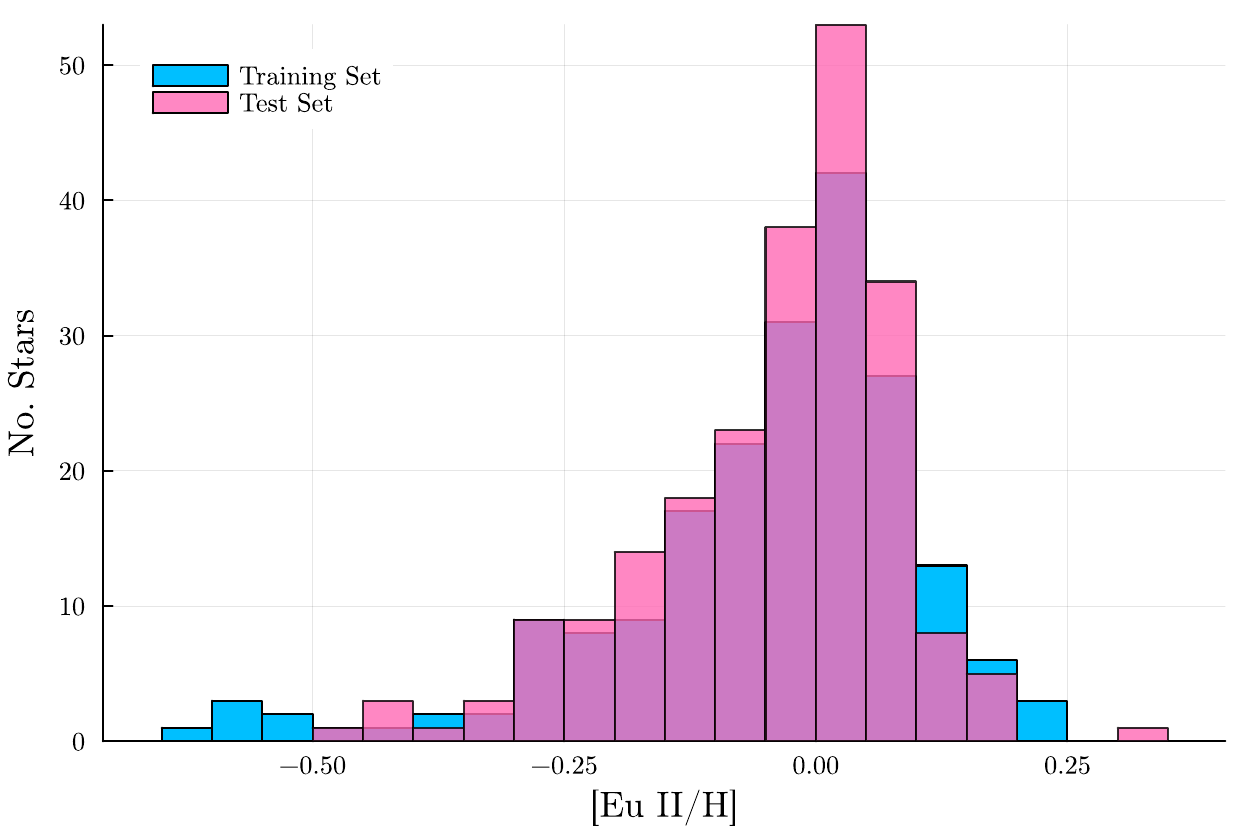}
    \caption{Distribution of the Eu II abundances in the training and test set for the 9-label \textsc{cannon} model.}
    \label{fig:Eu II distribution}
\end{figure}

\begin{figure}
	\includegraphics[width=\columnwidth]{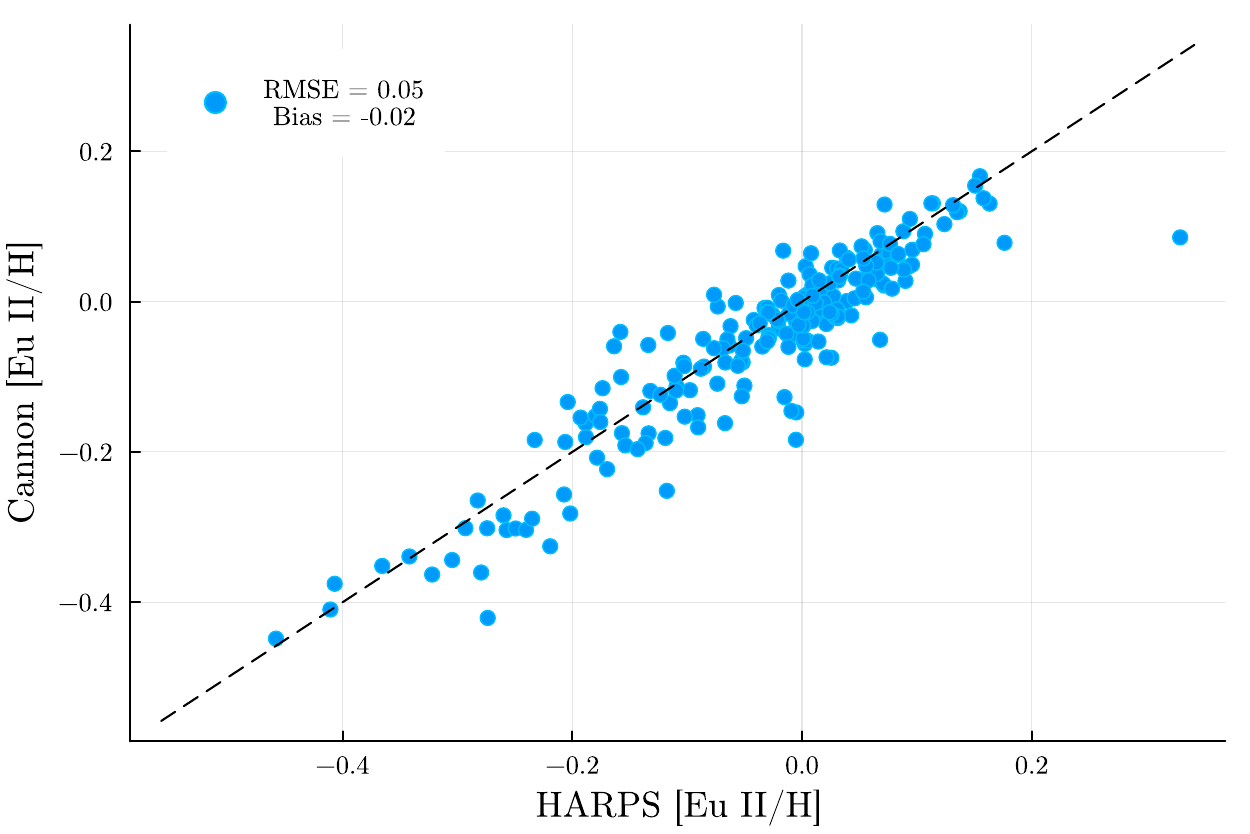}
    \caption{One-to-one plot of the \citet{aquilina_2026} Eu II abundances and those from the 9-label \textsc{cannon} model. The small bias indicates the lack of systematic trends and the RMSE is comparable to measurement uncertainties.}
    \label{fig:Eu II comp labels}
\end{figure}

Previous investigations have established that Eu exhibits a unique r-process scatter that distinguishes it from the other elements \citep[][]{battistini_neutron_2016,delgado_mena_chemical_2017}, limiting its ability to predict the their abundances in data-driven models \citep[][]{mead_regression_2025}. However, few studies have tested whether including an explicit r-process label improves the reconstruction of the spectrum itself (flux prediction) rather than directly predicting chemical abundances. We investigate the predictive power of Eu by training a 9-label \textsc{cannon} model including [Eu/H] as a fourth abundance label (\teff, \logg, [M/H], \vsini, \vmic, [Ti/H], [Y/H], [Nd/H] \& [Eu/H]). Figure \ref{fig:Eu II distribution} shows the distribution of the training and test set Eu abundances used in the 9-label model from the \citet{aquilina_2026} catalogue. We also include a one-to-one plot of the transferred Eu abundances from the 9-label \textsc{cannon} model and the reference set in Figure \ref{fig:Eu II comp labels} to demonstrate the fidelity of the model.

\noindent
\\\\
Including [Eu/H] does not improve the generative model of most elements; this yields a median normalised residual standard deviation of 1.3 across the 95 lines (similar to the median of 1.2 from the 8-label model). This includes elements such as Dy, Pr, Ru, Sm, Tb, Mo, Nd and Gd that have r-process contributions but do not show an improvement in their median normalised residual standard deviation in the 9-label model. The most significant recovery improvement is simply for Eu itself with the standard deviation of the normalised residuals reducing from 2.32 to 1.84 in Figure \ref{fig:line sigma int}. When focusing on our representative element lines in Figure \ref{fig:line sigma int}, we see improvements in the standard deviation of normalised residuals from the 8-label to the 9-label model of 1.12 to 0.90 in Ca, 1.91 to 1.64 in V, 2.32 to 1.84 in Na, and 18.38 to 14.48 in Li. The addition of Eu likely captures systematic trends that improve the recovery of Li and Na, while its nucleosynthetic correlation with the $\alpha$-elements (Section \ref{sec: results-nlte and uncorrelated}) facilitates the improved recovery of Ca and the Type II supernovae production of V. However, the marginal improvement to the global spectral reconstruction physically validates that Eu acts as a largely independent chemical dimension for Milky Way disk stars in the $-1.13 < \mathrm{[Fe/H]} < 0.29$~dex regime.

\subsection{Simulating The Cannon performance at medium-resolution (28000)}
\label{sec: discussion-galah resolution}
We investigate how spectral resolution limits the recovery of subtle chemical trends and restricts the observable dimensionality of chemical space. To do this, we calculate the line spread function at a resolving power of $R=28,000$ using the \textsc{compute\_lsf\_matrix} function in \textsc{korg} and apply this to both the observed HARPS spectra and the reconstructed \textsc{cannon} spectra of the 8-label model. The resolution of the convolved observed and reconstructed spectra matches that of the GALAH survey \citep[HERMES spectrograph,][]{DeSilva_2015,buder_galahdr4_2024} and approximates the regime of upcoming facilities like 4MOST \citep[$ 18000 < R < 21000$,][]{dejong_4most_2019}.
\\\\
The fine-scale residuals identified in Section \ref{sec:results} are suppressed after convolution to R=28,000, with the median MAD decreasing from 0.3\% to 0.2\%. However, the residual signal is distributed across several wavelength samples by the GALAH line-spread function and overall element detectability depends on the combined residual across the full line profile rather than the amplitude at a single wavelength pixel. For a feature sampled by (N) independent pixels, the detection significance would increase approximately as $\sqrt{N}$. The decrease in the residual amplitude indicates a loss of accessible chemical information, but this does not scale directly with the per-pixel mean residual change.
\\\\
We re-evaluate the recovery of the 95 lines with measured abundances in our reference set using the convolved data and flux uncertainties from the HARPS spectra. At $R=28,000$, the median standard deviation of the normalised flux residuals across the 95 lines decreases from 1.2 at high resolution to 0.7. The per element recovery remains largely similar to the high-resolution analysis: Li and Na remain the most difficult to reconstruct ($\sigma = 15.7$ and $2.0$ in normalised residuals respectively), while all other elements (excluding Eu) are recovered either better than expected or as expected from their corresponding measurement uncertainty ($\sigma < 1.2$ in normalised residuals). This improved recovery of specific absorption features (shown in Figure \ref{fig:line sigma int}) is likely driven by the smoothing of complex line blends and small wavelength mismatches that dominate our high-resolution residuals in Section \ref{sec: results-line characterisation}. However, Li, Al, V, Cr, Y, Zr, Mo, Nd and Eu show significant improvement in recovery in Figure \ref{fig:line sigma int} with the standard deviation decreasing by at least 0.5. This suggests an apparent reduction in chemical dimensionality, where the chosen labels appear sufficient only because the subtle spectral signatures of detailed nucleosynthesis are no longer resolvable.

\section{Conclusions}
\label{sec:concl}

It is now understood that abundances are highly correlated. This means that we are more restricted than optimistically imagined in the pursuit of chemical tagging. That is, reconstructing disk clusters using their nucleosynthetic fingerprints. This pursuit is penalised by the limited chemical space, in practice. However, patterns in the residuals from low-dimensional predictive models may reveal chemical information not captured by the dominant abundance correlations, and may ultimately help constrain the physical processes that generate these abundances.  that generate these abundances. These patterns may uncover links to discrete signatures of birth sites, interstellar conditions at different sites in the disk, or individual stellar conditions associated with planet or binary formation or mass transfer. To evaluate the fidelity of a low-dimensional chemical representation, we reconstruct we reconstruct the spectral flux of red giant stars with 5-, 8- and 9-label \textsc{cannon} models (\teff, \logg, [M/H], \vmic, \vsini, [Ti/H], [Y/H], [Nd/H] \& [Eu/H]) trained on HARPS ($R=115000$) spectra. All stellar parameters and abundances are from the high-precision \citet{aquilina_2026} catalogue of 426 red giant stars. 
\\\\
We demonstrate that an 8-label \textsc{cannon} model, utilising only stellar parameters (\teff, \logg and [M/H]) and three abundance labels (from different nucleosynthetic families), accurately predicts the spectral flux across a wide metallicity range ($-1.13 < \text{[M/H]} < 0.29$) in the Milky Way disk. This global reconstruction achieves a typical fidelity of $\mathrm{MAD}=0.3\%$, while the full residual distribution is slightly broader than expected from the observational noise. There are small, key flux deviations between the generated model and observations at the cores of many absorption features. Overall however, the successful reconstruction of the spectra using a physically incomplete label set demonstrates that much of the chemical information encoded in low-\(\alpha\) disk spectra is predictable from a low-dimensional chemical representation. The strong covariance between elements is directly demonstrated by the accurate recovery of individual absorption features from elements that are not included as training labels. Across 95 measured lines in \citet{aquilina_2026} within our wavelength region, we successfully recover the line fluxes to less than 0.02 for up to 94\% of test stars.  
\\\\
Despite the 8-label \textsc{cannon} model reproducing spectra to the sub-percent level, 1\% flux residuals can translate to abundance deviations of 0.01--0.05~dex for the elements Ba, Ca, Eu, Mo, Nd, Sr, and Zr. We directly quantify this to inform both measurement precision and SNR needed to tap into this conditional information away from a low dimensional model. The Eu 6646~\AA{} line exhibits the largest uncaptured variance: a residual of 0.04 corresponds to a 0.27~dex abundance difference. The monotonic relationship between test-set flux residuals and their corresponding abundance differences provides an empirical mapping between residual flux amplitude and the abundance-scale information not captured by the 8-label \textsc{cannon} model. High-resolution spectra (to avoid line blending) with high-SNR are required to reliably extract these departures from low-dimensional predictions. We find the most substantial flux residuals at the lightest element features (Li, with 54\% of residuals above SNR $>100$), molecular bands (CH with $\sigma = 0.01$ and MgH with $\sigma \leq 0.019$), and Diffuse Interstellar Bands ($\sigma \leq 0.015$). Including a light element label may improve the modelling of the Li I doublet and molecular features.
\\\\
We construct a 9-label \textsc{cannon} model to test the predictive power of Eu as an abundance label. The inclusion of Eu only marginally improves the global spectral reconstruction, with improvements to the standard deviation of the normalised residuals up to 0.3 (excluding Eu and Li absorption features). This indicates that Eu contributes relatively little independent predictive information beyond that already captured by the 8-label model. We observe marginally improvements to the distribution of normalised residuals for Ca, V, Na and Li in the 9-label model. These improvements suggest that [Eu/H] contains some predictive information through its correlations with other elements, but contributes little additional information to the global spectral reconstruction. The fact that adding Eu produces only localised improvements, rather than a substantial improvement in global spectral fidelity, reinforces the predictive sufficiency of the underlying 8-label model for this sample.
\\\\
We analyse the apparent dimensionality at medium resolution by convolving the observed and reconstructed spectra of the 8-label \textsc{cannon} model to (R=28000), comparable to GALAH and the 4MOST high-resolution arm. The median residual amplitude decreases from MAD=0.3\% to 0.2\% after convolution. At this resolution, the residual signal associated with individual chemical features is redistributed across the instrumental line profile, and the weaker abundance information identified at high resolution is less spectrally distinct. The corresponding absolute amplitude of the residual structure identified at HARPS resolution is therefore reduced and distributed over a broader wavelength range at lower resolution.
\\\\
Modelling dominant chemical variance in stellar spectra as we have done in this work provides a way to isolate and characterise the weaker spectral signatures associated with different element families. We have demonstrated that stellar parameters, including metallicity and only three abundance labels drawn from different nucleosynthetic families capturse most of the spectral variation in our low-$\alpha$ disk sample. Compared to lower resolution tests, the weaker abundance information that remains is most clearly resolved in the high-resolution HARPS spectra. At high resolution, individual residual features are less affected by blending and retain larger amplitudes. Extending the modeling approach introduced here over a broader metallicity range and to other evolutionary phases will allow us to investigate how the information carried by individual absorption features away from a dominant baseline varies with multiplicity \citep[][]{elbadry_2024}, planet-host status \citep[][]{ge_2022,michel_2024,delvaerny_2025}, age, and Galactic orbital parameters. The measured residual amplitudes provide an empirical benchmark for future observing strategies, in particular data quality and signal to noise requirements, and for assessing the chemical information that is accessible to large spectroscopic surveys such as 4MOST.

\section*{Acknowledgments}

MKN acknowledges support from the Australian Research Council under grant number DP260102813. SB acknowledges support from the Australian Research Council under grant number DE240100150.

Some text edits and language refinement during the preparation of this work were performed in highly supervised engagement with AI language models Gemini and ChatGPT.

Based on data obtained from the ESO Science Archive Facility with DOI: https://doi.org/10.18727/archive/33. This work also makes use of data from the European Space Agency (ESA) space mission Gaia. Gaia data are being processed by the Gaia Data Processing and Analysis Consortium (DPAC). Funding for the DPAC is provided by national institutions, in particular the institutions participating in the Gaia MultiLateral Agreement (MLA). The Gaia mission website is https://www.cosmos.esa.int/gaia. The Gaia archive website is https://archives.esac.esa.int/gaia.

\section*{Data Availability}
All input data used are publicly available or can be requested from the lead author.

\bibliographystyle{mnras}
\bibliography{mnemonic,main} 

\appendix
\section{Supplementary Label Validation}
\label{appendix:supplementary label validation}
\begin{figure*}
	\includegraphics[width=\textwidth]{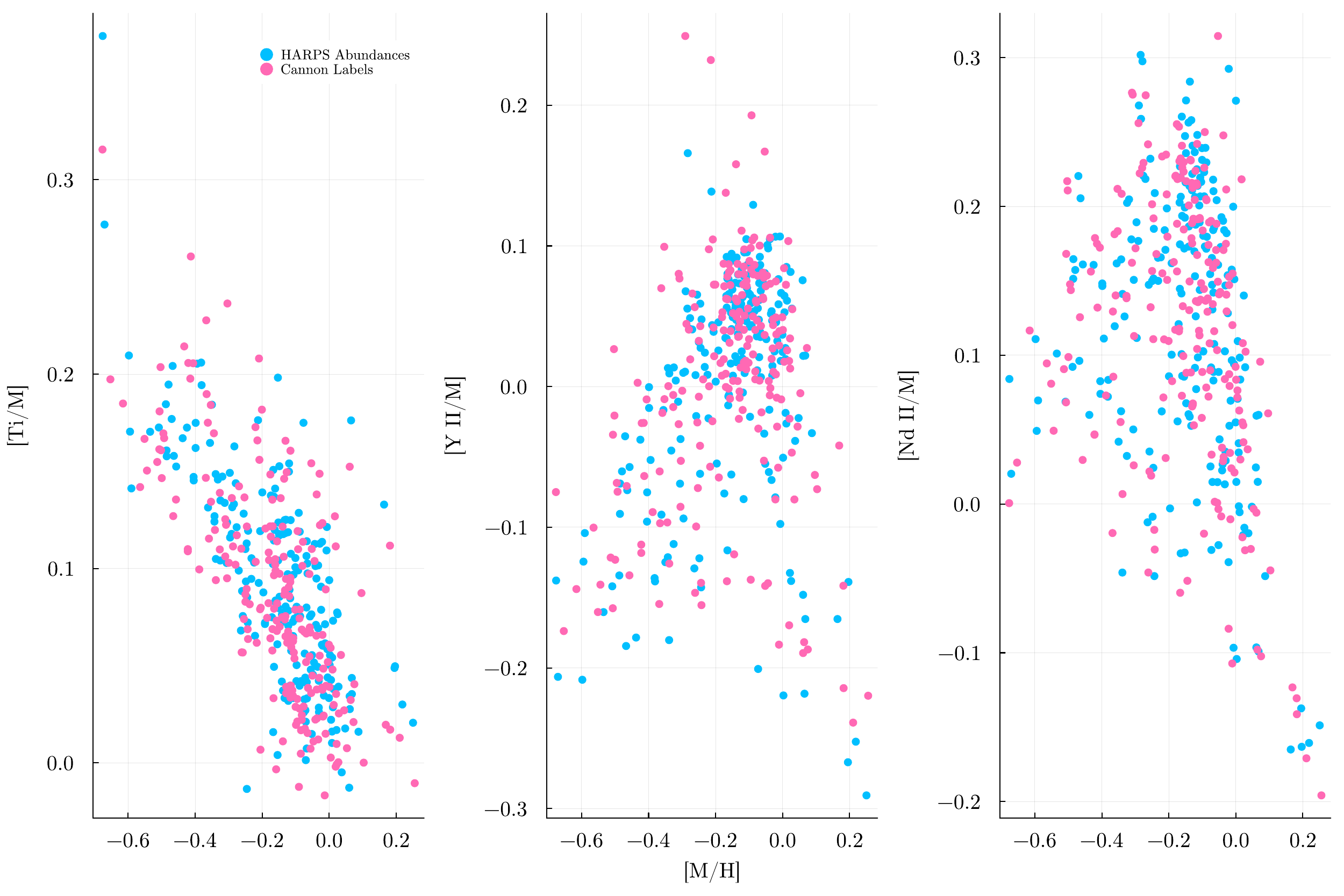}
    \caption{Abundances derived by \citet{aquilina_2026} in blue and the recovered abundances from the 8-label \textsc{cannon} model in pink for our labels Ti, Y and Nd with respect to [M/H]. The consistency between the abundance trends of the reference set and the transferred labels demonstrates the high-fidelity of our 8-label \textsc{cannon} model. See Section \ref{sec: results-label accuracy} for a one-to-one comparison.}
    \label{fig:xm label abundance trends}
\end{figure*}

To further demonstrate the high fidelity of our label transfer discussed in Section \ref{sec: results-label accuracy}, we include Figure \ref{fig:xm label abundance trends}. Figure \ref{fig:xm label abundance trends} shows the abundances derived by \citet{aquilina_2026} for our chosen labels (Ti, Y and Nd) and the recovered labels from the 8-label \textsc{cannon} model with respect to [M/H]. The trend of the recovered abundances labels is consistent with the trend in the reference set, demonstrating that the high-fidelity of our model.

\section{Supplementary Model Validation}
\label{appendix:label spectra reconstruction}

Following the validation of our label transfer (Section \ref{sec: results-label accuracy}) and global spectral reconstructions (Section \ref{sec: spectrum validation}), this appendix provides two supplementary diagnostics. First, we present full spectral reconstructions for a typical test star to visually demonstrate the high fidelity of the 8-label model across all wavelengths. Second, we analyse the  linear coefficients of the 8- and 9-label models to ensure that each label correctly traces its corresponding nucleosynthetic pathway ([Ti/H] for $\alpha$-elements, [Y/H] for light s-process, [Nd/H] for heavy s-process, and [Eu/H] for r-process elements). Together, these checks confirm that \textsc{the cannon} models are driven by true physical variations rather than unmodelled systematic artifacts.

\subsection{Full Spectral Reconstructions}
\begin{figure*}
	\includegraphics[width=\textwidth]{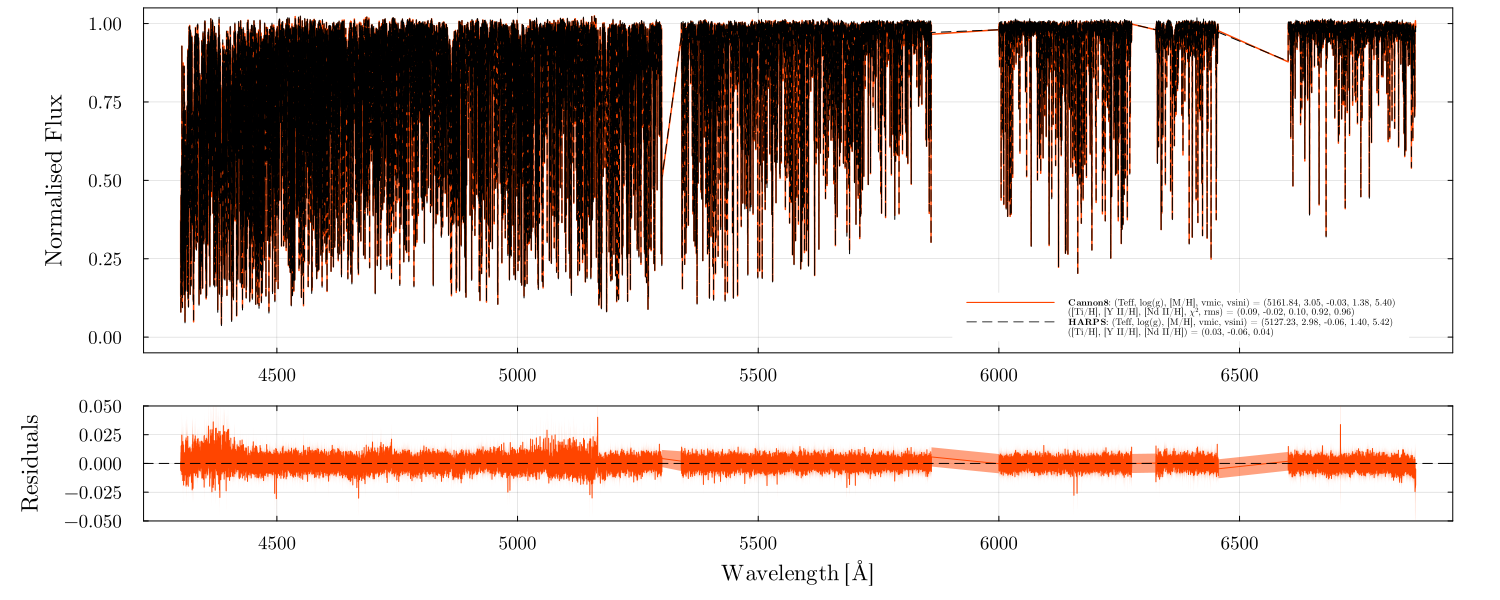}
    \caption{Top: Sample spectrum of HD 17793 ($\teff = 5127$~K, $\log g = 2.98$~dex, [M/H] $= -0.06$~dex) showing the HARPS data and 8-label \textsc{cannon} model spectra. Bottom: Residuals between HARPS and the 8-label \textsc{cannon} model. We include labels from the reference set and 8-label model for comparison.}
    \label{fig:Full example spectrum and cannon}
\end{figure*}

\begin{figure}
	\includegraphics[width=\columnwidth]{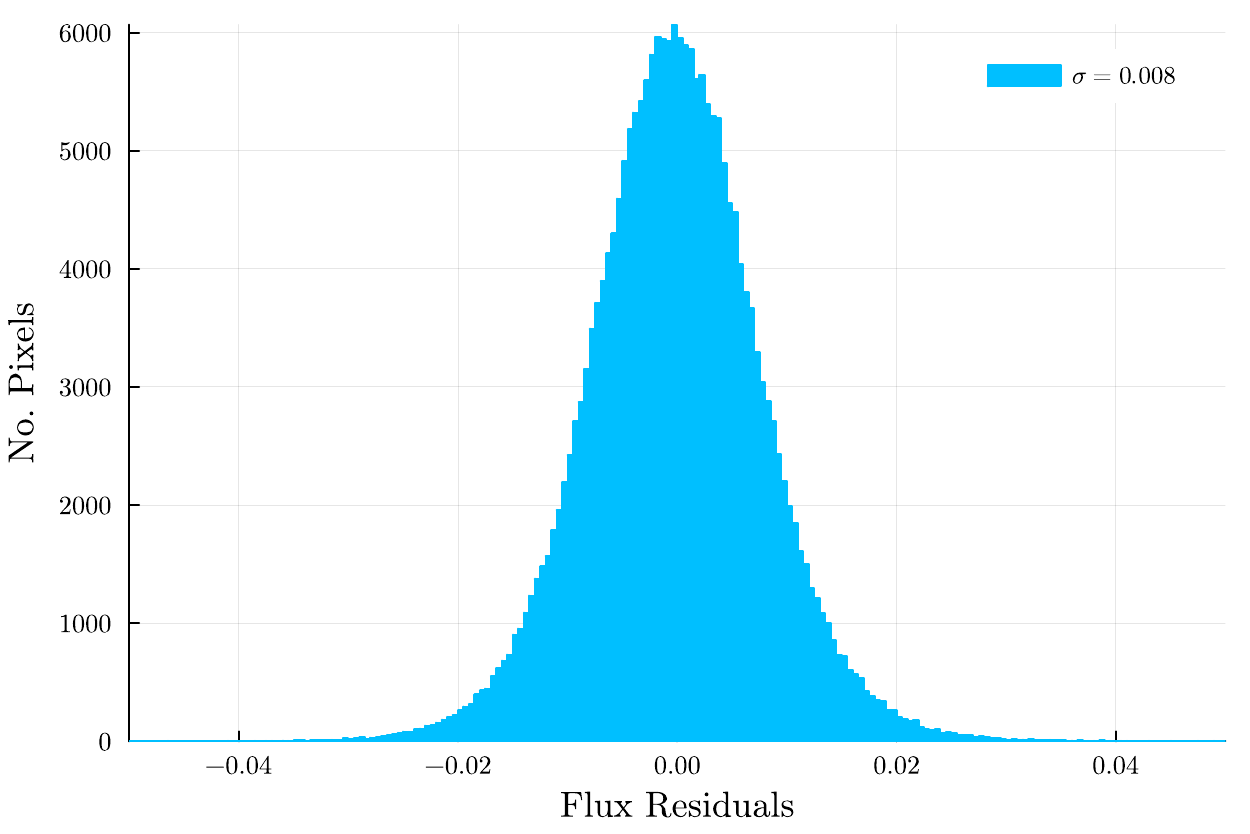}
    \caption{Distribution of flux residuals between the HARPS spectrum and 8-label reconstruction of HIP20226 ($\teff = 4722$~K, $\log g = 2.15$~dex, [M/H] $= -0.67$~dex). The standard deviation of 0.008 corresponds to a standard deviation of 1.101 when the flux residuals are normalised by the flux uncertainties.}
    \label{fig:HIP20226 residual distribution}
\end{figure}

\begin{figure}
	\includegraphics[width=\columnwidth]{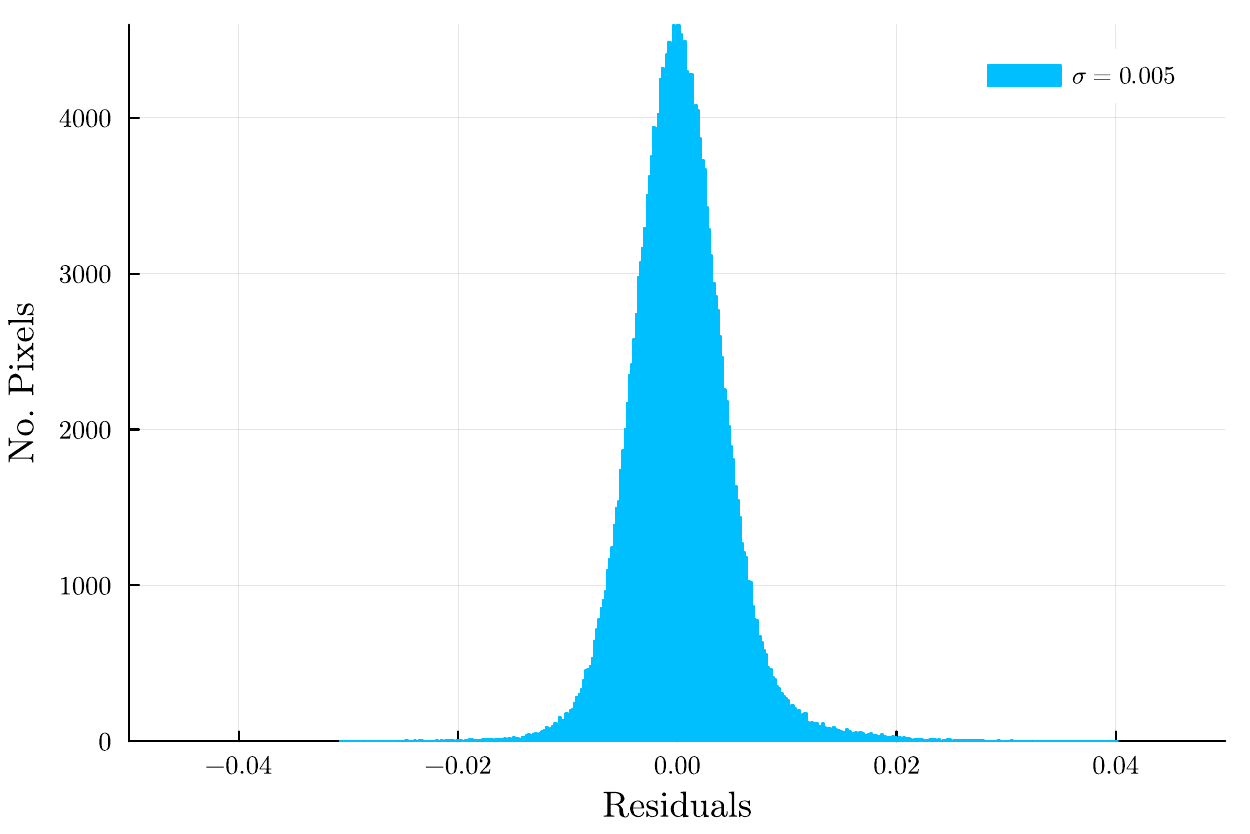}
    \caption{Distribution of flux residuals between the HARPS spectrum and 8-label reconstruction of HD 17793 ($\teff = 5127$~K, $\log g = 2.98$~dex, [M/H] $= -0.06$~dex). The standard deviation of 0.008 corresponds to a standard deviation of 1.051 when the flux residuals are normalised by the flux uncertainties.}
    \label{fig:HD17793 residual distribution}
\end{figure}

\begin{figure}
	\includegraphics[width=\columnwidth]{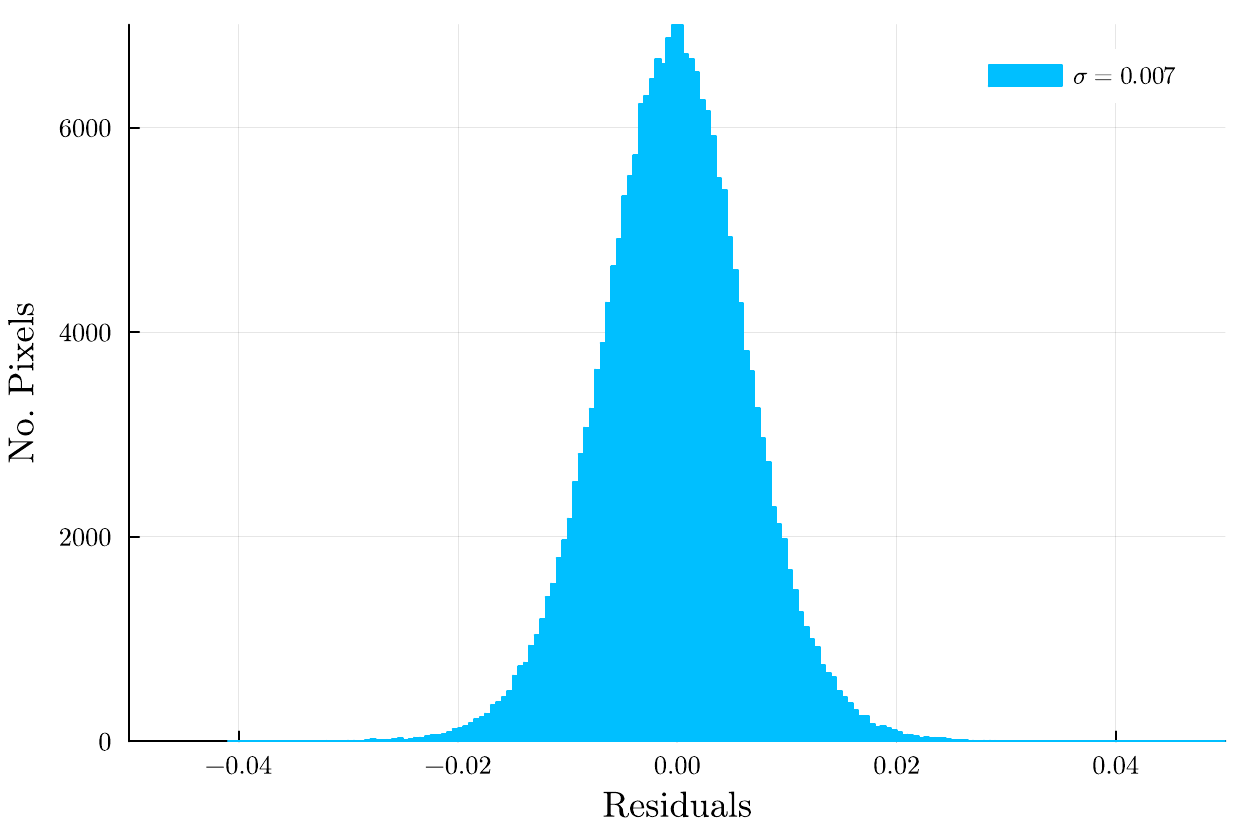}
    \caption{Distribution of flux residuals between the HARPS spectrum and 8-label reconstruction of HD 176704 ($\teff = 4457$~K, $\log g = 2.38$~dex, [M/H] $= 0.22$~dex). The standard deviation of 0.008 corresponds to a standard deviation of 0.912 when the flux residuals are normalised by the flux uncertainties.}
    \label{fig:HD176704 residual distribution}
\end{figure}

To further validate the 8-label \textsc{cannon} model across the entire wavelength range (4300 to 6900 \AA{}), Figure \ref{fig:Full example spectrum and cannon} presents the complete HARPS spectrum, the model reconstruction, and the corresponding residuals for a representative star HD 17793. We provide the HARPS spectrum, model reconstruction, residuals and normalised residuals for stars HIP 20226, HD 17793 and HD 176704 across 6100 to 6200~\AA{}. We also include distributions of the flux residuals for stars HIP 20226, HD 17793 and HD 176704 to show the success of the 8-label model across the entire wavelength range at different metallicities in Figures \ref{fig:HIP20226 residual distribution} to \ref{fig:HD176704 residual distribution}, respectively.

\begin{figure*}
	\includegraphics[width=\textwidth]{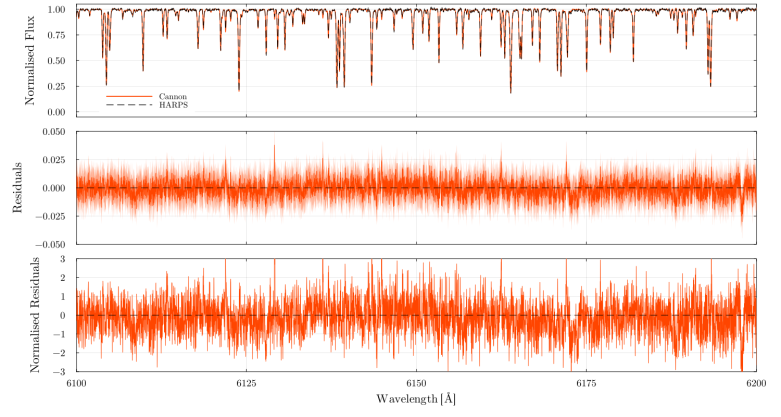}
    \caption{\textit{Top:} Sample spectrum from 6100~\AA{} to 6200~\AA{} of HIP20226 ($\teff = 4722$~K, $\logg = 2.15$~dex, [M/H] $= -0.67$~dex) showing the HARPS data and the successful spectral reconstruction of the 8-label \textsc{cannon} model. \textit{Middle:} Residuals between HARPS and the 8-label \textsc{cannon} model. \textit{Bottom:} Residuals between HARPS and the 8-label \textsc{cannon} model normalised by the flux uncertainties (see Equation \ref{eqn:normalised_residuals}) to demonstrate that most residuals are within 2$\sigma_{\lambda}$. Note that all wavelengths are in the vacuum rest frame.}
    \label{fig:Example spectrum and cannon 1 red}
\end{figure*}

\begin{figure*}
	\includegraphics[width=\textwidth]{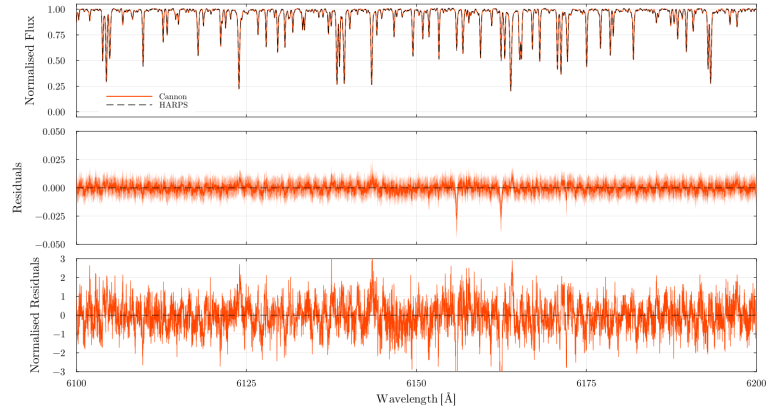}
    \caption{Format identical to Figure \ref{fig:Example spectrum and cannon 1 red} but instead showing the spectra, residuals and normalised residuals of HD 17793 ($\teff = 5127$~K, $\log g = 2.98$~dex, [M/H] $= -0.06$~dex)}
    \label{fig:Example spectrum and cannon 2 red}
\end{figure*}

\begin{figure*}
	\includegraphics[width=\textwidth]{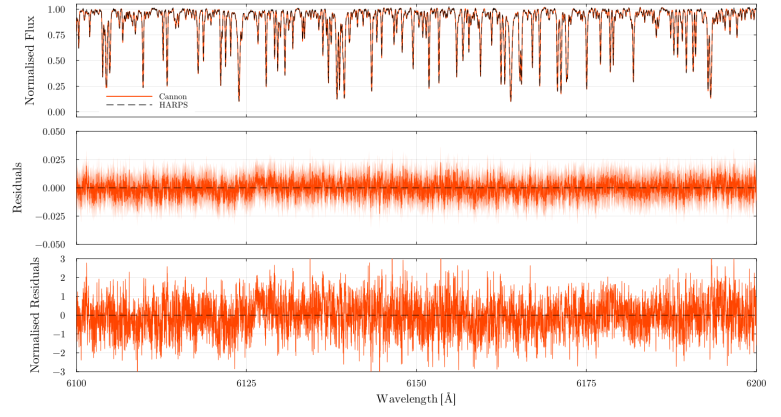}
    \caption{Format identical to Figure \ref{fig:Example spectrum and cannon 1 red} but instead showing the spectra, residuals and normalised residuals of HD 176704 ($\teff = 4457$~K, $\logg = 2.38$~dex, [M/H] $= 0.22$~dex).}
    \label{fig:Example spectrum and cannon 3 red}
\end{figure*}

\begin{figure*}
	\includegraphics[width=\textwidth]{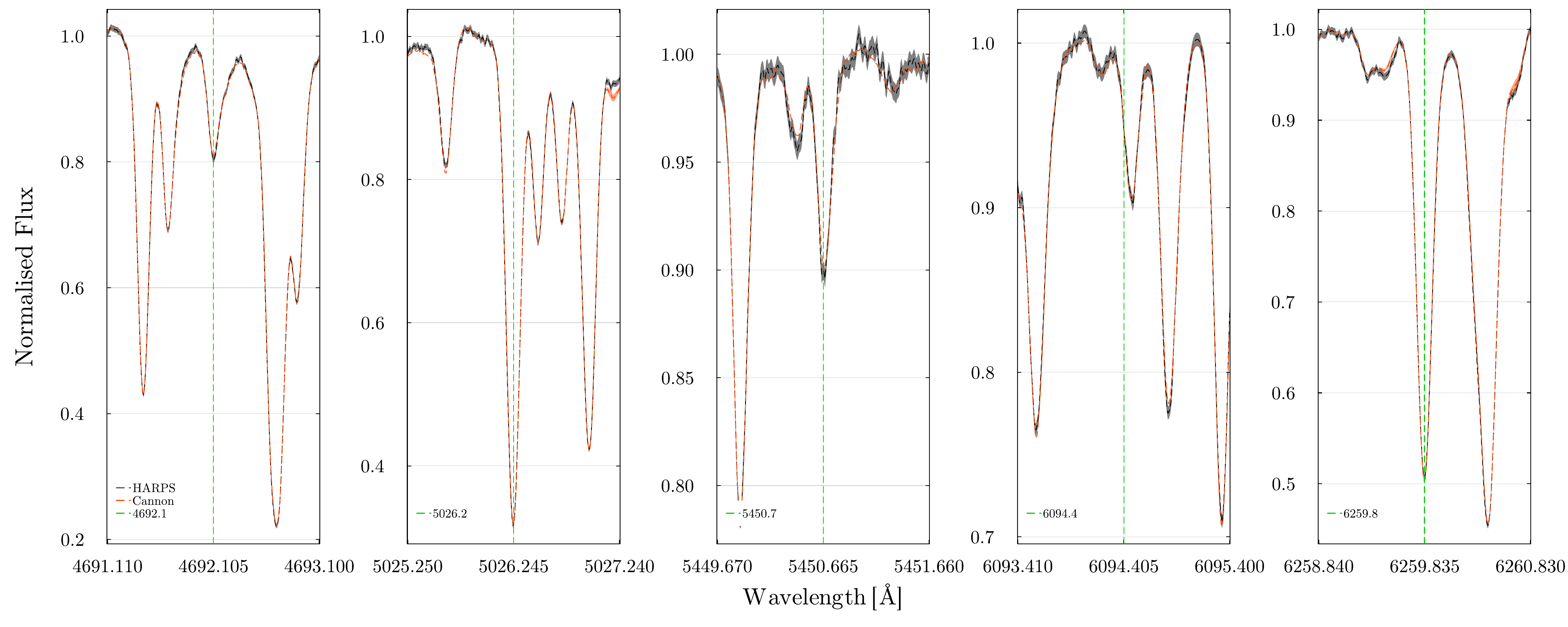}
    \caption{HARPS spectrum (black) and 8-label spectral reconstruction (red) around five Ti absorption lines in HD 17793. The green dashed line indicates the central wavelength of the line. Beneath each line are the flux residuals (data - reconstruction). Ti is one of our abundance labels in the 8-label model and should be reconstructed to a high fidelity.}
    \label{fig:Example spectrum Ti}
\end{figure*}

\begin{figure*}
	\includegraphics[width=\textwidth]{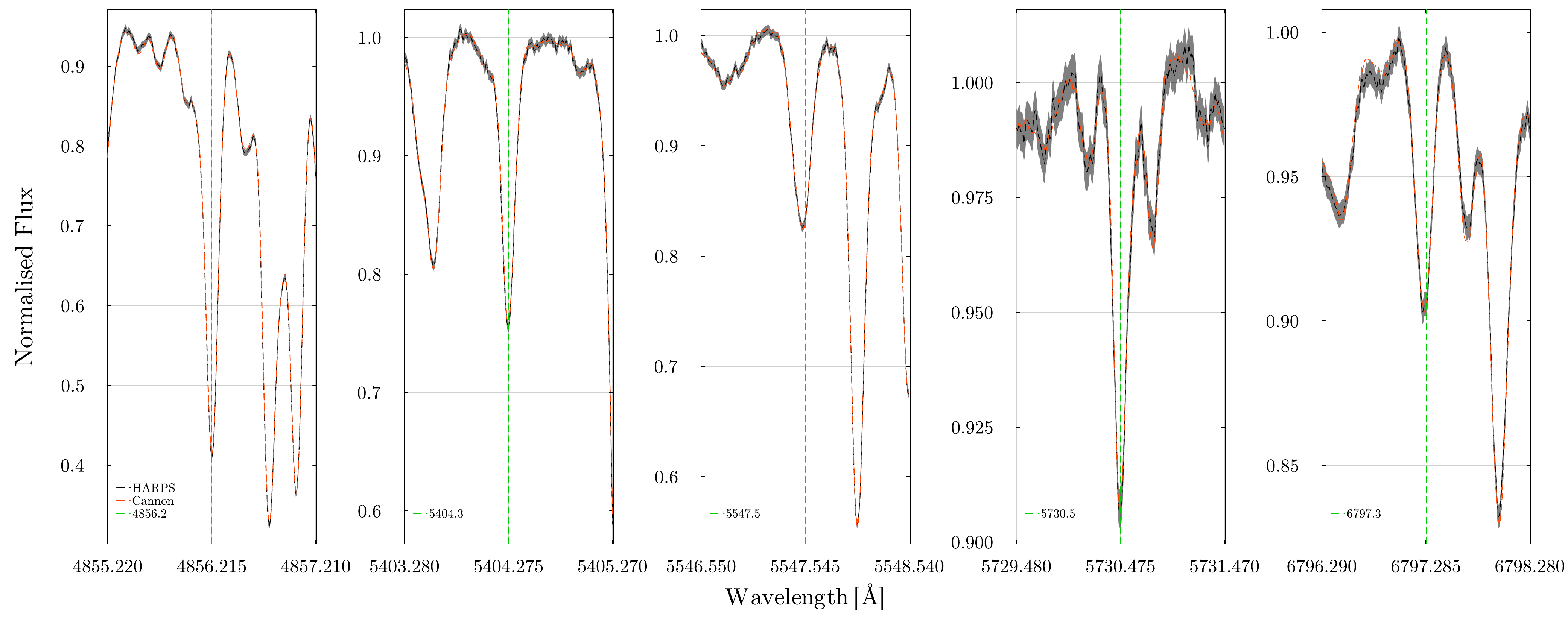}
    \caption{Format is identical to Figure \ref{fig:Example spectrum Ti} but instead demonstrates the high fidelity reconstruction for five Y lines.}
    \label{fig:Example spectrum Y II}
\end{figure*}

\begin{figure*}
	\includegraphics[width=\textwidth]{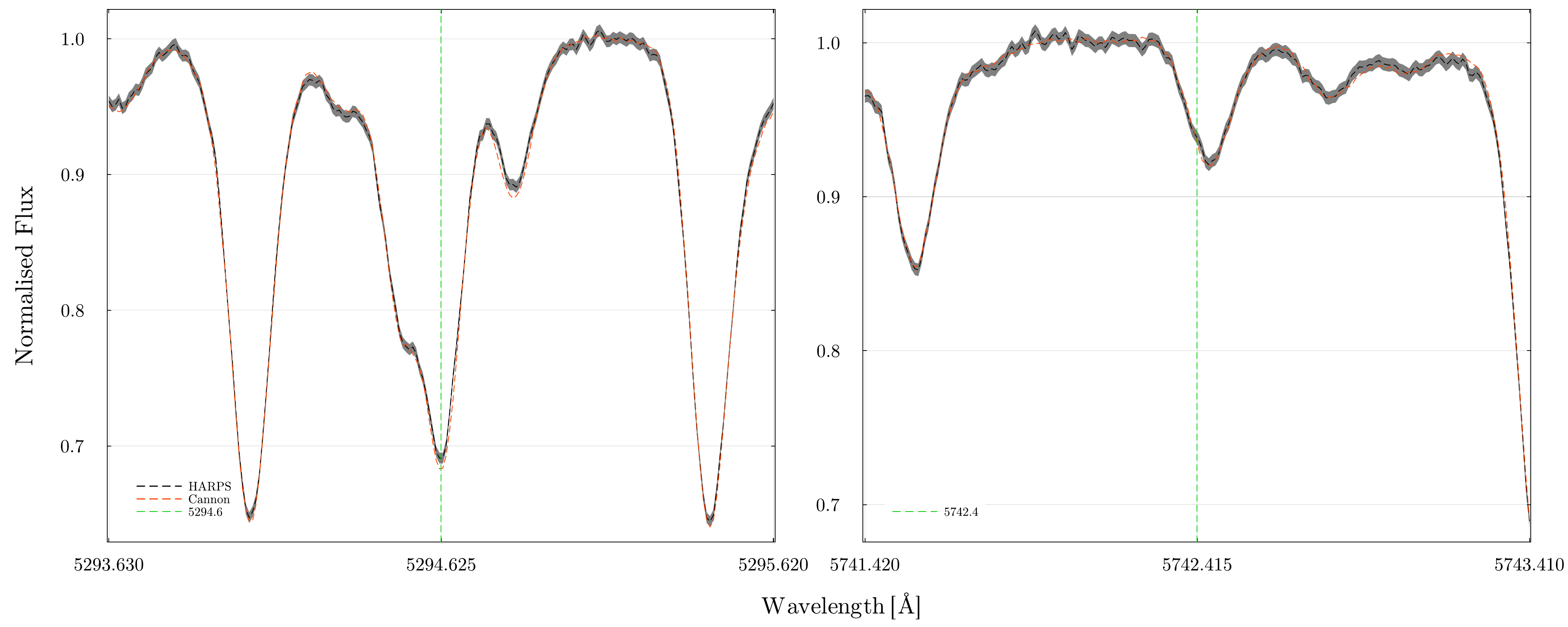}
    \caption{Format is identical to Figure \ref{fig:Example spectrum Ti} but instead demonstrates the high fidelity reconstruction for two Nd lines.}
    \label{fig:Example spectrum Nd II}
\end{figure*}

\noindent
\\\\
To demonstrate the model's fidelity around key nucleosynthetic features, Figures \ref{fig:Example spectrum Ti} through \ref{fig:Example spectrum Nd II} detail the reconstructions of additional absorption lines associated with our chosen abundance labels ([Ti/H], [Y/H], and [Nd/H]) for HD 17793.

\subsection{Linear Coefficients and Scatter}
\label{sec: results-linear coefficients}
\begin{figure*}
	\includegraphics[width=\textwidth]{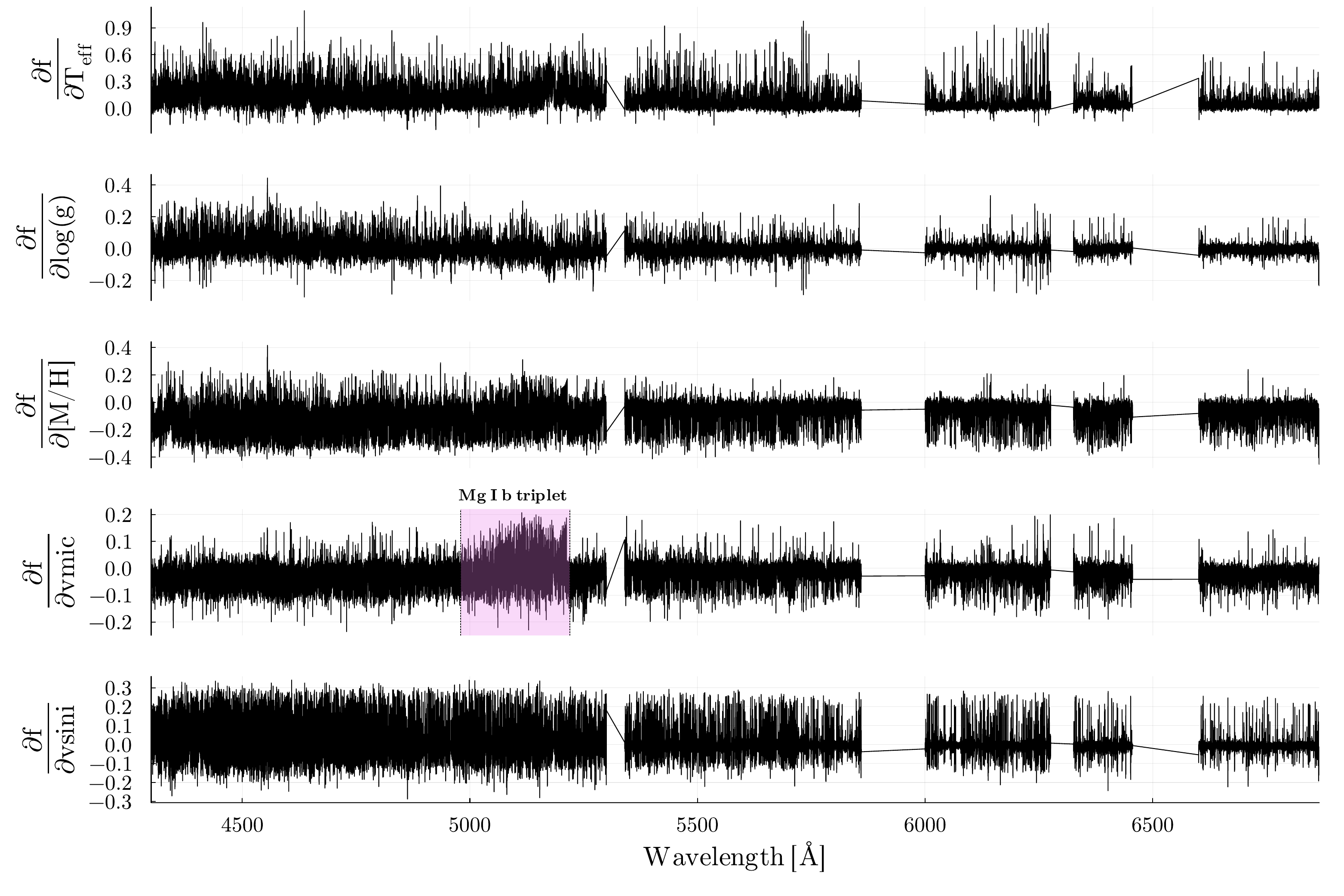}
    \caption{Linear coefficients of each stellar parameter label (\teff, \logg, [M/H],\vmic \& \vsini) in the 8-label \textsc{cannon} model. Peaks and troughs indicate correlations between specific abundances and atomic features. 
    Features located at these peaks and troughs that are associated with the corresponding labels (eg., strong lines that are known to be impacted by \vmic) are shaded in purple. Regions of no variation correspond to removed telluric regions and the HARPS CCD gap. Note that all wavelengths are in the vacuum rest frame.}
    \label{fig:linear coefficients params}
\end{figure*}

\begin{figure*}
	\includegraphics[width=\textwidth]{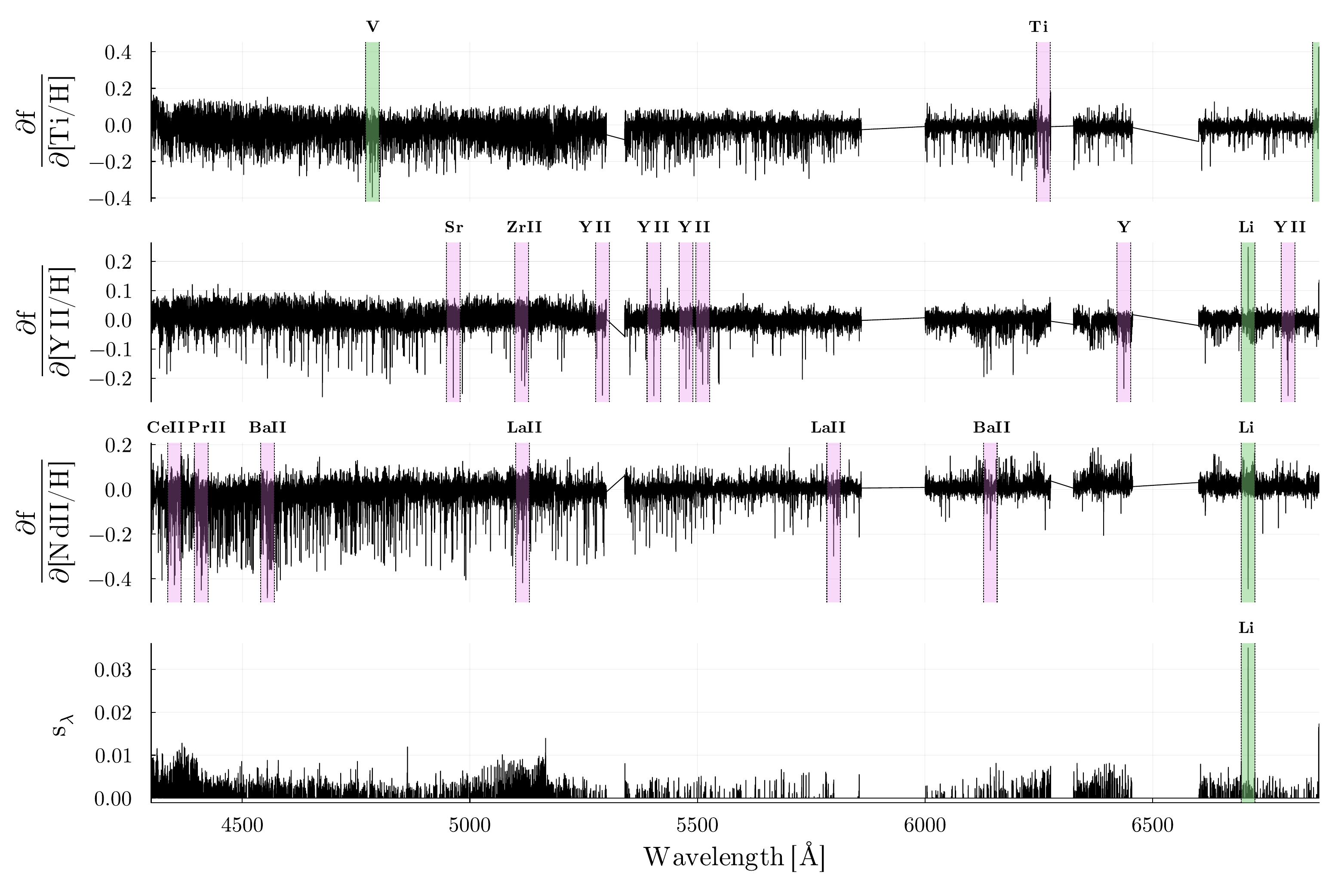}
    \caption{Linear coefficients of each abundance label ([Ti/H], [Y/H] \& [Nd/H]) as well as the scatter in the 8-label \textsc{cannon} model. Peaks and troughs indicate correlations between specific abundances and atomic features. Features located at these peaks and troughs that correspond to elements within the same nucleosynthetic family as each label are shaded in purple. Features with high coefficients that are not associated with the corresponding label are shaded in green. These features likely reflect shared systematics captured by the model within that label. Regions of no variation correspond to removed telluric regions and the HARPS CCD gap. Note that all wavelengths are in the vacuum rest frame.}
    \label{fig:linear coefficients X/H}
\end{figure*}

The linear coefficients of \textsc{the cannon} shown in Figures \ref{fig:linear coefficients params}, \ref{fig:linear coefficients X/H} and \ref{fig:linear coeff Eu II} reveal the sensitivity of the flux to each label in the 8- and 9-label models. This highlights learned correlations between specific abundances and atomic features in the vacuum rest frame. Peaks in the model scatter term in Figure \ref{fig:linear coefficients X/H} identify wavelengths (in the vacuum rest frame) where the current quadratic label set is insufficient to describe the spectral variance, pointing to missing physics or unconstrained chemical species. 

\subsubsection{8-label Model Coefficients}
The linear coefficients for the stellar parameters (\teff, \logg, [M/H], \vmic \& \vsini) exhibit widespread structure across the spectrum in Figure \ref{fig:linear coefficients params}. This indicates that these fundamental parameters influence nearly all spectral regions. Notably, the \vmic coefficient displays a significant increase in amplitude around the Mg b triplet. We shade this region in Figure \ref{fig:linear coefficients params}. Microturbulence alters line profiles, distributing the absorption over a wider wavelength range \citep[][]{palumbo_2026}. This effect is most pronounced in strong lines situated on the flat portion of the curve of growth \citep[][]{Gray_2005}, such as the Mg b triplet.
\\\\
The abundance linear coefficients in Figure \ref{fig:linear coefficients X/H} show localised peaks and troughs corresponding to specific atomic transitions. Absorption features with high coefficients that are associated with the corresponding labels  (e.g., troughs around Ti lines in the Ti coefficient) are shaded in purple in Figure \ref{fig:linear coefficients X/H}. However, note that while correlations may be physical (ie., tracing element production) they could also be caused by learned systematics (e.g., from incorrect ab initio stellar models used to derive the training labels or tellurics). Absorption features that have high linear coefficients but are not associated with the abundance label are shaded in green in Figure \ref{fig:linear coefficients X/H}. We detect distinct features in the [Y/H] and [Nd/H] coefficient vector around light s-process (Sr, Zr \& Y) and heavy s-process element (Ce, Ba, La \& Pr) absorption features, respectively. This demonstrates that \textsc{the cannon} is learning information directly from absorption features associated with each specific element label. We do also observe large variations in the linear coefficients of [Y/H] and [Nd/H] around the Li doublet at 6709.5~\AA{}. The peak and trough around Li have been shaded in green in Figure \ref{fig:linear coefficients X/H}, to highlight that the Li line depth is positively correlated with Y and anti-correlated with Nd, respectively. As Li is primarily produced by intermediate-mass AGB stars after the initial big bang nucleosynthesis \citep[][]{vanraai_2012,wang_2025}, Y by low-mass and intermediate-mass AGB stars, and Nd primarily by low-mass AGB stars \citep[][]{contursi_2024}, these correlations likely reflect the mass dependence of the progenitor stars responsible for their production.
\\\\
Compared to the other abundance labels, the [Ti/H] linear coefficient vector displays fewer prominent features across our wavelength region. This relative lack of prominent features may be due to the density of $\alpha$-element lines broadly distributed throughout the spectrum. However, we do note a trough in the [Ti/H] coefficients around a V line at 4785.8~\AA{} and a peak at 6785.9~\AA{}. These two lines have been shaded in green in Figure \ref{fig:linear coefficients X/H} and a Ti line shaded in purple. The trough around the V line may be reflective of the Type Ia supernovae contributions to Ti production, linking our [Ti/H] to the production of some iron-peak elements as well. The peak at 6785.9~\AA{} is likely a non-physical artifact since this coincides with the edge of our wavelength grid following the masking of telluric regions. 

\subsubsection{9-label Model Coefficients}

\begin{figure}
	\includegraphics[width=\columnwidth]{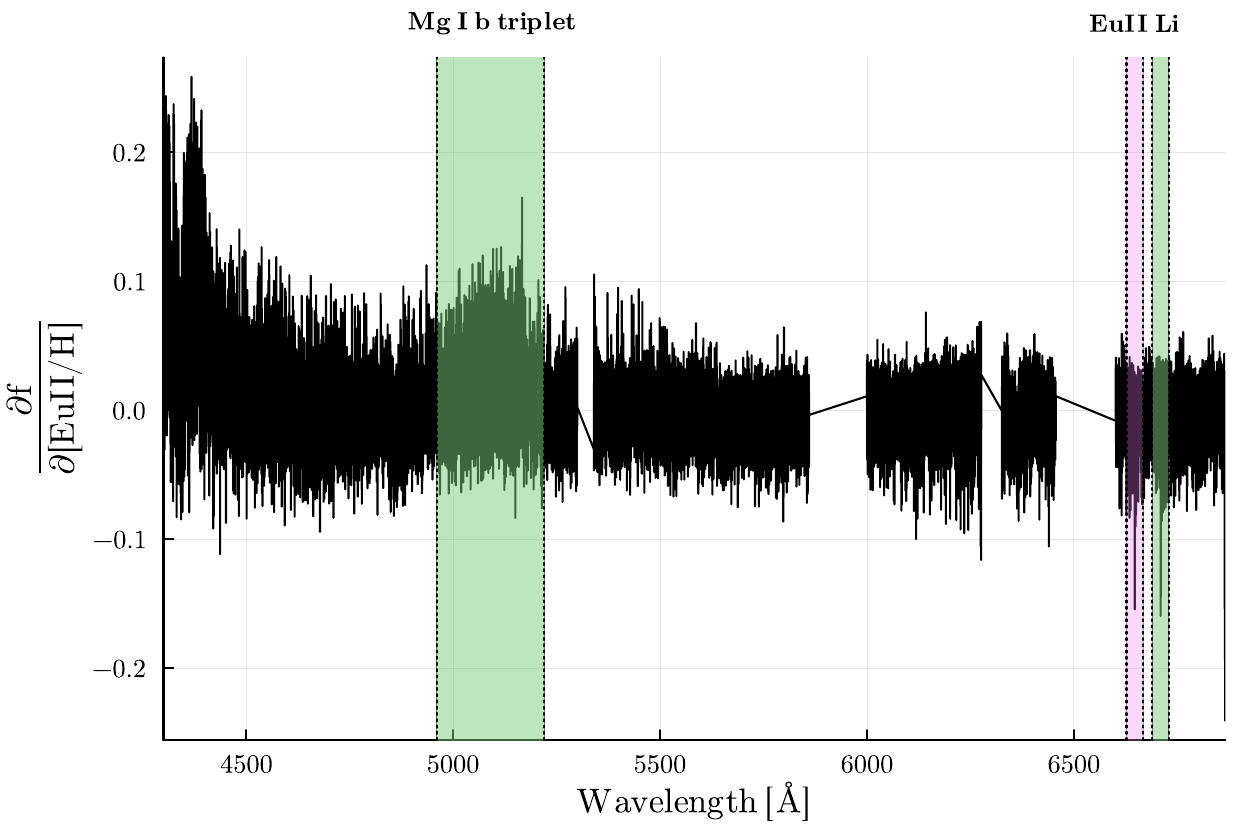}
    \caption{Eu linear coefficient in the 9-label \textsc{cannon} model. Unmodelled systematics around molecular features and the Mg b I triplet appear in the [Eu/H] coefficient in the 9-label model. The Mg b triplet and the Li doublet have been shaded in green, since it is information not associated with our Eu label. The Eu line at 6646~\AA{} is shaded in purple to show that the model is learning information from features we expect to be captured by the Eu label. Regions of no variation correspond to removed telluric regions and the HARPS CCD gap. Note that all wavelengths are in the vacuum rest frame.}
    \label{fig:linear coeff Eu II}
\end{figure}

The addition of [Eu/H] as an abundance label significantly alters the internal structure of the model's coefficients, and shows minimal changes on the generative model line recovery for features other than that of Eu (see Section \ref{sec: discussion-Eu II label}). For example, the strong \logg coefficient peaks observed at Ba absorption lines in the 8-label model are reduced by up to 42\%. We see prominent peaks in the [Eu/H] coefficient around molecular features below 4500~\AA{} and MgH bands surrounding the Mg b triplet in Figure \ref{fig:linear coeff Eu II}. The Mg b triplet and the Li doublet have been shaded in green, and the Eu line at 6646~\AA{} shaded in purple. This implies that while \textsc{the cannon} is learning information from the absorption features associated with our Eu label, it has also found correlations with features not associated with the [Eu/H] abundance label. The minimal predictive power of Eu in reconstructing the absorption features of other elements indicates that it is largely decoupled from other elements, as previously shown by \citet{mead_regression_2025}.

\section{Residual Correlations}

\begin{figure*}
	\includegraphics[width=\textwidth, height=\textheight, keepaspectratio]{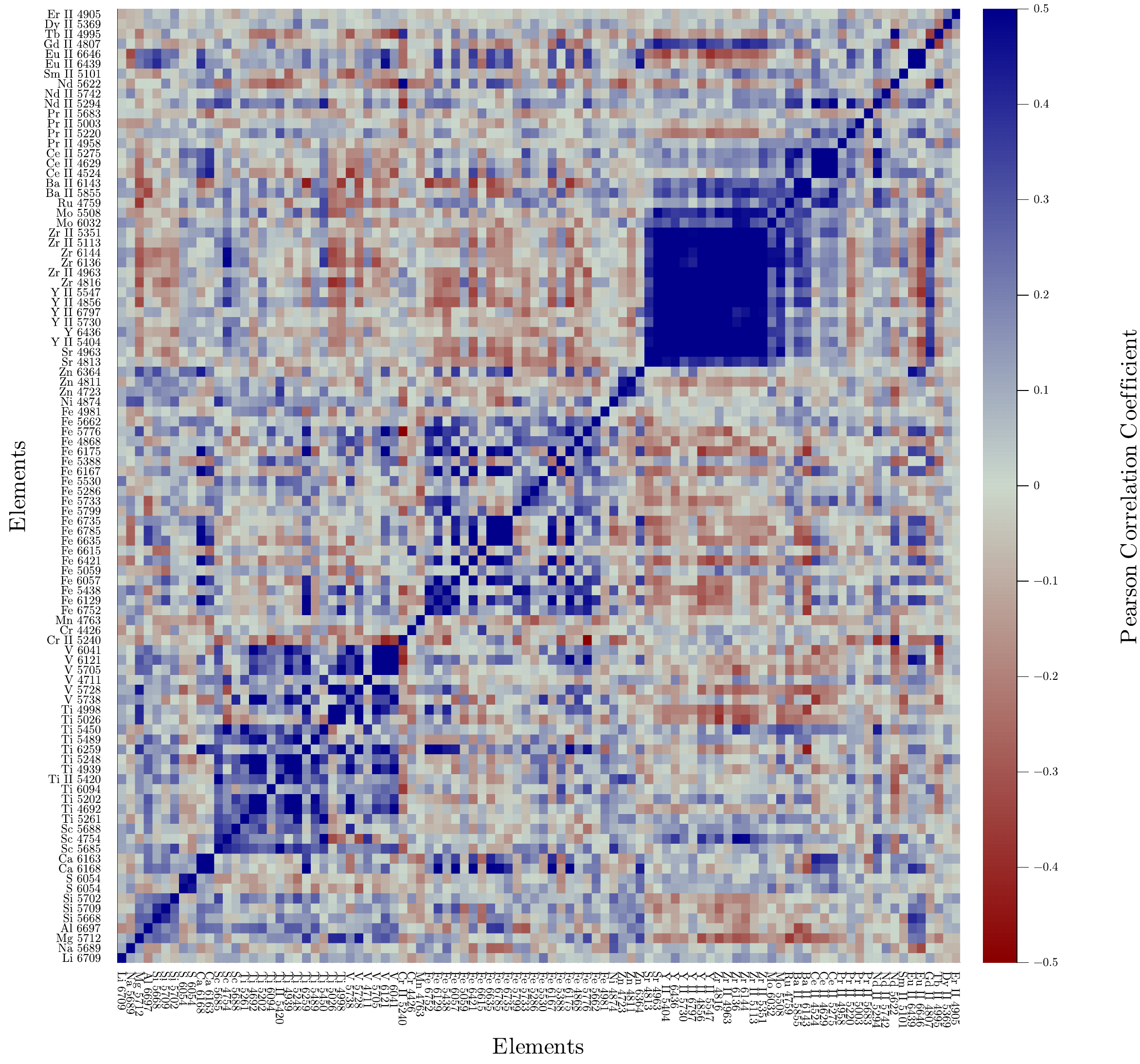}
    \caption{Correlations between the flux residuals of 95 absorption features (from 34 elements and ionisation states) for 220 test stars. Flux residuals are those between the HARPS spectra and 8-label \textsc{cannon} model.}
    \label{fig:residual correlations}
\end{figure*}

In Section \ref{sec:residual correlations}, we discuss the correlations and anti-correlations between the flux residuals of 95 absorption features that span 34 elements/ionisation states. The correlations between lines of the same elements as well as lines of elements within the same nucleosynthetic family are shown in Figure \ref{fig:residual correlations}. In Figure \ref{fig:residual correlations}, we also note correlations/anti-correlations between r-process elements (Eu, Gd and Tb) and light s-process elements (Sr, Y and Zr).

\bsp	
\label{lastpage}
\end{document}